\documentclass[botnum, fleqn, final]{unmeethesis}
\pdfoutput=1

\usepackage{amsmath,amsfonts,amssymb,amstext}
\usepackage{graphicx}
\usepackage[bookmarks = true, pdfpagemode = None, pdfstartview = FitH, colorlinks = true, urlcolor = blue,hyperfootnotes=false]{hyperref}
\usepackage{hypcap}
\usepackage{multirow}
\usepackage{listings} 

\begin{document}
\frontmatter

\title{Dispersion, controlled dispersion, and three applications}

\author{Douglas H. Bradshaw}

\degreesubject{Ph.D., Optical Science and Engineering}

\degree{Doctorate of Philosophy \\ Optical Science and Engineering}

\documenttype{Dissertation}

\previousdegrees{B.S., Applied Physics, Brigham Young University, 2000 \\
                 M.A., Biophysics, Johns Hopkins University, 2002}

\date{May, 2010}

\maketitle

\makecopyright

\begin{dedication}

To Gina
   
\end{dedication}

\begin{acknowledgments}

I have had the good fortune of learning from and studying under four exceptional mentors. Under Rudolph Wolfgang, I learned about the nature of optical pulses, about spectrometers and the power of seeing the world in reciprocal space.  Under Ivan Deutsch, I learned some of the fundamentals of quantum mechanics and quantum optics.  Under Michael Di Rosa, I benefited from a rare combination of freedom and involvement, along with encouragement to take my ideas and run and make my own mistakes.  Finally, from Peter Milonni I have been slowly learning what it means to be a theoretical physicist who focuses on optics.

Much of the most interesting physics for me happens not at a computer or in the lab but with a colleague near a chalk board.  I'd like to acknowledge many pleasurable and fruitful discussions with Denis Seletskiy at the University of New Mexico.  Our talks in the ``Center for Amature Studies'' form great memories around several interesting ideas.  I'd like to acknowledge many similarly engaging discussions at Los Alamos National Laboratory with Michael Di Rosa, John Mcguire, and Peter Milonni.

I've already mentioned Gina in my dedication, but she is the part of this whole experience that keeps on amazing me when I think about it.  Somehow, Gina, you had the courage to embark on this long journey with me and our little growing family.  I don't think that either of us realized going into this how long a Ph.D. can actually take.  But your patience, your faith, and your persistant encouragement continue to amaze me.

Finally, much of this dissertation was written while I was supported by a Los Alamos Laboratory Directed Research and Development grant.  I am genuinely grateful for this support.

 \end{acknowledgments}

\maketitleabstract

\begin{abstract}
Causality dictates that all physical media must be dispersive.  (We will call a medium \emph{dispersive} if its refractive index varies with frequency.)  Ordinarily, strong dispersion is accompanied either by strong absorption or strong gain.  However, over the past 15 years several groups have demonstrated that it is possible to have media that are both strongly dispersive and roughly transparent for some finite bandwidth.  In these media, group and phase velocities may differ from each other by many orders of magnitude and even by sign.  Relationships and intuitive models that are satisfactory when it is reasonable to neglect dispersion may then fail dramatically.

In this dissertation we analyze three such cases of failure.  Before looking at the specific cases, we review some basic ideas relating to dispersion.  We review some of the geometric meanings of group velocity, touch on the relationship between group velocity and causality, and give some examples of techniques by which the group velocity may be manipulated.  We describe the interplay between group velocity and energy density for non-absorbing dispersive media.  We discuss the ideas of temporary absorption and emission as dictated by an instantaneous spectrum.  We then apply these concepts in three specific areas.

First, non-dispersive formulations for the momentum of light in a medium must be adjusted to account for dispersion.  For over 100 years, there has been a gradual discussion of the proper form for the per-photon momentum.  Two forms, each of which has experimental relevance in a `dispersionless' medium, are the Abraham momentum, and the Minkowski momentum.  If $\omega$ is the angular frequency, $n$ is the refractive index, $\hbar$ is Planck's constant, and $c$ is the speed of light, then these reduce in a dispersionless medium to per-photon momenta of $\hbar \omega/(nc)$, and $n\hbar \omega/c$ respectively.  A simple generalization of the two momenta to dispersive media entails multiplying each per-photon momentum by $n/n_g$, where $n_g$ is the group refractive index.  The resulting forms are experimentally relevant for the case of the Abraham momentum, but not for the Minkowski momentum.  We show how dispersion modulates the displacement of a sphere embedded in a dispersive medium by a pulse.

Second, pulse transformation in a nonstationary medium is modulated by the presence of dispersion.  Dispersion may enhance or mitigate the frequency response of a pulse to a changing refractive index, and if dispersion changes with time, the pulse bandwidth must change in a compensatory fashion.  We introduce an explicit description of the kinetics of dispersive nonstationary inhomogeneous media.  Using this description, we show how the group velocity can modulate the frequency response to a change in the refractive index and how Doppler shifts may become large in a dispersive medium as the velocity of the Doppler shifting surface approaches the group velocity.  We explain a simple way to use existing technology to either compress or decompress a given pulse, changing its bandwidth and spatial extent by several orders of magnitude while otherwise preserving its envelope shape.  We then introduce a dynamic descriptions of two simple media--one dispersive and one nondispersive.  We compare the transformation of basic quantities like photon number, momentum density, and frequency by a temporal change in the refractive index in a specific non-dispersive medium to those wrought by a temporal change in the group refractive index in a specific dispersive medium. The differences between to media are fundamental and emphasize the salience of dispersion in the study of nonstationary media.

Finally, we note that the nature of a single optical cavity quasimode depends on intracavity dispersion.  We show that the quantum field noise associated with a single cavity mode may be modulated by dispersion.  For a well-chosen mode in a high-Q cavity, this can amount to either an increase or a decrease in total vacuum field energy by several orders of magnitude.  We focus on the ``white light cavity,'' showing that the quantum noise of an ideal white light cavity diverges as the cavity finesse improves.

\clearpage 
\end{abstract}

\tableofcontents
\listoffigures
\listoftables

\begin{glossary}{Longest  string}
\item[$\alpha$]
Absorption coefficient (m$^{-1}$)
\item[$c$]
Vacuum speed of light ($299792458$ m/s)
\item[$k$]
Angular wave number ($\omega n(\omega)/c$)
\item[$L$]
Round trip length of an optical cavity
\item[$n$]
Real refractive index ($Re[\sqrt{\epsilon_r \mu_r}]$)
\item[$n_g$]
Group refractive index or just group index ($c/v_g$)
\item[$\vec r$]
Position
\item[$t$]
Time
\item[$u$]
Energy density
\item[$V_{eff}$]
Effective mode volume
\item[$v_b$]
Beat velocity (approaches $v_g$ as spectral distance between beating components approaches zero)
\item[$v_g$]
Group velocity ($dk/d\omega$)
\item[$\epsilon_0$]
Permittivity of free space ($1/(c^2 \mu_0)$)
\item[$\epsilon_r$]
Relative permittivity ($\epsilon/\epsilon_0$)
\item[$\lambda$]
Wavelength in a particular medium ($c/(n \nu)$)
\item[$\lambda_0$]
Vacuum wavelength ($c/\nu$)
\item[$\mu_0$]
Permeability of free space ($4\pi \times 10^{-7}$N/A$^2$)
\item[$\mu_r$]
Relative permeability ($\mu/\mu_0$)
\item[$\nu$]
Frequency ($\omega/(2 \pi)$)
\item[$\sigma_0$]
Vacuum wave number
\item[$\omega$]
Angular frequency (real and positive, unless otherwise specified)

\end{glossary}

\mainmatter

\chapter{Introduction}\label{ch:introduction-and-outline}
The concept of group velocity has an interesting role in physics.  On the one hand, its importance has been appreciated for a well over a century.  For example, group velocity was lucidly described by Lord Rayleigh as early as the late 1800s \cite{Rayleigh-1877, Rayleigh-1881}.  In a letter to \emph{Nature} entitled ``On the velocity of light'' he explained \cite{Rayleigh-1881}, 
\begin{quote}
 It is evident however that in the case of light, or even of sound, we have no means of identifying a particular wave so as to determine its rate of progress.  What we really do in most cases is to impress some peculiarity, it may be of intensity, or of wave-length, or of polarization, upon a part of an otherwise continuous train of waves, and determine the velocity at which this \emph{peculiarity} travels.  Thus in the experiments of Fizeau and Cornu, as well as in those of Young and Forbes, the light is rendered intermittent by the action of a toothed wheel; and the result is the velocity of the group of waves, and not necessarily the velocity of an individual wave.
\end{quote}
Group velocity also played a critical role in de Broglie's theory of matter waves \cite{deBroglie-1924}.  While the phase velocities of de Broglie waves associated with a moving object are superluminal, de Broglie showed that the group velocity of the waves matches the velocity of the object, relating the group velocity to the propagation of mass/energy through space.  The important role that group velocity and group index play in the performance of spatial interferometers was also recognized generations ago \cite{Candler-1946}.

On the other hand, differences between the group velocity and the phase velocity are often neglected.  In optics, this is often reasonably practical.  Strong dispersion is generally accompanied by strong absorption or strong gain, while transparent media tend to be only weakly dispersive.  Since the bulk of optical experiment and theory (sensibly) involves light traveling through roughly transparent media, dispersion is often negligible.  We can illustrate this with a numerical example.  Suppose that we are concerned with an etalon or an optical cavity that consists of BK7.  At a vacuum wavelength of 633 nm, BK7 has a refractive index of 1.515 and a group refractive index of 1.537.  The formula for the free spectral range represented in some standard introductory optics texts (see, for example, Verdeyen or Guenther \cite{Verdeyen-1994, Guenther-1990}) is $\Delta \nu_{FSR}=c/(2 n d)$, where $n$ is the refractive index and $2d$ is the round-trip cavity length.  A formula that takes first order dispersion into account (but in its turn neglects higher orders of dispersion) would be $\Delta \nu_{FSR}=c/(2 n_g d)$, where $n_g$ is the group index.  For BK7 at 633 nm, the expressions differ from each other by 1.4\%.
From this perspective, several results of the last decade or so are interesting because they combine strong dispersion with reasonably small absorption for select, narrow bandwidths.  In 1999, Hau, Harris, Dutton, and Behroozi reported that they had succeeded in slowing the propagation of a pulse of light to a mere $17$ m/s \cite{Hau-Harris-Dutton-Behroozi-1999}.  If this medium could be placed in a cavity then neglect of dispersion could lead to a larger error in the estimate of the free spectral range than in the case of BK7--instead of an error of 1.4\% it would lead to an estimate that is too large by a factor of over 17 million for the relevant wavelengths.  

The 1999 demonstration of slow light in a weakly absorbing media was just one in a string of experiments showing increasing control over dispersion.  In 2000, Wang, Kuzmich, and Dogariu demonstrated nearly lossless pulse propagation through a medium with negative group velocity, implying that the peak of a pulse left the far side of the negative group velocity medium before the incoming pulse peak actually entered the medium \cite{Wang-Kuzmich-Dogariu-2000}.  This corresponds to a negative group index.  Finally, in 2001 Liu, Dutton, Behroozi, and Hau reported that they had completely stopped a pulse of light and then had recovered it and sent it on its way  \cite{Liu-Dutton-Behroozi-Hau-2001}.  This corresponds to ``stopped'' light that is later ``started'' again.  It not only demonstrated that the group velocity could be brought to $0$, but that it could be controlled dynamically while a pulse was propagating through the medium.  All three of these feats, those of slow, fast, and stopped light, were performed in the absence of substantial absorption or gain or pulse deformation.  Together, they represent newly opened physical regimes where dispersive effects in roughly transparent media are not only non-negligible, but may dominate the answers to particular questions.  The opportunity represented by the opening up of these new regimes is a primary motivation for this dissertation.

\section{Dissertation outline}

In this dissertation we seek to do two things.  First, we lay a foundation for understanding the physics of dispersion.  In Chapter \ref{ch:group-velocity}, we review the definition of group velocity.  We look at some of the geometric meanings of group velocity.  We distinguish the group velocity from the phase velocity, the signal velocity, and the energy velocity.    We also briefly review some of the work that has been done on the relationship between anomalous dispersion and causality.  Following this we give a brief review of some of the techniques whereby strong dispersion with minimal gain and/or absorption is brought about.  In Chapter \ref{ch:energy-density}, we look at the interplay between energy density and dispersion.  Following the work of Peatross et al. on the role of the instantaneous spectrum in pulse propagation through dispersive media \cite{Peatross-Ware-Glasgow-2001}, we also explore the relationship between dispersion and temporary absorption and emission in ``absorptionless'' media.

Second, we apply these ideas to three specific questions.  The relationship between dispersion and absorption requires that dispersion have an impact on momentum exchange between field and medium. In Chapter \ref{ch:momentum}, we explore the relationship between dispersion and photon momentum, specifying how the nondispersive forms for the Abraham and Minkowski momentum densities are altered when dispersion is taken into account. We predict that by modulating the group velocity of a pulse in a medium it should be possible to modulate the displacement of a small particle embedded in that medium.  

One interesting aspect of controllably dispersive media is that in many cases the optical properties of a medium may be altered while a wave is propagating through it.  This suggests that it may be profitable to look at controllably dispersive media from the perspective of nonstationary electromagnetics.  In Chapter \ref{ch:nonstationary}, we ask what effects dispersion can have on the transformations wrought upon a pulse traveling through a nonstationary medium.  We begin with a general kinetic approach based on the preservation of discrete translational symmetries.  Using this approach we find that dispersion may modulate the frequency response of a pulse to temporal changes in refractive index, and derive simple analytical expressions for Doppler shifts in dispersive media when group velocity dispersion may be neglected.  These show that dispersion may lead to large reflective Doppler shifts when the group velocity grows close to the velocity of a moving surface. Using the kinetic formalism, we also show how temporal and spatial control of the group velocity of a medium may be combined to scale pulse duration (and so pulse bandwidth and longitudinal extent) over many orders of magnitude without otherwise altering the pulse envelope.  This explains and generalizes a recent numerical proposal for the compression of pulses using magnetized dispersive nonstationary plasmas \cite{Avitzour-Shvets-2008}.  We then supplement these general considerations with a comparison of two specific transformations, one by a temporal change in the refractive index in a non-dispersive medium, and the other by a temporal change in the group index in a dispersive medium.  In order to make a full comparison, we introduce boundary conditions for each transformation.  For the non-dispersive case, the boundary conditions are taken from a classic work by Morgenthaler \cite{Morgenthaler-1958}.  For the dispersive case, we use an idealized version of the boundary conditions that were apparent in the stopped-light experiment of Liu, Dutton, Behroozi, and Hau \cite{Liu-Dutton-Behroozi-Hau-2001}.  We compare the effect of these two different transformations on 20 different fundamental quantities associated with a propagating pulse.

In Chapter \ref{ch:cavity}, we address the effect of anomalous intracavity dispersion on the quantum field noise associated with a single (pseudo-)mode in an optical cavity.
We do so using two approaches.  The first approach follows earlier work by Drummond \cite{Drummond-1990} and Milonni \cite{Milonni-1995}.  In this approach we assume a lossless dispersive cavity and show via the classical dispersive energy density that the electromagnetic field strengths all scale with the square root of the group velocity.  Associating a harmonic oscillator which each mode leads to noise terms whose associated fields must similarly scale with the square root of the group velocity.  When the dispersion is anomalous, the classical expression for the energy density implies that the energy associated with the fields alone is larger than the total modal energy.  This is because the total modal energy also includes a term that represents temporary absorption by the medium.  In lossless, anomalously dispersive media this term is negative, implying temporary gain, in accordance with the dictates of the instantaneous spectrum.  In practical terms, this means that the medium energy is negative because it has temporarily donated energy to the electric field.  When the average group index approaches zero as in the ``white light cavity'' \cite{Wicht-Danzmann-Rinkleff-1997}, the total classical energy density goes to zero unless the field energy goes to infinity.  Thus the field energy associated with even the vacuum state of such a cavity diverges.  These considerations motivate us to calculate the quantum field noise of anomalously dispersive cavities, and particularly of white light cavities, from a second perspective.  We use a simple approach where the pseudo-modes of an open cavity may be probed by true modes or ``modes-of-the-universe \cite{Lang-Scully-Lamb-1973}.''  This second approach allows us to corroborate the results of the first without relying on an expression for the electromagnetic energy density of an anomalously dispersive medium.  Because this approach allows for an open cavity, it not only reveals the amount of vacuum field noise associated with a particular resonance but also shows the effect of dispersion on spectral distribution of that noise.  We first show analytically that the second approach leads to the same conclusion as the first approach if we assume perfect finesse.  We then relax this assumption and show numerically in a specific physical model how the quantum field noise depends on finesse.  The divergence associated with a white light cavity for the case of infinite finesse is relaxed when reasonable physical assumptions are made.  However, the quantum field noise associated with an anomalously dispersive cavity mode remains substantially larger than that of the corresponding evacuated cavity.

These last three paper-like chapters represent three papers.  Chapter \ref{ch:momentum}, on momentum, is taken from a paper \cite{Bradshaw-Shi-Boyd-Milonni-2010} that I coauthored with Zhimin Shi and Robert Boyd (University of Rochester), and with  Peter Milonni (Los Alamos and University of Rochester), who was the primary author.  Professor Boyd's group is currently planning experiments to test some aspects of the theory developed in that paper, which has been published in a special issue of \emph{Optics Communications} in memory of Krzysztof Wodkiewicz.  The work in this paper has been cited in two manuscripts by Rodney Loudon, FRS, and Stephen
Barnett, FRS, to be submitted for publication in \emph{Physical Review Letters} and \emph{Proceedings of the Royal Society of London}.  Chapters \ref{ch:nonstationary}, on pulse transformations by a dispersive nonstationary medium, and \ref{ch:cavity}, on the interplay between dispersion and the vacuum field energy associated with a particular cavity mode, are papers which I have coauthored with Michael D. Di Rosa.  I am the primary author of these last two chapters.

\begin{table}
\begin{center}
\begin{tabular}{|c|c|c|}
\hline
 Journal Reference                            & Coauthors                       &  Chapter\\
\hline 

Optics Communications
Volume 283, Pages 650-656     & Z.~Shi, R.~W.~Boyd,             &  Ch.~\ref{ch:momentum}\\
                                              & and P.~W.~Milonni&\\
\hline

Ready for submission                                & M.~D.~Di Rosa                   &  Ch.~\ref{ch:nonstationary}      \\
\hline

Ready for submission				      & M.~D.~Di Rosa			&  Ch.~\ref{ch:cavity}\\
\hline
\end{tabular}
\caption[Table of published/prepared work.]{Table of published/prepared work with location in text.}
\end{center}
\end{table}
\chapter{Group velocity, the group refractive index, and controlled dispersion}\label{ch:group-velocity}
When we use the word ``dispersive'' to describe a medium, we mean that the refractive index of that medium varies with frequency.  Another way to say this is that in a dispersive medium the group velocity differs from the phase velocity and the group index differs from the refractive index.  In order to understand the implications of controlled dispersion, we need to understand the meaning of the group velocity and of its companion, the group index.

In a letter to Nature entitled ``On the velocity of light,'' Lord Rayleigh wrote \cite{Rayleigh-1881},
\begin{quote}
 I have investigated the general relation between the group velocity $U$ and the phase velocity $V$.  It appears that if $k$ be inversely proportional to the wavelength,
\[
U=\frac{d(kV)}{dk},
\]
and is identical with $V$ only when $V$ is independent of $k$, as has hitherto been supposed to be the case for light in vacuum.
\end{quote}

The definition of the group velocity employed by Lord Rayleigh in 1877 and 1881 is the one we use today.  Writing $k=n \omega/c$ and $V=c/n$, we write the group velocity in its more standard form,
\begin{equation}\label{Eq:vg-def}
 v_g=\frac{d\omega}{dk}.
\end{equation}

The group index is defined analogously to the phase index:
\begin{equation}
 n_g=\frac{c}{v_g}=c\frac{dk}{d\omega}.
\end{equation} 

The purpose of this chapter is to explore some of the fundamental ideas associated with the group velocity, the group index, and their control.

\section{Some geometric meanings of group velocity and the group index}
The definition of the group velocity as given in Eq.~(\ref{Eq:vg-def}) is really a geometrical definition.  It says that the group velocity is the ratio of infinitessimal changes in temporal periodicity to infinitessimal changes in spatial periodicity.  In this section we review some simple geometrical consequences of dispersion that have practical optical consequences.
\subsection{The group velocity is the velocity of fixed phase relationships between copropagating plane waves with infinitesimal spectral separation}
Electromagnetic disturbances propagating through a lossless, homogeneous, isotropic, linear medium may be decomposed as a summation of plane waves.  The behavior of the disturbance over time depends on the interference pattern formed by the waves, which is dictated by the relative phases between the many components at each point in space and time.  If two plane waves have phases $\phi_1=k_1 x-\omega_1 t+\theta_1$ and $\phi_2=k_2 x-\omega_2 t+\theta_2$, the relative phase may be taken to be $\phi_2-\phi_1=(k_2-k_1)x-(\omega_2-\omega_1)t+\theta_2-\theta_1$.  The velocity ($v_b$) of a beat between these two waves is simply the velocity of a plane of constant relative phase.  This velocity is given by 
\begin{equation}
 v_b=\frac{\omega_2-\omega_1}{k_2-k_1}.
\end{equation}
Taking the limit for the beat velocity as the waves grow closer to each other in temporal and spatial frequency yields the group velocity.  To the extent that the beat velocity is constant for the phase relationships between all of the salient frequency components of a pulse, the pulse will propagate without distortion.

\subsubsection{The velocity of fixed phase relationships between plane waves with finite spectral separation may be found through an averaged group index}\label{Sss:ngbar}
Interestingly, for finite frequency differences, the beat velocity is not that given by the spectral average of the group velocity between the two frequencies.  Rather, it is determined by the a spectral average over the group index, where the group index is defined as 
\begin{equation}
 n_g=c \frac{dk}{d\omega}.
\end{equation} 
We can see this by writing $k_2$ in terms of $k_1$ as 
\[
k_2=k_1+\int_{\omega_1}^{\omega_2}\frac{dk}{d\omega}d\omega=k_1+\int_{\omega_1}^{\omega_2}\frac{n_g(\omega)}{c} d\omega.
\]
Thus,
\begin{equation}\label{Eq:ngbar}
 k_2-k_1=(\omega_2-\omega_1)\frac{\overline n_g}{c},
\end{equation} 

where the line over $n_g$ is used to denote a spectral average, so that 

\begin{equation}\label{Eq:ngbar-def}
\overline n_g=\frac{\int_{\omega_1}^{\omega_2}n_g(\omega)d \omega}{\omega_2-\omega_1}.                                                                                       \end{equation} 

Thus, the beat velocity may be written in terms of the spectrally averaged group index as
\begin{equation}\label{Eq:vbeat}
 v_b=\frac{\omega_2-\omega_1}{k_2-k_1}=\frac{c}{\overline n_g}.
\end{equation}
This definition, based on a spectral average over all frequencies between the frequencies of interest, accounts for group velocity dispersion as well as all higher orders of dispersion.  If these terms are negligible for the bandwidth of interest, $\overline n_g$ may be replaced with $n_g$.

A pulse will propagate through a medium without distortion only to the extent that $v_b$ is similar for each pair of frequency components.

\subsubsection{Relating the velocity of fixed phase relationships to the velocity of pulse propagation in a lossless medium.}
Having established that $n_g$ and $v_g$ can be interpreted in terms of the speed of constant phase relationships, we now explore the relationship between the speed of a fixed phase relationship and the speed with which a pulse will travel through a medium.

Briefly, the velocity of a pulse is generally considered to be the velocity of its envelope.  The envelope of a pulse is defined by the phase relationships between the frequency components of that pulse.  Thus, the velocity of those phase relationships is the velocity of the pulse.

We now make this argument in more detail.  First, we introduce the group and phase velocities as limits of two velocities that naturally arise when different frequency components are summed.  In doing so, we find that although these velocities are, in practice, applied to pulses with finite bandwidths, they acquire exact meanings only over spectra sufficiently narrow that changes in $n$ (for the case of the phase velocity) and $n_g$ (for the case of the group velocity) can be neglected across the relevant range.  We then find that if these changes can be neglected then the pulse envelope is defined by the phase relationships of the frequency components of the pulse.  Thus, the speed of these phase relationships defines the speed of the pulse.

\paragraph{The phase and group velocities are only well-defined over frequencies ranges sufficiently narrow such that $n$ and $n_g$ can be considered to be constant.}
A pulse can be thought of as a sum of plane waves.
If we consider propagation in a single dimension, we can go further and consider a pulse in terms of a sum of frequency components. To begin to see how the group velocity works, it is sufficient to start with the very simple example of a sum of two sinusoids of equal amplitude.
We might represent such a sum as 
\begin{equation}
 E=\cos(\omega_1 t-k_1 x+\phi_1)+\cos(\omega_2 t -k_2 x+\phi_2).
\end{equation} 
We can rewrite this sum as a multiple of two sinusoidal terms by using the identity
\[
 \cos(A)+\cos(B)=2\cos\left(\frac{A+B}{2}\right)\cos\left(\frac{A-B}{2}\right).
\]
This gives
\begin{equation}
 E=2\cos\left(\bar\omega t-\bar k x+\bar \phi \right)
    \cos\left(\Delta \omega t-\Delta k x+\Delta \phi \right),
\end{equation}
where $\bar \omega=\left(\omega_2+\omega_1\right)/2$ and $\Delta\omega=\left(\omega_2-\omega_1\right)/2$ and similarly for the $k$ and $\phi$ terms.

If $\omega_2$ and $\omega_1$ are similar in size, to each other, the difference between them ($2 \Delta \omega$) is much smaller than their sum ($2 \bar \omega$).  The slowly varying term becomes an envelope gradually damping out and reviving the oscillation of the second term.

The velocity associated with a constant value for the rapidly oscillating term is $\bar \omega/\bar k$, while that associated with the slowly oscillating term is $\Delta \omega/\Delta k$.
Rewriting these terms by expanding them in terms of values for $\omega$ and $n$ gives
\begin{equation}
 v_f=\frac{\omega_2+\omega_1}{\omega_2 n_2/c +\omega_1 n_1/c}
\end{equation}
and 
\begin{equation}
 v_s=\frac{\omega_2-\omega_1}{\omega_2 n_2/c +\omega_1 n_1/c},
\end{equation}
where $v_f$ is the velocity of the fast oscillations and $v_s$ is the velocity of the slow oscillations.  While these expressions are related to the group velocity and the phase velocity, they are not exactly equal to them.  To relate them to the group and phase velocities, we must either make assumptions about $n(\omega)$ or about the frequencies.  Taking the limit of the two velocities as $(\omega_2-\omega_1)\rightarrow 0$ gives $v_f=c/n=v_p$ and $v_s=c/n_g=v_g$.

Here we see clearly that the group and the phase velocity are only well defined as functions of frequency.  Since any pulse must have components taken from a range of frequencies, these velocities as applied to that pulse are approximate to the extent that $n$ and $n_g$ change across those frequencies.

\paragraph{Changes in the envelope of spectrally narrow pulses in lossless media can come only from changes in phase relationships between its frequency components.}
A pulse can be thought of as a sum of sinusoids.  That is,
\begin{equation}
 E(t)=\int d\omega E(\omega) e^{-i \omega t}d\omega,
\end{equation} 
where $E(\omega)=|E(\omega)|e^{-i\phi(\omega)}$.  If the amplitudes are constant then only the phases change.  The absolute phase determines the position of the envelope as a whole.  The relative phases determine the shape of the envelope.

\subsubsection{The group index and the free spectral range of an optical cavity}
In Chapter \ref{ch:introduction-and-outline} we used the free spectral range of an optical cavity as an example of a case where dispersion is generally ignored but can have real consequences.  A commonly used form \cite{Verdeyen-1994,Guenther-1990} for the free spectral range is 
\begin{equation}\label{Eq:fsr-non}
 \Delta \nu_{FSRnon}=\frac{c}{2 L n},
\end{equation} 
where $2 L$ is the round trip cavity length.  We asserted that a more correct form for the free spectral range would be
\begin{equation}\label{Eq:fsr-dis}
 \Delta \nu_{FSRdis}=\frac{c}{2 L n_g},
\end{equation} 
but acknowledged that this is also an approximation that ignores higher order dispersion.

However, we are now in a position to argue that the difference between Eqs.~(\ref{Eq:fsr-non}) and ~(\ref{Eq:fsr-dis}) is more fundamental in nature than the difference between Eq.~{\ref{Eq:fsr-dis}} and forms that account for higher order dispersion.

The round trip phases associated with two neighboring cavity resonances must differ from each other by $2\pi$.  Thus,
\begin{equation}\label{Eq:2pi}
(k_{n+1}-k_{n}) (2 L)=2\pi.
\end{equation}
We can relate the difference in wave numbers in Eq.~(\ref{Eq:2pi}) to a difference in frequency using Eq.~(\ref{Eq:ngbar}), giving
\[
\omega_{n+1}-\omega_n=\frac{2 \pi c}{2 \overline n_g L},
\]
so that the free spectral range, $\Delta \nu_{FSR}$, is given by
\begin{equation}
 \Delta \nu_{FSR}=\frac{c}{2 \overline n_g L},
\end{equation} 
where $\overline n_g$ is given by Eq.~(\ref{Eq:ngbar-def}).  This expression is exact.  Thus, accounting for higher orders of dispersion leads to taking a spectral average of the group index between the two neighboring resonances.  On the other hand, $n_g$ has no fixed relationship with the spectral average of the refractive index between the two resonances.

In summary, there is a difference between the kind of approximation we make when we ignore dispersion and the kind of approximation we make when we ignore only group velocity dispersion and higher orders of dispersion.  If we ignore dispersion, the formula for the free spectral range becomes $\Delta \nu_{FSR} \approx c/2nd$.  Taking only first order dispersion into account, we get $\Delta \nu_{FSR} \approx c/2n_g d$.  Finally, we have just seen that if we account for all orders of dispersion we get $\Delta \nu_{FSR}=c/2 \overline n_g  d$.  As a cavity grows longer, the free spectral range becomes smaller and $\overline n_g \rightarrow n_g$ whether or not higher orders of dispersion are significant.  In a dispersive medium, $n_g$ will not make a similar approach to $n$ regardless of cavity length.  Intuitively, we can understand the more fundamental relationship between $n_g$ and the spectral separation between neighboring resonances in terms of the defintion of the group index, $n_g=c(dk/d\omega)$.  

\subsection{Group optical path length}
A pulse that propagates through a spatial interface from a medium with one group velocity to a medium with a second group velocity is scaled longitudinally \cite{Harris-Hau-1999} but not temporally (when a pulse crosses an interface with a temporal component, for example a moving interface, this is no longer true--see Chapter \ref{ch:nonstationary}).  If the pulse duration is given by $\tau_p$ and the longitudinal extent of the pulse by $l$ then (in the absence of distortion) the two quantities may be related by the group velocity according to $\tau_p=l/v_g=n_g l/c$.  Just as the pulse duration is conserved at a stationary interface, so is the product $n_g l$.

Following Candler \cite{Candler-1946}, we now extend the idea of the optical path length to define a group optical path length.
The optical path length associated with a distance $d$ in a homogeneous medium is just the distance multiplied by the appropriate refractive index,
\begin{equation}
 L_{OP}=n d.
\end{equation}
The phase accrued by a wave in traveling a distance $d$ in a medium with a refractive index $n$ is the same as that accrued by a wave of the same frequency traveling through a vacuum distance $nd$.

The group optical path length associated with a distance $d$ in a homogeneous medium is just that distance multiplied by the appropriate group index,
\begin{equation}
 L_{GOP}=n_g d.
\end{equation}
The difference between phases accrued by two infinitesimally different frequencies as they travel along a path with a group optical path length of $L_{GOP}$ is the same as the difference they would accrue if they traveled an actual distance of $L_{GOP}$ through the vacuum.  In this sense a milimeter of propagation through a slow light medium may be equivalent to a kilometer of propagation through vacuum if $n_g=10^6$.

\subsection{Group optical path length and spatial interferometers used as spectral filters}
Many interferometers that are used for spectral separation actually depend on wavelength.  Fabry-Perot interferometers, Young's double slit experiment, and diffractive gratings, for example, may all be analyzed in terms of wavelength.  If the mapping between wavelength and frequency is altered, the frequency response of such interferometers is altered while the wavelength response is unchanged.  To emphasize this distiction, we will refer to such interferometers as \emph{spatial} interferometers.  

The interference patterns of spatial interferometers depend directly upon wavelength and only indirectly upon frequency.  The spectral resolving power of a spatial interferometer comes from phase differences in optical path lengths for different colors.  That difference is related to the group optical path length:
\[
\frac{d(k L)}{d\omega}=L\frac{dk}{d\omega}=L \frac{n_g}{c}.
\]
Thus, as was pointed out by Candler \cite{Candler-1946}, the spectral resolving power of spatial filters is conveniently given in terms of the group optical path length.  We apply this idea to an optical cavity in greater detail in Chapter \ref{ch:cavity} and particularly in Appendix \ref{appen:Lg}.

A second way to understand the function of the group index in spatial interferometers is by thinking of the group index as a map between relative changes in wavelength and changes in frequency.  To see this more clearly, we can rewrite the group index as
\begin{equation}
 n_g=-n\frac{d \ln\lambda}{d\ln\omega}
\end{equation}  
(see Appendix \ref{appen:ng}).
In vacuum, $n_g=1$, and $n=1$, meaning that a .01\% increase in frequency leads to roughly a .01\% decrease in wavelength.  In a moderately slow medium, $n_g$ might be $100$ while $n$ remains at approximately $1$.  In this case a .01\% increase in frequency leads to roughly a 1\% decrease in wavelength.  Thus, with no change to the spatial resolution of an interferometer, its spectral resolution may be enhanced 100-fold.

\section{Group velocity, information velocity, energy velocity, and causality}
Lord Rayleigh argued that in measurements of the speed of light, it is not typically the phase velocity that is measured, but the velocity of a \emph{peculiarity} or a change impressed on that light \cite{Rayleigh-1877,Rayleigh-1881}.  That peculiarity would travel at the group velocity.  Later, in de Broglie's thesis on the wave nature of matter, de Broglie derived superluminal phase velocities for the component waves of a moving particle.  Referring to Rayleigh's work, he showed that although the phase velocities associated with his matter waves would be superluminal, the group velocity of those waves would match the velocity of the particle \cite{deBroglie-1924}.  

From Lord Rayleigh's work comes the idea that the group velocity may be used as a signal velocity.  From de Broglie's work we see that the group velocity can function as a velocity of energy transport.  However, equating the group velocity with either the signal velocity or the energy velocity becomes problematic when the group velocity exceeds the speed of light.  Superluminal group velocity occurs even in ordinary media; the group velocity at the spectral center of an absorbing resonance is generally superluminal (see Section \ref{se:controlled-dispersion}).

The subtleties associated with superluminal group velocity have been grounds for careful thought for over a century and still drive a large portion of the discussion of dispersive media; a valuable dissertation (or perhaps several) could still be written on the topic.  However, because the relationship between causality and dispersion is beyond the scope of this work, we make just a few points and refer the reader to \cite{Milonni-2004} for a more detailed introduction.

First, causality is built into the response functions of the media used to obtain superluminal group velocities.  Thus, the question is not whether causality is broken by such media--it cannot be.  Rather, the goal is to understand how causality is upheld.  

Second, the task of understanding how causality is upheld is not as simple as it may first seem because it requires that we develop precise definitions of quantities like the information velocity and the energy velocity.  In many practical situations the group velocity works perfectly well as a proxy for each of these.  However, the reality of absorptionless anomalous dispersion shows that both velocities are in general independent of the group velocity.

Third, the difficulty of the problem is in fact an opportunity to come to a better understanding of what we mean when we reference the energy velocity or the information velocity of a disturbance.  This is the lens through which the rest of the points in this section should be seen.

\subsection{Information velocity}
Around 1910, the question of causality in the presence of superluminal group velocity was the subject of several conferences.  Responding to the topic,
\begin{quote}
Sommerfeld demonstrated theoretically that the velocity of the
front of a square-shaped pulse propagating through any medium is
identically equal to c and hence relativistic causality is preserved. In
a follow-up study, Brillouin suggested that the group velocity is not
physically meaningful when the dispersion is anomalous because the
pulse becomes severely distorted \cite{Stenner-Gauthier-Neifeld-2004}.
\end{quote}
An English summary of this part of the history can be found in Brillouin's \emph{Wave propagation and group velocity} \cite{Brillouin-1960}.

Brillouin's suggestion that a pulse will become severely distorted as it propagates under anomalous dispersion is likely to be correct if the distance through which a pulse propagates under anomalous disperion is sufficiently long.  However, it is demonstrably incorrect for propagation over shorter distances.  Garrett and McCumber showed via simulations that a pulse passing through a finite anomalously dispersive medium may retain its overall shape while its average position advances faster than the speed of light for a limited distance \cite{Garrett-McCumber-1970}.  Later analysis and experiment have demonstrated superluminal pulse propagation without gross distortion in active media \cite{Chaio-1993,Steinberg-Chiao-1994,Bolda-Garrison-Chiao-1994,Wang-Kuzmich-Dogariu-2000,Stenner-Gauthier-Neifeld-2004}.  Pulse distortion does not therefore provide satisfactory explanation for the preservation of causality.

Sommerfeld's statement on the propagation of the front of a square-shaped pulse leads us to consider what it means to have information communicated by the electromagnetic field.  In his simple model, he associates that information with the moment at which a pulse is turned on.  The spectrum associated with this instant transition is infinite, and any physical response function must therefore dictate that the signal associated with this moment moves at $c$.

This concept has been generalized by Garrison et al \cite{Garrison-Mitchell-Chiao-Bolda-1998}, who propose that a signal is associated with a point of non-analyticity.  Because any such point will have an infinite spectrum, it will experience a total group index of $1$ and move at $c$.  If the information velocity can be associated with such points, then our problem is solved.

However, this definition of group velocity leaves us without a way to understand the superluminal transmission of a smooth pulse.  It may be impossible to generate a pulse that is smooth for all time because such a pulse would have an infinite extent.  However, it certainly seems possible to generate, for example, a Gaussian pulse that is smooth for that portion of the pulse which has a power over some minimum threshold.  Because of the fast decay of the wings of a Gaussian, that threshold could be very low without making the duration of the analytic region unreasonably long.  For such a pulse we may need a different definition for information velocity.

Kuzmich et al. \cite{Kuzmich-Dogariu-Chiao-2001} used signal-to-noise ratios to analyze the faster-than-light propagation associated with single photon pulses propagating through a non-lossy superluminal medium.  They showed that the speed of propagation of the photon as given by its expected arrival time at a detector could actually exceed $c$.  However, the signal velocity, defined operationally as using the signal to noise ratio, actually decreased because of the anomalously dispersive medium due to the effect of quantum fluctuations.

This slower signal velocity, as judged by the signal to noise ratio, was later verified in practice (see, for example, Stenner et al. \cite{Stenner-Gauthier-Neifeld-2004}).

\subsection{Energy velocity}

Just as causality prevents a signal from travel faster than the $c$, it is also prevents energy from traveling faster than the speed of light.  However, as we have just mentioned, Kuzmich et al. showed that the speed of propagation of the photon as given by its expected arrival time at a detector could indeed exceed $c$ \cite{Kuzmich-Dogariu-Chiao-2001}.

Evidently, the fact that energy may not propagate faster than $c$ does not mean that the energy \emph{associated with a particular wave} cannot travel faster than light if the wave is not a closed system.  That is, if energy is added to the front of a moving pulse and taken from the back of the pulse then the electromagnetic field energy associated with that pulse may have an average velocity that is faster than the speed of light.  This is allowed through energy exchange with the medium through which the light is traveling.  

An analogy is useful here.  There is a limit to how fast an orange may move around the globe . . . perhaps it is currently about the speed of a fast jet.  However, if your brother is in London and you live in Albuquerque, it is still possible for you to give him an orange faster than this so long as there are oranges in London.  You must simply call someone in London and ask them to give your brother an orange.  You can promise to send another orange to repay them following the next flight out of Albuquerque.

One interesting point with regard to this metaphor is that if you are to give your brother the orange early, you must still be able to send the message somehow. That message cannot move faster than the speed of light.  In the case of a pulse propagating through a superluminal medium, the message corresponds to the leading edge of the pulse, which must be amplified.  The pulse can never pass this leading edge.

The metaphor also brings the question of how the exchange of energy between pulse and medium is mediated.  We will address this question in Section \ref{se:temporary-absorption}.
\section{Controlled dispersion}\label{se:controlled-dispersion}
Controlling dispersion means controlling the slope of the real part of the refractive index.  Because the real and imaginary parts of the refractive index are linked via the Kramers-Kronig relation, we can alter the real part by changing the imaginary part.  We will consider three media from the perspective of their linear electric susceptibility, which we will call $\chi$.  

In a linear medium, the complex refractive index, $\eta$, is given by $\eta=\sqrt{\epsilon_r \mu_r}=\sqrt{(1+\chi)(1+\chi_m)}$, where $\chi_m$ is the magnetic linear susceptibility.  For $\chi<<1$ in a nonmagnetic medium, the real and imaginary parts of $\chi$ are related to the real and imaginary parts of the complex refractive index, $\eta$, via $\eta_r \approx 1+\chi_r/2$, and $\eta_i=\chi_i/2$.  In this case we may use the electric susceptibility to conveniently quantify the dispersion and absorption of different media. A dispersive but roughly transparent medium will have a high ratio $\omega (d\chi_r/d\omega)/\chi_i$.

\subsection{Two-level resonance}
\begin{figure}
	\centerline{\includegraphics[width=12cm,clip]{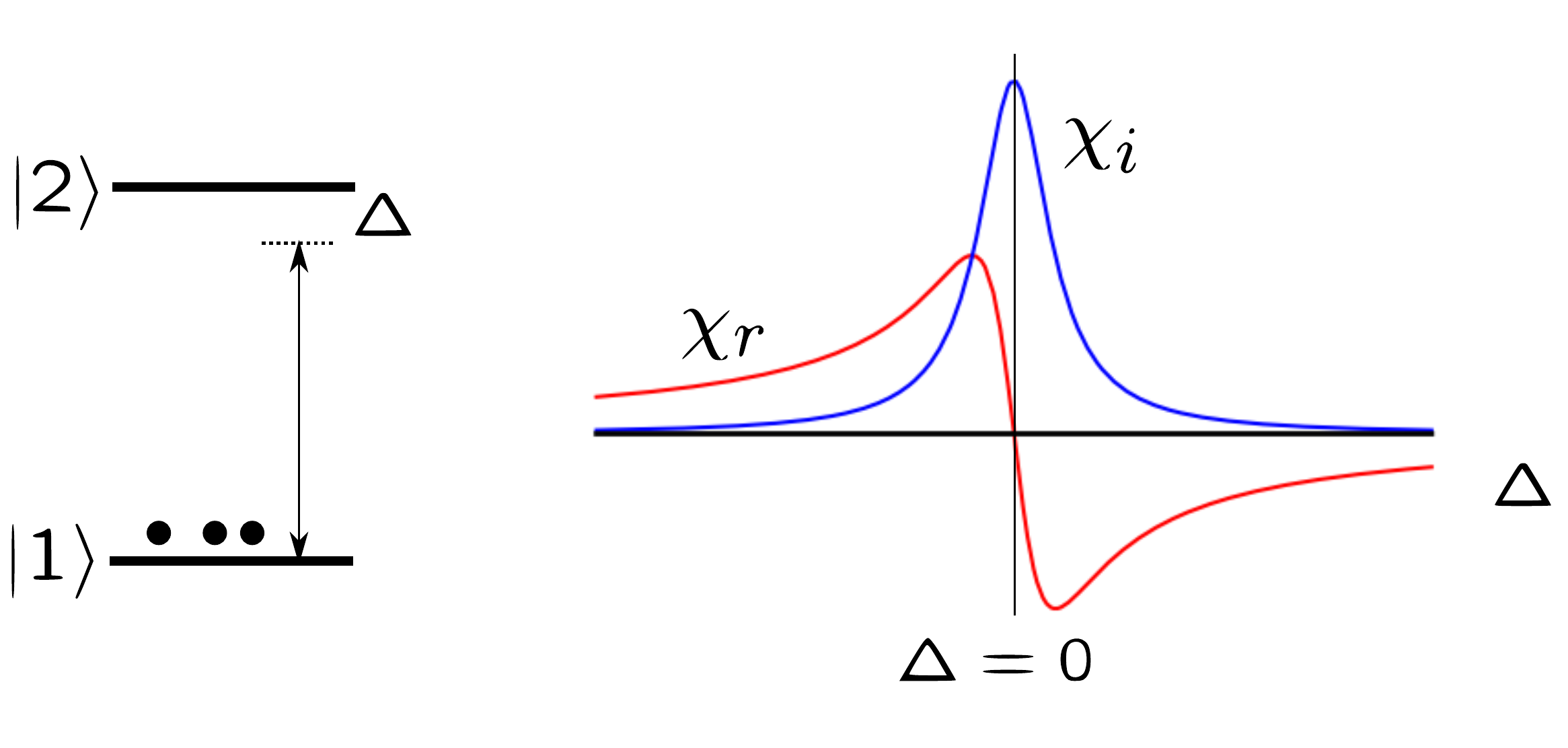}}
	\caption{\label{Fig:2-level} At zero detuning ($\Delta=0$), a simple 2-level absorptive resonance combines strong anomalous dispersion and strong absorption.}
\end{figure}
Figure \ref{Fig:2-level} shows the real and imaginary susceptibilities, $\chi_r$ and $\chi_i$, associated with a probe weakly interacting with a 2-level absorptive transition.  (In a sufficiently dilute medium, $|\chi|<<1$ and the real and imaginary susceptibilities are proportional to the refractive index and absorption coefficients.)  In Figure \ref{Fig:2-level}, the peak of absorption corresponds to strong negative dispersion at $\Delta=0$.  The relationship between the real and imaginary parts of analytic function, as given by the Kramers-Kronig relationship is roughly similar to that of a derivative.  Thus, the peak of $\chi_i$ corresponds to the maximum slope of $\chi_r$, while the peaks of $\chi_r$ correspond to the maximum slope of $\chi_i$.  Using this simple heuristic, it is easy to guess at the consequence of placing two absorptive resonances side by side.
\subsection{Two absorptive resonances side by side}
\begin{figure}
	\centerline{\includegraphics[width=12cm,clip]{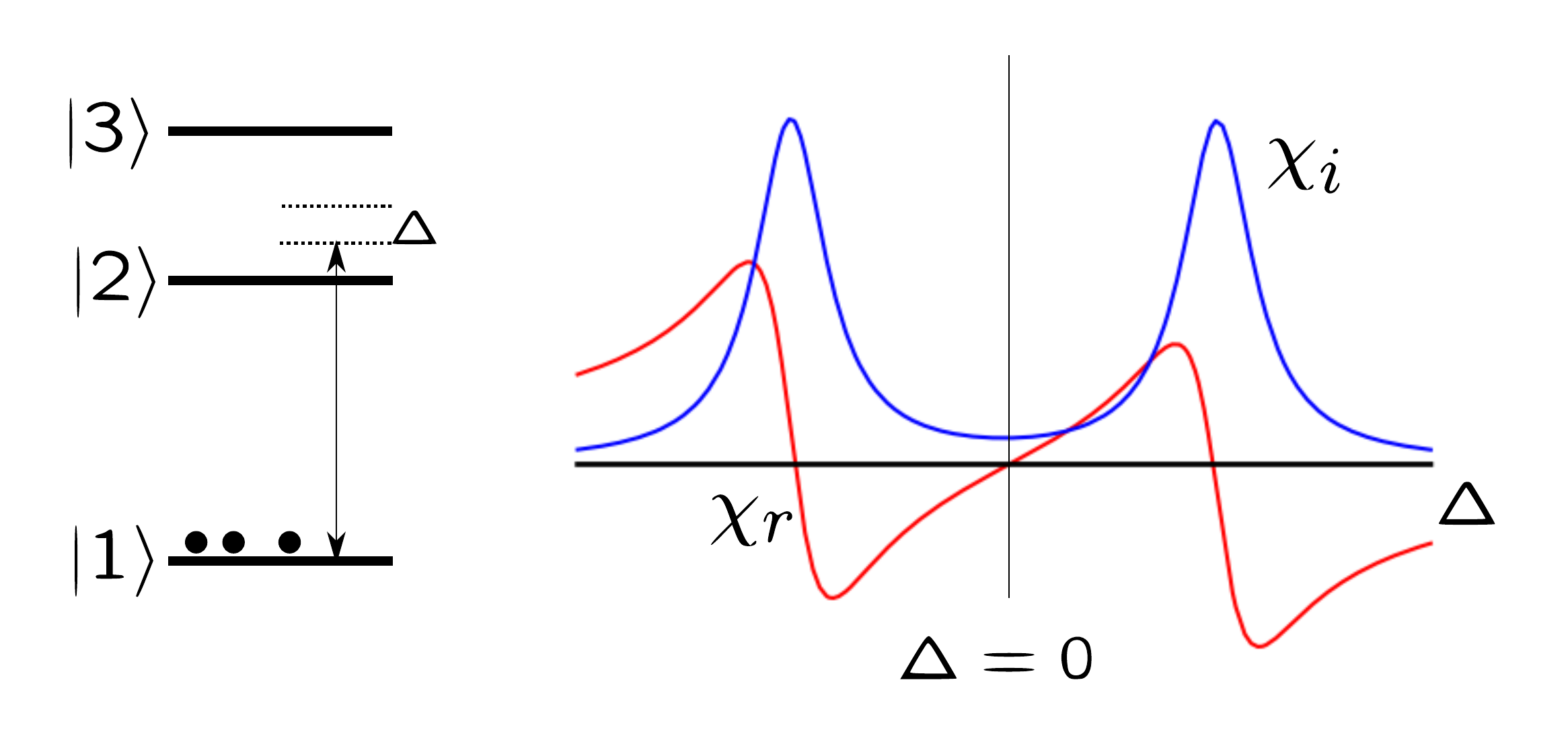}}
	\caption{\label{Fig:2-lines} In between two absorptive resonances placed side-by-side, there is strong normal dispersion and group velocity dispersion is minimized.}
\end{figure}

As is represented in Figure \ref{Fig:2-lines}, two absorption lines may be combined to form a slow-light medium with only moderate absorption.  If this profile is combined with a second medium that exhibits broad-band gain, the overall absorption may be mitigated without severely dampening the dispersion.  As symmetry between the two absorption lines is approached, all even derivatives of $\chi_r$ go to zero.  However, this arrangement is no better than the wing of a single resonance in terms of maximizing the slope of the real part of the susceptibility while minimizing the magnitude of the imaginary part.

\subsection{Strong, controllable normal dispersion through Electromagnetically Induced Transparency}
\begin{figure}
	\centerline{\includegraphics[width=12cm,clip]{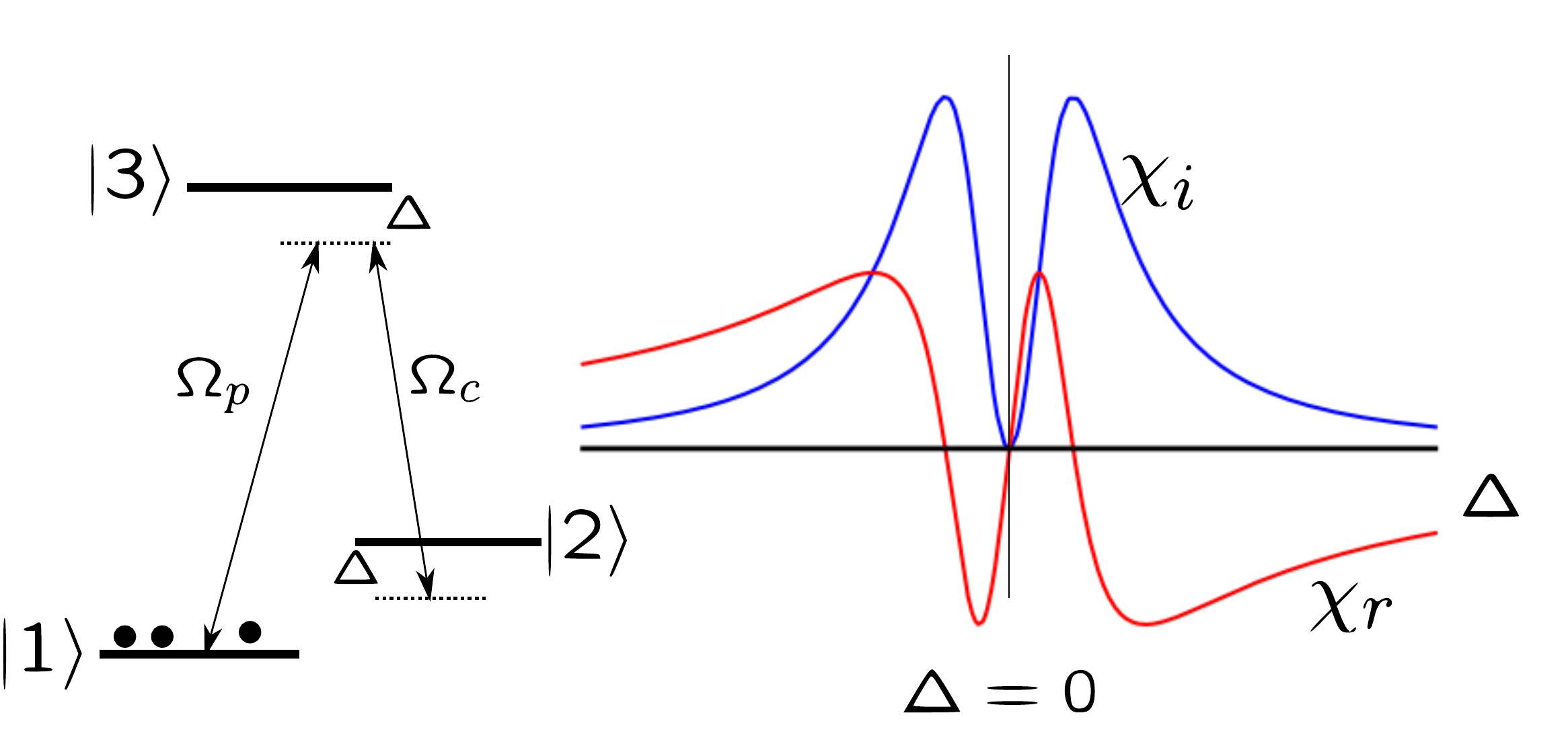}}
	\caption{\label{Fig:EIT} EIT.  Together, a coupling beam ($\Omega_c$) and an atom form an effective medium as seen by the probe beam ($\Omega_p$).  Destructive quantum interference between two alternative absorption pathways allows for minimal absorption in combination with strong dispersion at $\Delta=0$.}
\end{figure}
A more powerful approach utilizes Electromagnetically Induced Transparency (EIT) \cite{Harris-Field-Kasapi-1992, Fleischhauer-Imamoglu-Marangos-2005}.  In this approach, diagrammed in Figure \ref{Fig:EIT}, quantum interference between two absorptive pathways leads to the potential for stronger dispersion coupled with less absorption.  The EIT configuration has the additional advantages of more perfect symmetry (leading to a complete disappearance of even derivatives of $\chi_r$ at $\Delta=0$) and controllability.  While an ideal EIT system would have no absorption, in practice, loss due to decay, inhomogeneous broadening, spatial inhomogeneity, and various other sources of decoherence is always present.  EIT configurations have been used to achieve vary slow and even vanishing group velocities \cite{Hau-Harris-Dutton-Behroozi-1999,Liu-Dutton-Behroozi-Hau-2001,Zhang-Hernandez-Zhu-2008}.

\subsection{Controllable anomalous dispersion via a double Raman scheme}
\begin{figure}
	\centerline{\includegraphics[width=12cm,clip]{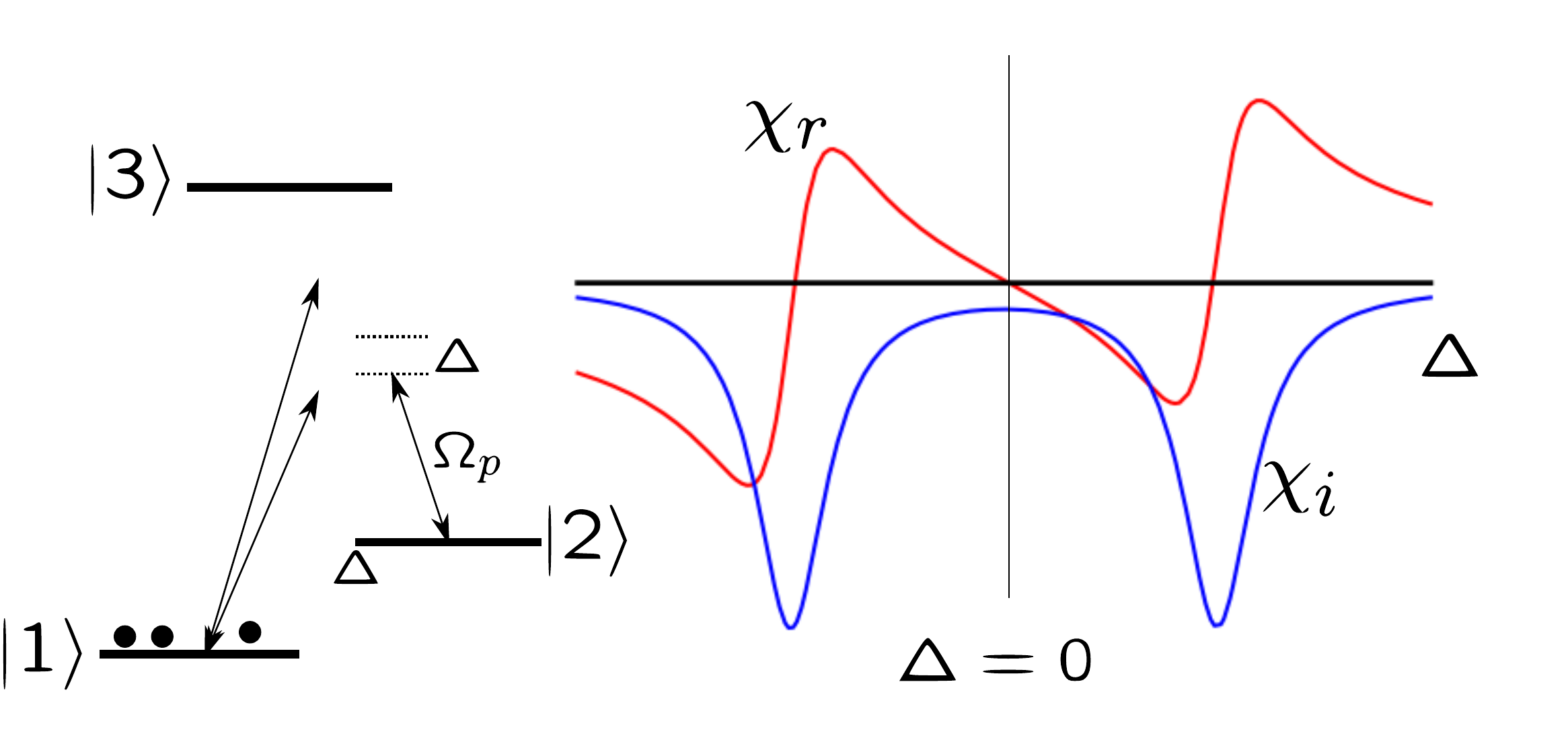}}
	\caption{\label{Fig:2-gains} At zero detuning ($\Delta=0$), a simple 2-level absorptive resonance combines strong anomalous dispersion and strong absorption.}
\end{figure}
Although an anomalously dispersive counterpart to EIT has not yet been devised, two gain lines may be combined, yielding controllable anomalous dispersion via the double Raman scheme represented in Figure \ref{Fig:2-gains}.  The two gain lines lead to anomalous dispersion coupled with moderate gain.  These may be combined with a broad-band absorber to yield a medium with no net gain but with strong anomalous dispersion.  This technique has been used to obtain strong anomalous dispersion with minimal gain \cite{Wang-Kuzmich-Dogariu-2000,Stenner-Gauthier-Neifeld-2004,Pati-Salit-Salit-Shahriar-2007}.
\chapter{Energy density and temporary absorption in lossless dispersive media}\label{ch:energy-density}
In 1999, Hau and Harris published a paper that concretely described some of the features of a pulse that is introduced into a slow-group-velocity medium \cite{Harris-Hau-1999}.  One of the most striking features of this transition is the spatial compression of the pulse.  When a pulse leaves a vacuum and enters a slow-light medium, it may be compressed longitudinally by many orders of magnitude.  However, in the case they examined, that of a pulse entering an EIT medium (see Section \ref{se:controlled-dispersion}), the electric permittivity and the magnetic permeability in the slow light medium are very close to their vacuum values for fields near EIT resonance.  Therefore the values of the squared electric and magnetic fields remain continuous across the boundary even as the pulse is compressed.  The total field energy associated with the pulse then scales with the group velocity.  When that velocity is slowed, very little of the original pulse energy remains in the field.  Rather, it is coherently stored in the slow-light medium.

The partitioning of energy between field and medium is, in general, a subtle subject.  However, in the special case where absorption may be ignored it becomes simpler.  In this chapter, we examine the effect of dispersion on the energy density of a dispersive, lossless medium.

\section{Energy density in a lossless, dispersive medium}\label{se:energy-density}
Jackson \cite{Jackson-1998} credits Brillouin \cite{Brillouin-1960} for the original formulation of the energy density of a of a spectrally narrow wave traveling through a dispersive medium with negligible loss.  We write this expression, considering the case where the imaginary portions of $\epsilon$ and $\mu$ can be ignored, as
\begin{equation}
	u=\frac{1}{2}\frac{d (\omega \epsilon)}{d \omega}(\omega_0)\langle \mathbf E \cdot \mathbf E \rangle+\frac{1}{2}\frac{d (\omega \mu)}{d \omega}(\omega_0)\langle \mathbf H \cdot \mathbf H\rangle,
\end{equation}
where the angled brackets denote a cycle average.

In an isotropic medium, $\mathbf D=\epsilon \mathbf E$ and $\mathbf B=\mu \mathbf H$, and we can rewrite this expression as
\[
	u=\left(1+\frac{d\ln\epsilon}{d\ln\omega}\right) \frac{\langle \mathbf D \cdot \mathbf E \rangle}{2} +\left(1+\frac{d\ln\mu}{d\ln\omega}\right) \frac{\langle \mathbf B \cdot \mathbf H \rangle}{2} .
\]
For a plane wave propagating through a lossless medium, $|\mathbf H|=\sqrt{\epsilon/\mu} |\mathbf E|$, and $\mu$ and $\epsilon$ are real so that 
\[
\langle \mathbf D \cdot \mathbf E \rangle=\langle \mathbf B \cdot \mathbf H \rangle.
\]
Using this fact and the identity 
\[
\frac{n_g}{n}=1+\frac{1}{2} \frac{d\ln\epsilon}{d\ln\omega}+\frac{1}{2} \frac{d\ln \mu}{d\ln\mu},
\] 
which is valid when the imaginary parts of $\epsilon$ and $\mu$ are negligible (see the end of Appendix ~\ref{appen:ng} and Eq.~(\ref{Eq:ng-over-n})), we can write
\[
u=\frac{n_g}{n} \left[\frac{\langle \mathbf D \cdot \mathbf E \rangle}{2}+\frac{\langle \mathbf B \cdot \mathbf H \rangle}{2}\right].
\]
But the term in brackets is just the cycle average of the energy density of an electromagnetic field in a linear, nondispersive medium.  We may therefore write
\begin{equation}\label{Eq:dispersive-u}
	u=\frac{n_g}{n}u_0,
\end{equation}
where $u_0$ is the cycle-averaged energy density in the case where there is no dispersion, written
\begin{equation}\label{Eq:non-dispersive}
 u_0=\frac{\langle \mathbf D \cdot \mathbf E \rangle}{2}+\frac{\langle \mathbf B \cdot \mathbf H \rangle}{2}.
\end{equation}

Note that the ratio between the magnitude of the Poynting vector and the energy density for a traveling quasimonochromatic wave in a dispersive medium is therefore
\begin{equation}
 \frac{|\mathbf S|}{u}=\frac{n}{n_g}\frac{|\mathbf E\times \mathbf H|}{u_0}=\frac{c}{n_g}.
\end{equation}  

Expression (\ref{Eq:dispersive-u}) for the energy density in a dispersive absorptionless medium is intuitive and matches our expectations based on the 1999 paper by Harris and Hau \cite{Harris-Hau-1999}, and comparisons between $u$ and $u_0$ lead naturally to a partitioning between energy that is stored because of dispersion and the total energy.  When $n \approx 1$, as was the case in their paper, the energy can be neatly partitioned into a fraction, $1/n_g$, that is stored in the form of field energy and a fraction, $(n_g-1)/n_g$, that is stored by the medium.  When $n \neq 1$, there seem to be two different forms of storage by the medium, one that is related to dispersion, and another that is related to the phase velocity.  We can form a partition between a dispersively stored energy fraction ($((n_g/n)-1)/(n_g/n)$) and non-dispersively stored energy fraction ($1/(n_g/n)$).  However, the non-dispersively stored energy fraction can no longer be unequivocally assigned to the field.

Expression ~(\ref{Eq:dispersive-u}) is also interesting because it brings up the possibility of a negative total energy density being associated with a wave in a dispersive medium.  Although the idea of real electromagnetic fields being associated with a negative total energy density may seem unphysical, we will see in the next two sections that it has a clear and simple interpretation.

\section{Anomalous dispersion and temporary emission}\label{se:anomalous-dispersion}
\begin{figure}
	\centerline{\includegraphics[width=15cm, clip]{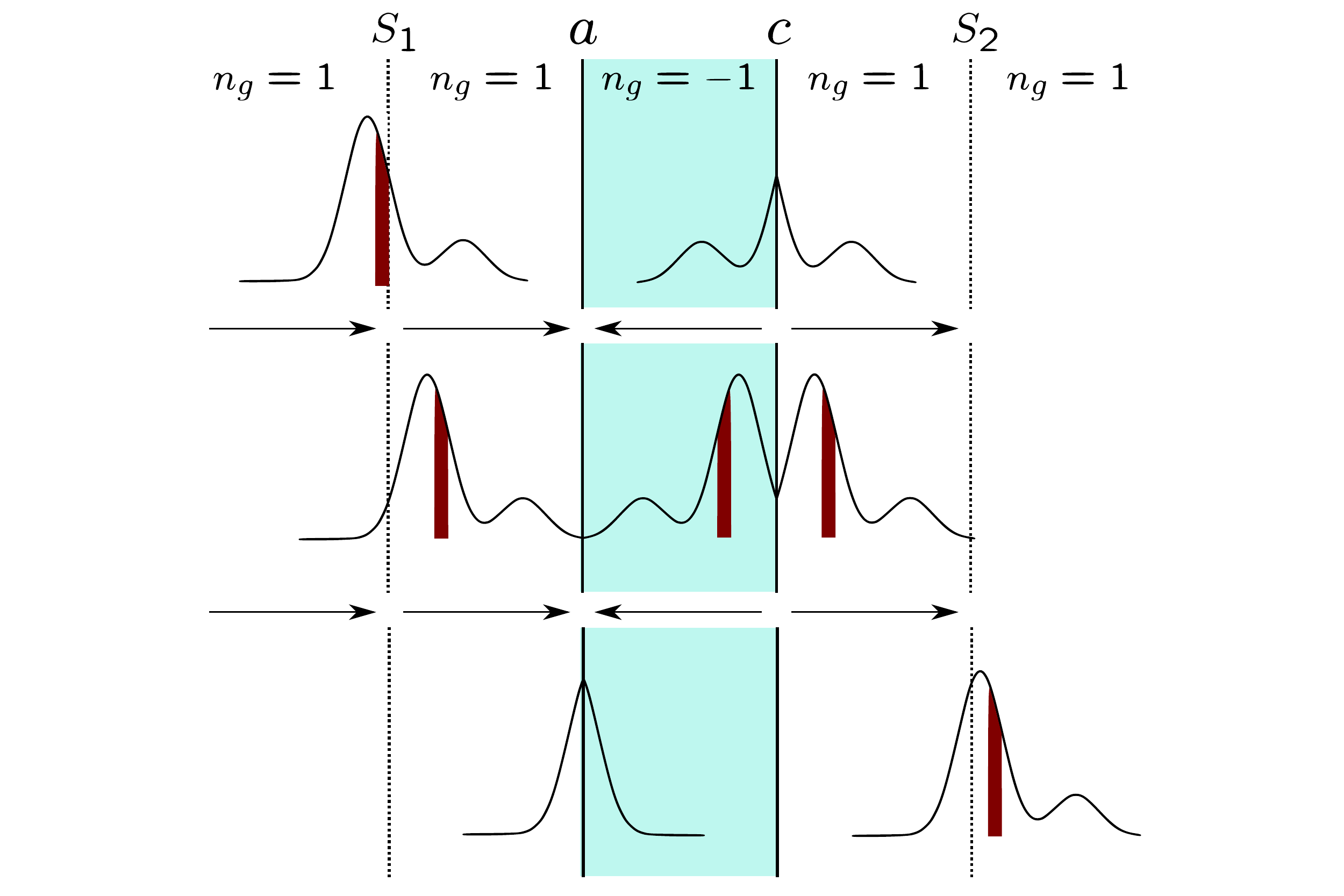}}
	\caption{\label{Fig:neg-slab} A pulse interacts with an anomalously dispersive slab with a group index of $-1$ and with negligible absorption or gain.  The two solid lines labeled $a$ and $c$ delimit changes in the group index.  They also represent lines of symmetry.  The magnitude of the Poynting vector is symmetric in form around line $a$ between the dotted line $S_1$ and the solid line $c$.  There is a similar symmetry about line $c$ between lines $a$ and $S_2$.  The portions of the pulse that are outside of the dotted lines have no copies.  Those portions inside the dotted lines are represented in triplicate. Portions of the pulse between $a$ and $c$ move backwards.  Watching the pulse propagate, it seems as if two pulses are created at the solid line labled $c$ and then two pulses are later annihilated at the solid line labeled $a$.}
\end{figure}
Figure \ref{Fig:neg-slab} depicts a planar pulse approaching and crossing an anomalously dispersive slab at normal incidence.  For simplicity, we take the slab to have the properties $\alpha \approx 0$, $\epsilon_r \approx 1$, $\mu_r \approx 1$, and $n_g \approx -1$ over the narrow bandwidth of the incident pulse. If we watch one particular segment of the pulse (for example that segment colored red in Figure \ref{Fig:neg-slab}) we see that when it crosses $S_1$, the first dotted line, two copies of it appear downstream.  One is beyond the dispersive medium and moving in the original direction of propagation.  The other is in the anomalously dispersive medium and moves upstream.  Eventually, the original upstream segment collides with its backward copy and the two dissappear, leaving only one downstream copy, which is now beyond the second dotted line, $S_2$.

If we look at the total \emph{field} energy associated with any given segment of the pulse, we see that it changes with time.  Before the pulse segment reaches the first dotted line, it has an initial field energy, $U_0$.  As it crosses the dotted line, two copies are made.  Counting the field energy associated with all three copies of the segment,  we get a total field energy of $3 U_0$.  As the most advanced copy crosses the second dotted line, the two lagging copies annihilate each other and the total field energy goes back to $U_0$.  Conservation of energy requires that the extra $2 U_0$ be accounted for by an equivalent energy deficit somewhere.  Since the only other entity postulated besides the original pulse is the medium, it must come from the medium.  The net effect is that the medium donates energy to the field for a limited time and is later paid back.  In Section \ref{se:temporary-absorption}, we will see that the medium is only paid back because the overall pulse has the narrow spectrum that we originally assigned to it.

So far, we have found that the extra field energy that is present when a segment of the pulse is propagating through the anomalously dispersive medium suggests an energy deficit in the medium.  We now seek to localize that energy deficit.  In our model, energy is moved by electromagnetic fields, as represented by the Poynting vector.  We can find the energy deficit associated with a particular slice of the medium by comparing the total energy that has entered it to the total energy that has left it.  Because the nondispersive quantities associated with the medium have roughly their vacuum values, we know that the Poynting vector points downstream in the medium just as it does in the vacuum.  

Thus, to find the energy deficit associated with our slice, we compare the time integrated Poynting vector at the downstream surface of our slice, which corresponds to the energy which has left the slice, to the time integrated Poynting vector at the upstream surface of our slice, which corresponds to the energy which has entered the slice.  The difference between these two quantities gives the total energy deficit in the medium between the two surfaces.

The Poynting vector history associated with a surface in the medium is easy to visualize.  If the surface were in the vacuum, the history would be given by the form of the pulse downstream from the surface.  In our special medium, it corresponds to the form of the pulse that is upstream of the surface.  In the geometry of Figure \ref{Fig:neg-slab}, the correspondence also ends at the line $a$, where annihilation destroys the record.

When we subtract the energy that has crossed the first surface from the energy that has crossed the second surface, what we are left with corresponds to the energy between the two surfaces.  In other words, the total energy deficit in our slice is equal to the field energy in that slice.  Applying this reasoning to a thin slice suggests that if the field energy density is $u_{field}$ and the density of energy absorbed by the medium is $u_{exchange}$, then we know for our scenario that 
\[
 u_{field}+u_{exchange}=-u_{field},
\]
or that $u_{exchange}=-2 u_{field}$.

If $n_g=-2$ (but all other quantities remain unaltered), the field energy relates to the Poynting vector history differently because a given portion of the pulse takes two times as long to cross a surface.  Taking this into account, we find for this case that
\[
u_{field}+u_{exchange}=-2 u_{field}.
\]
Allowing $n_g$ to take an arbitrary value gives
\[
u_{field}+u_{exchange}=n_g u_{field}.
\]
If we allow $n$ to take values other than $1$, we can apply this same logic.  The longitudinal scaling of a pulse associated with $n_g$ must now be normalized by $n$ and instead of $u_{field}$ and $u_{exchange}$ we can work in terms of a non-dispersive energy density ($u_0$), given by Eq.~\ref{Eq:non-dispersive}, and a dispersive energy density ($u_{disp}$), corresponding to the energy dispersively stored (or donated) by the medium.  Then we can write
\[
u_0+u_{disp}=\frac{n_g}{n}u_0.
\]
Denoting the total energy associated with the propagating wave as $u$, we can rewrite this as
\[
 u=\frac{n_g}{n}u_0,
\]
which is exactly Eq.~\ref{Eq:dispersive-u}.

\section{Temporary absorption and emission and the instantaneous spectrum in a lossless dispersive medium}\label{se:temporary-absorption}

When we calculate a pulse spectrum, we take an integral over the entire pulse.  However, an atom responding to the pulse can only respond to that portion that has interacted with up to any particular point in time.  Thus, the atom sees and interacts with an ``instantaneous spectrum'' that differs from the spectrum of the pulse as a whole.  Peatross et al. put it this way \cite{Peatross-Ware-Glasgow-2001}:
\begin{quote}
 The principle of causality requires a medium experiencing
a pulse to be prepared for an abrupt termination of
the field at any moment, in which case further exchange
of energy with the field cannot take place. Such a termination
produces a truncated waveform that generally contains
a wider range of spectral components than are
present in the pulse taken in its entirety. This momentary
spectrum can lap onto nearby absorbing or amplifying
resonances. The medium accordingly attenuates or
amplifies this perceived spectrum. As the medium experiences
the waveform, it continually reassesses the spectrum
and thereby treats the front and the rear of the
pulses differently.
\end{quote}

In this section we briefly review their results and seek to apply them to absorptionless dispersion in the narrow-band limit.  
\subsection{Instantaneous spectrum and energy exchange in a dispersive medium}
Assuming, as we did in the previous section, that energy propagates only via the Poynting vector, we may write \cite{Peatross-Ware-Glasgow-2001}
\begin{equation}
 \nabla \cdot \mathbf S + \frac{\delta u}{\delta t},
\end{equation} electromagnetic fields, 
where $u$, the total energy density, is given by 
\begin{equation}
 u(t)=u_{field}+u_{exchange}+u(-\infty).
\end{equation} 
Here $u_{field}$ represents the energy density of the electromagnetic fields, $u_{exchange}$ represents the density of total energy absorbed or emitted by the medium, and $u(-\infty$) represents the density of energy stored in the medium before the arrival of the field.  Assuming that the medium is non-magnetic, Peatross et al. show, by imposing the requirement of causality on the linear electric susceptibility, that 

\begin{equation}
u_{exchange}=\epsilon_0 \int_{-\infty}^\infty |E_t(\omega)|^2 \omega \chi_i(\omega) d\omega, 
\end{equation} 
where
\begin{equation}\label{Eq:E_t}
 E_t(\omega)\equiv \frac{1}{\sqrt{2 \pi}}\int_{-\infty}^t dt^\prime E(t^\prime)\exp(i \omega t^\prime),
\end{equation} 
and $E(\omega)$ and $\chi(\omega)$ had previously been defined in a way that matches the Fourier transform convention of Eq.~(\ref{Eq:E_t}).
In other words, the $u_{exchange}$ is completely determined by the interaction of the instantaneous power spectrum with the imaginary portion of the susceptibility.  

A narrow band pulse traveling through a dispersive, absorptionless medium has an instantaneous spectrum that narrows as the pulse amplitude decreases and broadens as it increases.  Although the medium is absorptionless with respect to the pulse as a whole, the fact that the medium is dispersive means in the spectral wings the medium is active (in the case of fast light), or absorbing (in the case of slow light).

\subsection{The instantaneous spectrum and narrow band pulses in dispersive absorptionless media}
In Section \ref{se:energy-density} we found a form for the total energy density associated with a pulse in a dispersive absorptionless medium and noted that the total energy density could be negative.  In Section \ref{se:anomalous-dispersion} we showed that the nature of the kinetic propagation of a pulse through an absorptionless anomalously dispersive medium suggests that the medium must temporarily donate energy to the electromagnetic field.  We now see that the instantaneous spectrum provides the mechanism by which this occurs.

An important point is that temporary absorption and temporary emission as dictated by the spontaneous spectrum are really just ordinary absorption and ordinary emission.  The fact that they are temporary is specific to the way that later portions of the pulse interact with the medium and is guaranteed only when the overall pulse spectrum remains within the narrow bandwidth for which the absorption is close to zero.  If a pulse is truncated, the energy that has been either absorbed or emitted remains either absorbed or emitted.

This equivalence between dispersive energy exchange with the medium and absorption and emission is useful in answering simple questions.  For example, what kind of momentum exchange should we expect to see between field and medium when a pulse propagates through a dispersive medium?  Because we know how momentum exchange works under absorption and emission and because we know that light removed from the field or added to the field by dispersion has been temporarily absorbed or emitted, we can understand this momentum exchange in simple terms.

Chapter \ref{ch:momentum} discusses momentum in a dispersive medium in more detail.
\chapter{Momentum in a dispersive medium}\label{ch:momentum}
\newcommand{\bd}{{\bf d}}
\newcommand{\bv}{{\bf v}}
\newcommand{\hbp}{\hat{\bp}}
\newcommand{\hbx}{\hat{\bx}}
\newcommand{\hq}{\hat{q}}
\newcommand{\hp}{\hat{p}}
\newcommand{\ha}{\hat{a}}
\newcommand{\ad}{a^{\dag}}
\newcommand{\hsig}{{\hat{\sigma}}}
\newcommand{\nt}{\tilde{n}}
\newcommand{\itf}{\sl}
\newcommand{\eps}{\epsilon}
\newcommand{\bsig}{\pmb{$\sigma$}}
\newcommand{\beps}{\pmb{$\eps$}}
\newcommand{\bmu}{\pmb{$\mu$}}
\newcommand{\balpha}{\pmb{$\alpha$}}
\newcommand{\bbeta}{\pmb{$\beta$}}
\newcommand{\bgamma}{\pmb{$\gamma$}}
\newcommand{\bu}{{\bf u}}
\newcommand{\bpi}{\pmb{$\pi$}}
\newcommand{\bSig}{\pmb{$\Sigma$}}
\newcommand{\be}{\begin{equation}}
\newcommand{\ee}{\end{equation}}
\newcommand{\bea}{\begin{eqnarray}}
\newcommand{\eea}{\end{eqnarray}}
\newcommand{\sss}{_{{\bf k}\lambda}}
\newcommand{\ssss}{_{{\bf k}\lambda,s}}
\newcommand{\dip}{\langle\sigma(t)\rangle}
\newcommand{\dipp}{\langle\sigma^{\dag}(t)\rangle}
\newcommand{\sigz}{\langle\sigma_z(t)\rangle}
\newcommand{\sig}{{\sigma}}
\newcommand{\sigd}{{\sigma}^{\dag}}
\newcommand{\ra}{\rangle}
\newcommand{\la}{\langle}
\newcommand{\om}{\omega}
\newcommand{\Om}{\Omega}
\newcommand{\pa}{\partial}
\newcommand{\bR}{{\bf R}}
\newcommand{\bx}{{\bf x}}
\newcommand{\br}{{\bf r}}
\newcommand{\bE}{{\bf E}}
\newcommand{\bH}{{\bf H}}
\newcommand{\bB}{{\bf B}}
\newcommand{\bP}{{\bf P}}
\newcommand{\bD}{{\bf D}}
\newcommand{\bA}{{\bf A}}
\newcommand{\bek}{{\bf e}\rmk}
\newcommand{\rmk}{_{{\bf k}\lambda}}
\newcommand{\bsij}{{\bf s}_{ij}}
\newcommand{\bk}{{\bf k}}
\newcommand{\bp}{{\bf p}}
\newcommand{\epso}{{1\over 4\pi\eps_0}}
\newcommand{\BB}{{\mathcal B}}
\newcommand{\AAA}{{\mathcal A}}
\newcommand{\NN}{{\mathcal N}}
\newcommand{\mm}{{\mathcal M}}
\newcommand{\RR}{{\mathcal R}}
\newcommand{\bS}{{\bf S}}
\newcommand{\bL}{{\bf L}}
\newcommand{\bJ}{{\bf J}}
\newcommand{\bI}{{\bf I}}
\newcommand{\bF}{{\bf F}}
\newcommand{\bsub}{\begin{subequations}}
\newcommand{\esub}{\end{subequations}}
\newcommand{\baline}{\begin{eqalignno}}
\newcommand{\ealine}{\end{eqalignno}}
\newcommand{\isat}{{I_{\rm sat}}}
\newcommand{\Is}{I^{\rm sat}}
\newcommand{\Ip}{I^{(+)}}
\newcommand{\Imm}{I^{(-)}}
\newcommand{\Inu}{I_{\nu}}
\newcommand{\bInu}{\overline{I}_{\nu}}
\newcommand{\bN}{\overline{N}}
\newcommand{\qnu}{q_{\nu}}
\newcommand{\oqn}{\overline{q}_{\nu}}
\newcommand{\qsat}{q^{\rm sat}}
\newcommand{\Iout}{I_{\nu}^{\rm out}}
\newcommand{\topt}{t_{\rm opt}}
\newcommand{\crr}{{\mathcal{R}}}
\newcommand{\cE}{{\mathcal{E}}}
\newcommand{\cH}{{\mathcal{H}}}
\newcommand{\epsoo}{\epsilon_0}
\newcommand{\muo}{\mu_0}
\newcommand{\ombar}{\overline{\om}}
\newcommand{\bPi}{{\bf \Pi}}
\newcommand{\hz}{\hat{z}}

\section{Abstract}
When the effects of dispersion are included, neither the Abraham nor the Minkowski expression for electromagnetic momentum  in a dielectric medium gives the correct recoil momentum for absorbers or emitters of radiation. The total
momentum density associated with a field in a dielectric medium has three contributions: (i) the Abraham momentum density of the field, (ii) the momentum density associated with the Abraham force, and (iii) a momentum density arising from the dispersive part of the response of the medium to the field, the 
latter having a form evidently first derived by D.F. Nelson [Phys. Rev. A{\bf 44}, 3985 (1991)]. All three 
contributions are required for momentum conservation in the recoil of an absorber or emitter in a dielectric
medium. We consider the momentum exchanged and the force on a polarizable particle (e.g., an
atom or a small dielectric sphere) in a host dielectric when a pulse of light is incident upon it, including the dispersion of the dielectric medium as well as a dispersive component in the response of the particle to the field. 
The force can be greatly increased in slow-light dielectric media. 

\section{Introduction}
Electromagnetic momentum in a dielectric medium is a subject with a very extensive literature, especially
in connection with its different formulations. The two most favored forms by far are those of Abraham
and Minkowski; as aptly remarked in a recent paper \cite{hinds}, ``There is ... a bewildering array of experimental studies and associated theoretical analyses which appear to favor one or other of these momenta or, indeed, others." 
An aspect of this subject that has received surprisingly little attention concerns the effects of dispersion on the
Minkowski and Abraham momenta and on the electromagnetic forces on polarizable particles. The intent of the present paper is to address such effects, which might help to clarify the physical interpretation of the Abraham and Minkowski momenta and the distinction between them. 

We first review briefly the Abraham and Minkowski momenta for the situation usually
considered---a dielectric medium assumed to be dispersionless and non-absorbing at a 
frequency $\om$. The Abraham and Minkowski momentum densities are respectively
\be
\bP_A = \frac{1}{c^2}\bE\times\bH \ \ \ \ \ {\rm and} \ \ \ \ \ \bP_M=\bD\times\bB
\label{momA}
\ee
in the standard notation for the fields on the right-hand sides. We will take the 
permeability $\mu$ to be equal to its vacuum
value $\muo$, which is generally an excellent approximation at optical frequencies. For single
photons the magnitudes of the Abraham and Minkowski momenta are given by (see Section II)
\be
p_A=\frac{1}{ n}\frac{\hbar\om}{ c} \ \ \ \ \ {\rm and} \ \ \ \ \ p_M=n\frac{\hbar\om}{ c}, 
\label{photA}
\ee
where $n$ is the refractive index at frequency $\om$. From $\bD=\epsoo n^2\bE$ it follows that
\be
\frac{\pa\bP_M}{\pa t}=\frac{\pa\bP_A}{\pa t}+{\bf f}^A,
\label{momM}
\ee
where
\be
{\bf f}^A=\frac{1}{ c^2}(n^2-1)\frac{\pa}{\pa t}(\bE\times\bH)
\label{abedensity}
\ee
is the {\sl Abraham force density}. For single-photon fields the momentum $p^A$ associated with the Abraham force
is $[(n^2-1)/n]\hbar\om/c$, and (\ref{momM}) becomes $p_M=p_A+p^A$.

The Abraham momentum is generally regarded as the correct momentum of the electromagnetic field \cite{jackson},
whereas the Minkowski momentum evidently includes the momentum of the dielectric medium as well as that of the field. Ginzburg \cite{ginz} calls $p_M$ the momentum of a 
``photon in a medium," and notes that
its use, together with energy and momentum conservation laws, yields correct results for Cerenkov radiation as
well as the Doppler shift. Experiments appear by and large to indicate that it is the momentum $n\hbar\om/c$
per photon that provides the recoil and radiation pressure experienced by an object immersed in a dielectric medium
\cite{jonesand}. However, when dispersion ($dn/d\om$) is accounted for, $n\hbar\om/c$ is not the Minkowski
momentum of a photon, as we review in the following section.

This paper is organized as follows. In the following section we briefly discuss the generalization of the Abraham and Minkowski momenta to the case of a dispersive dielectric medium \cite{garrison} and consider two examples: (i) the Doppler
shift in a dielectric medium \cite{fermi} and (ii) the displacement of a dielectric block on a frictionless surface due to the passage of a single-photon field through it \cite{nandor}. A consistent description of
momentum transfer in these examples requires that we account for momentum imparted to the medium. In Section  
\ref{forces} we calculate the force exerted by a quasimonochromatic plane wave on a polarizable particle
and on a dispersive dielectric medium modeled as a continuum, and obtain a dispersive contribution to the
latter in agreement with an expression that, to the best of our knowledge, was first derived, in a rather
different way, by Nelson \cite{nelson}. In Section \ref{momentumex} we consider the momentum exchange between
a plane-wave pulse and an electrically polarizable particle immersed in a nonabsorbing dielectric medium, and show that this momentum depends on both the dispersion of the medium and the variation with frequency of the polarizability; in particular, in slow-light media it can be large and in the direction opposite to that in which the field
propagates. Section \ref{sec:rayleigh} presents derivations of some results relevant to Section \ref{sec:sphere},
where we generalize the results of Section \ref{momentumex} to include absorption and discuss the forces 
exerted by a pulse on a small dielectric sphere in a host slow-light medium. Section \ref{sec:conclusions}
briefly summarizes our conclusions.

\section{Abraham and Minkowski Momenta for Dispersive Media}\label{abemink}
We first recall the expression for the total cycle-averaged energy density when a plane-wave monochromatic 
field [$\bE=\bE_{\om}e^{-i\om t}$, $\bH=\bH_{\om}e^{-i\om t}$, $\bH_{\om}^2=(\eps/\muo)\bE_{\om}^2$] propagates in a dispersive dielectric at a frequency $\om$ at which absorption is negligible \cite{landaulif}:
\be
u={1\over 4}\left[{d\over d\om}(\eps\om)\bE_{\om}^2+\muo\bH_{\om}^2\right],
\label{energydensity}
\ee
or equivalently, in terms of $\bE_{\om}$ and the group index $n_g=d(n\om)/d\om$,
\be
u={1\over 2}\epsoo nn_g\bE_{\om}^2.
\label{ueq}
\ee
When the field is quantized in a volume $V$, $u$ is in effect replaced by $q\hbar\om/V$, where $q$ is the expectation value of the photon number in the volume $V$;  therefore, from (\ref{ueq}), $\bE_{\om}^2$ is effectively $2\hbar\om/(\epsoo nn_gV)$ per photon. Thus, for single photons, the Abraham momentum defined by (\ref{momA}) is 
\be
p_A={n\over c}{1\over 2}\epsoo{2\hbar\om\over\epsoo nn_gV}V={1\over n_g}{\hbar\om\over c}.
\label{p1}
\ee
Similarly,
\be
p_M={n^2\over n_g}{\hbar\om\over c},
\label{p2}
\ee
which follows from the definition in (\ref{momA}) and the relation $\bD=\epsoo n^2\bE$;
thus $p_M=n^2p_A$. These same expressions for $p_A$ and $p_M$ can of course be obtained more formally by quantizing the fields $\bE$, $\bD$, $\bH$, and $\bB$ in a dispersive medium \cite{garrison}.

Two examples serve to clarify the differences among the momenta involved in the momentum exchange between light and matter. The first example
is based on an argument of Fermi's that the Doppler effect is a consequence of this momentum exchange \cite{fermi}, as 
follows. 
Consider an atom of mass $M$ inside a host dielectric medium with refractive index $n(\om)$. The atom has a sharply defined transition frequency $\om_0$ and is initially moving with velocity $v$ away from a source of light of frequency 
$\om$. Because the light in the atom's reference frame has a Doppler-shifted frequency $\om(1-nv/c)$ determined by the phase velocity ($c/n$) of light in the medium, the atom can absorb a photon
if $\om(1-nv/c)=\om_0$, or if
\be
\om\cong\om_0(1+nv/c). 
\label{dopp1}
\ee
We denote the momentum associated with a photon in the medium by $\wp$
and consider the implications of (nonrelativistic) energy and momentum conservation. The initial energy 
is $E_i=\hbar\om+{1\over 2}Mv^2$, and the final energy, after the atom has absorbed a photon, is
${1\over 2}Mv'^2+\hbar\om_0$, where $v'$ is the velocity of the atom after absorption. The initial momentum
is $\wp+Mv$, and the final momentum is just $Mv'$. Therefore
\be
{1\over 2}M(v'^2-v^2)\cong Mv(v'-v)=Mv(\wp/M)=\hbar(\om-\om_0),
\ee
or $\om\cong\om_0+\wp v/\hbar$. From (\ref{dopp1}) and $\om\cong\om_0$ we conclude that 
\be
\wp=n{\hbar\om\over c}.
\label{wp}
\ee
Thus, once we accept the fact that the Doppler shift depends on the refractive index of the medium according
to Eq. (\ref{dopp1}), we are led by energy and momentum conservation to conclude that an atom in the medium 
must recoil with momentum (\ref{wp}) when it absorbs (or emits) a photon of energy $\hbar\om$. Momentum
conservation in this example is discussed in more detail below.

In our second example we consider, following Balazs \cite{nandor}, a rigid block of mass $M$, refractive index $n$, and 
length $a$, initially sitting at rest on a frictionless surface. A single-photon pulse of frequency $\om$ passes through the block, which is assumed to be nonabsorbing at frequency $\om$ and to have anti-reflection coatings on its front and back surfaces. The length $a$ of the block is presumed to be much larger than the length of the pulse. If the photon momentum is $\wp_{\rm in}$ inside the block and $\wp_{\rm out}$ outside,
the block picks up a momentum $MV=\wp_{\rm out}-\wp_{\rm in}$ when the pulse enters. If the space outside the block is
vacuum, $\wp_{\rm out}=mc$, where $m=E/c^2=\hbar\om/c^2$. Similarly
$\wp_{\rm in}=mv_p$, where $v_p$ is the velocity of light in the block. Without dispersion,
$v_p=c/n$ and the momentum of the photon in the block is evidently $\wp_{\rm in}=mc/n=\hbar\om/nc$. 
The effect of dispersion is to replace $v_p=c/n$ by $v_g=c/n_g$ and $\wp_{\rm in}=\hbar\om/nc$ by $\wp_{\rm in}=\hbar\om/n_gc$. With or without dispersion, this example suggests that the photon momentum in the medium has the Abraham form. Note that the essential feature of Balazs's argument is simply that the velocity of light in the medium is $v_p$ (or, more generally, $v_g$). This, together with momentum conservation, is what leads him to conclude that
the momentum of the field has the Abraham form.

This prediction can in principle be tested experimentally. Conservation of momentum requires, according
to Balazs's argument, that $MV=m(c-v_g)$. When the pulse exits the block, the block recoils and comes to rest, and is left with a net displacement
\be
\Delta x=V\Delta t={m\over M}(c-v_g){a\over v_g}={\hbar\om\over Mc^2}(n_g-1)a
\ee
as a result of the light having passed through it. This is the prediction for the net displacement based on
the momentum $p_A$ given in (\ref{p1}). If the photon momentum inside the block were assumed to have the Minkowski form $n^2\hbar\om/cn_g$ given in (\ref{p2}), however, the displacement of the block would in similar fashion be predicted to be 
\be
\Delta x={\hbar\om\over Mc^2}a(n_g-n^2),
\ee
and if it were assumed to be $n\hbar\om/c$, as in Eq. (\ref{wp}), the prediction would be that the net
displacement of the block is
\be
\Delta x={\hbar\om\over Mc^2}an_g(1-n).
\ee
These different assumptions about the photon momentum can lead to different predictions not only for the
magnitude of the block displacement but also for its direction.

The first (Doppler) example suggests at first thought that the momentum of the photon is $n\hbar\om/c$ [Eq. (\ref{wp})], while the second (Balazs) example indicates that it is $\hbar\om/n_gc$. Let us consider more carefully the first example. There
is ample experimental evidence that the Doppler shift is $nv\om/c$ regardless of dispersion, 
as we have assumed, but does this imply that the momentum of a photon in a dielectric is in fact $n\hbar\om/c$?
We will show in the following section that the forces exerted by a plane monochromatic wave on the polarizable particles of a dielectric result in a momentum density of magnitude
\be
p_{\rm med}={\epsoo\over 2c}n(nn_g-1)E_{\om}^2=(n-{1\over n_g}){\hbar\om\over c}{1\over V};
\label{pmedium}
\ee
the second equality applies to a single photon, and follows from the replacement of $\bE_{\om}^2$ by 
$2\hbar\om/(\epsoo nn_gV)$, as discussed earlier. Now from the fact that
the Doppler shift implies that an absorber (or emitter) inside a dielectric 
recoils with momentum $n\hbar\om/c$, all
we can safely conclude from momentum conservation is that a momentum $n\hbar\om/c$ is taken from (or given to) 
the {\sl combined system of field
and dielectric}. Given that the medium has a momentum density (\ref{pmedium}) due to the force exerted on it by
the propagating field, we can attribute to the field (by conservation of momentum) a momentum density
\be
n{\hbar\om\over c}{1\over V}-P_{\rm med}={1\over n_g}{\hbar\om\over c}{1\over V}=p_A.
\ee
That is, the momentum of the field in this interpretation is given by the Abraham formula, consistent with the conclusion of the Balazs thought experiment. The recoil momentum $n\hbar\om/c$, which in general differs from both the Abraham and the Minkowski momenta, evidently gives the momentum not of the field as such but of the combined system of field plus dielectric. It is the momentum density equal to the {\sl total} energy density $u=\hbar\om/V$
for a monochromatic field divided by the phase velocity $c/n$ of the propagating wave. As already mentioned, experiments on the recoil of objects immersed in dielectric media have generally indicated that the recoil momentum is $n\hbar\om/c$ per unit of energy $\hbar\om$ of the field, just as in the Doppler effect. But this should not be taken to mean that $n\hbar\om/c$ is the momentum of a ``photon" existing independently of the medium in which the field propagates. Regardless of how this
momentum is apportioned between the field and the medium in which it propagates, the important thing for the theory,
of course, is that it correctly predicts the {\sl observable forces} exerted by electromagnetic fields. We next turn our attention specifically to the forces acting on polarizable particles in applied electromagnetic fields.

\section{Momenta and Forces on Polarizable Particles}\label{forces}
We will make the electric dipole approximation and consider field frequencies
such that absorption is negligible. Then the induced electric dipole moment of a particle in a field of frequency $\om$ is $\bd=\alpha(\om)\bE_{\om}\exp(-i\om t)$, and the polarizability $\alpha(\om)$
may be taken to be real. With these assumptions we now consider the forces acting on such particles in applied,
quasi-monochromatic fields. 

We begin with the Lorentz force on an electric dipole moment $\bd$ in an electromagnetic field \cite{barloud}:
\bea
\bF&=&(\bd\cdot\nabla)\bE+\dot{\bd}\times\bB\nonumber \\
&=&(\bd\cdot\nabla)\bE+\bd\times(\nabla\times\bE)+{\pa\over\pa t}(\bd\times\bB)\nonumber \\
&\equiv&\bF_E+\bF_B,
\label{dipforce}
\eea
where we define
\be
\bF_E=(\bd\cdot\nabla)\bE+\bd\times(\nabla\times\bE),
\label{forceE}
\ee
\be
\bF_B={\pa\over\pa t}(\bd\times\bB).
\label{forceB}
\ee
In writing the second equality in (\ref{dipforce}) we have used the Maxwell equation $\pa\bB/\pa t=-\nabla\times\bE$.
The dipole moment of interest here is induced by the electric field. Writing
\be
\bE=\cE_0(\br,t)e^{-i\om t}=e^{-i\om t}\int_{-\infty}^{\infty}d\Delta\tilde{\cE}_0(\br,\Delta)e^{-i\Delta t},
\ee
in which $|\pa\cE_0/\pa t|\ll\om|\cE_0|$ for a quasi-monochromatic field, we approximate $\bd$ as follows:
\bea
\bd(\br,t)&=&\int_{-\infty}^{\infty}d\Delta\alpha(\om+\Delta)\tilde{\cE}_0(\br,\Delta)e^{-i(\om+\Delta)t}\nonumber \\
&\cong& \int_{-\infty}^{\infty}d\Delta[\alpha(\om)+\Delta\alpha'(\om)]\tilde{\cE}_0(\br,\Delta)e^{-i(\om+\Delta)t}
\nonumber \\
&=&\left[\alpha(\om)\cE_0(\br,t)+i\alpha'(\om){\pa\cE_0\over\pa t}\right]e^{-i\om t}.
\label{eqd}
\eea
Here $\alpha'=d\alpha/d\om$ and we assume that higher-order dispersion is sufficiently weak that 
terms $d^m\alpha/d\om^m$ can be neglected for $m\ge 2$. Putting (\ref{eqd}) into (\ref{forceE}), we obtain 
after some straightforward manipulations and cycle-averaging the force
\be
\bF_E=\nabla\left[{1\over 4}\alpha(\om)|\cE|^2\right]+{1\over 4}\alpha'(\om)\bk{\pa\over\pa t}|\cE|^2,
\label{force11}
\ee
where $\cE$ and $\bk$ are defined by writing $\cE_0(\br,t)=\cE(\br,t)e^{i\bk\cdot\br}$. Since the refractive index $n$ of a medium in which local field corrections are negligible is given in terms of $\alpha$ by $n^2-1=N\alpha/\epsoo$, $N$ being the density
of dipoles in the dielectric, we have $\alpha'=(2n\epsoo/N)(dn/d\om)$ and
\be
\bF_E=\nabla\left[{1\over 4}\alpha(\om)|\cE|^2\right]+{\epsoo\over 2N}\bk n{dn\over d\om}{\pa\over\pa t}|\cE|^2.
\label{force1}
\ee
The first term is the ``dipole force" associated with the energy $W=-{1\over 2}\alpha(\om)\bE^2$ involved in inducing
an electric dipole moment in an electric field:
\be
W=-\int_0^{\bE}\bd\cdot d\bE=-\alpha(\om)\int_0^{\bE}\bE\cdot d\bE= -{1\over 2}\alpha(\om)\bE^2.
\ee
The second term in (\ref{force1}) is nonvanishing only because of dispersion
($dn/d\om\ne 0$). It is in the direction of propagation of the field, and implies for a uniform density $N$ of
atoms per unit volume a momentum density of magnitude
\be
P_D={1\over 2}\epsoo n^2{dn\over d\om}{\om\over c}|\cE|^2={1\over 2}{\epsoo\over c}n^2(n_g-n)|\cE|^2,
\label{nels}
\ee
since $k=n(\om)\om/c$. This momentum density comes specifically from the dispersion ($dn/d\om$) of
the medium.

The force $\bF_B$ defined by (\ref{forceB}), similarly, implies a momentum density $\bP^A$
imparted to the medium:
\be
\bP^A=N\bd\times\bB.
\ee
As the notation suggests, this momentum density is associated with the Abraham force density (\ref{abedensity}).
The result of a straightforward evaluation of $\bP^A$ based on (\ref{eqd}) and $\nabla\times\bE=-\pa\bB/\pa t$ is 
\be
\bP^A={1\over 2}\epsoo(n^2-1){\bk\over\om}|\cE|^2, \ \ \ \ P^A={1\over 2}{\epsoo\over c}n(n^2-1)|\cE|^2,
\label{fm}
\ee
when we use $\bk\cdot\bE=0$ and our assumption that $|\dot{\cE}_0|\ll\om|\cE_0|$. The magnitude of the total momentum density in the medium due to the force of the field on the dipoles is therefore
\bea
P_{\rm med}=P_D+P^A&=&{\epsoo\over 2c}\left[n^2(n_g-n)+n(n^2-1)\right]|\cE|^2\nonumber \\
&=&{\epsoo\over 2c}n(nn_g-1)|\cE|^2
\label{pmed}
\eea
in the approximation in which the field is sufficiently uniform that we can ignore the dipole force $\nabla[{1\over 4}\alpha|\cE|^2]$. 

The complete momentum density for the field and the medium is obtained by adding to (\ref{pmed}) the Abraham momentum 
density $P_A$ of the field. According to (\ref{momA}), $P_A=(\epsoo/ 2c)n|\cE|^2$, and so the total momentum density is
\be
P_A+P_D+P^A={\epsoo\over 2c}[n+n(nn_g-1)]|\cE|^2={\epsoo\over 2c}n^2n_g|\cE|^2
\label{tot}
\ee
if the dipole force is negligible.
To express these results in terms of single photons, we again replace $|\cE_0|^2$ by $2\hbar\om/(\epsoo nn_gV)$; 
then (\ref{tot}) takes the form
\be
p_A+p_D+p^A=n{\hbar\om\over c}{1\over V},
\label{ppmed}
\ee
consistent with the discussion in the preceding section. This is the total momentum density per photon, assuming that the dipole force is negligible. The momentum density of the medium per photon follows from (\ref{pmed}):
\bea
p_{\rm med}&=&p_D+p^A={\epsoo\over 2c}n(nn_g-1){2\hbar\om\over nn_g\epsoo V}\nonumber \\
&=&(n-{1\over n_g}){\hbar\om\over c}{1\over V},
\eea
as stated earlier [Eq. (\ref{pmedium})]. 

Consider the example of spontaneous emission by a guest atom in a host dielectric 
medium. The atom loses energy $\hbar\om_0$, and the quantum (photon in the
medium) of excitation carries away from the atom not only this energy but also a linear momentum 
$n\hbar\om/c$ [Eq. (\ref{ppmed})]. The atom therefore recoils with momentum $n\hbar\om/c$ \cite{pwmboyd}.

The momentum density (\ref{nels}) was obtained by Nelson \cite{nelson} in a rigorous treatment of a deformable
dielectric based on a Lagrangian formulation; in the present paper a dielectric medium is
treated as an idealized rigid body. From a microscopic perspective, this part of the momentum density
of the medium is attributable directly to the second term on the right-hand side of (\ref{eqd}), i.e., to the
part of the induced dipole moment that arises from dispersion. In the Appendix the relation of this term to 
the formula (\ref{energydensity}) for the total energy density is reviewed; the term is obviously a
general property of induced dipole moments in applied fields. Consider, for example, a two-level atom driven by
a quasi-monochromatic field with frequency $\om$ far-detuned from the atom's resonance frequency $\om_0$.
In the standard $u,v$ notation for the off-diagonal components of the density matrix in the rotating-wave
approximation \cite{jhe},
\be
u(t)-iv(t)\cong {1\over\Delta}\chi(t)+{i\over\Delta^2}{\pa\chi\over\pa t} + ... \ ,
\label{tla}
\ee
where $\chi(t)$ is the Rabi frequency and $\Delta$ is the detuning. The polarizability is proportional
to $1/\Delta$ in this approximation, and therefore (\ref{tla}) is just a special case of (\ref{eqd}).
 
\section{Momentum Exchange between a Light Pulse and an Induced Dipole}\label{momentumex}
We next consider the momentum exchange between a plane-wave {\sl pulse} and a single polarizable particle. We will assume again that the particle is characterized by a real polarizability $\alpha(\om)$ and that it is surrounded by a host medium with refractive index $n_b(\om)$. The electric field is assumed to be
\be
\bE(z,t)=\cE(t-z/v_{bg})\cos(\om t-kz),
\label{pulse}
\ee
with $k=n_b(\om)\om/c$ and group velocity $v_{bg}=c/n_{bg}$, $n_{bg}=(d/d\om)(\om n_{b})$.

The force acting on the particle is $\bF_E+\bF_B$. $\bF_B$ reduces to 
${1\over 2}\alpha(\om)(\bk/\om)(\pa/\pa t)|\cE|^2$, obtained by multiplying (\ref{fm}) by a volume $V$ 
describing the pulse, replacing $n^2-1$ by $N\alpha/\epsoo$ with $NV=1$ for the single particle, and
differentiation with respect to time. $\bF_E$ follows from (\ref{force11}). Then the force acting on the particle is in the $z$ direction and has the (cycle-averaged) magnitude 
\bea
F&=&{1\over 4}\alpha(\om){\pa\over\pa z}\cE^2+{1\over 4}\alpha'(\om)n_b(\om){\om\over c}{\pa\over\pa t}\cE^2\nonumber \\
&&\mbox{}+{1\over 2c}\alpha(\om)n_b(\om){\pa\over\pa t}\cE^2,
\label{forcey}
\eea 
where now we retain the dipole force, given by the first term on the right-hand side.
The momentum of the particle at $z$ at time $T$ is
\bea
p&=&\int_{-\infty}^T Fdt={1\over 4}\alpha\int_{-\infty}^T{\pa\over\pa z}\cE^2(t-z/v_{bg})dt\nonumber \\
&&\mbox+{1\over 4c}\alpha'n_b\om
\int_{-\infty}^T{\pa\over\pa t}\cE^2(t-z/v_{bg})dt\nonumber \\
&&\mbox{}+{1\over 2c}\alpha n_b\int_{-\infty}^T{\pa\over\pa t}\cE^2(t-z/v_{bg})dt\nonumber \\
&=&-{1\over 4}\alpha{1\over v_{bg}}\cE^2+{n_b\over 4c}\alpha'\om\cE^2+{1\over 2}\alpha{n_b\over c}\cE^2\nonumber \\
&=&{1\over 4c}[(2n_b-n_{bg})\alpha+n_b\om\alpha']\cE^2(T-z/v_{bg}).
\label{for1}
\eea

Hinds and Barnett \cite{Hinds-Barnett-2009} have considered the force on a two-level atom due to a pulse of light in free space.
In this case $n_b=n_{bg}=1$ and (\ref{for1}) reduces to
\be
p={1\over 4c}[\alpha+\om\alpha']\cE^2.
\ee
Following Hinds and Barnett, we argue that a pulse occupying the volume $V$ in the neighborhood of the atom
in free space corresponds to a number $q={1\over 2}\epsoo\cE^2V/\hbar\om$ of photons, so that
\be
p={1\over 2c}[\alpha+\om\alpha']{\hbar\om\over\epsoo V}q.
\ee
$\alpha=\epsoo(n^2-1)/N$, where $n$ is the refractive index in the case of $N$ polarizable particles per unit 
volume. Then
\bea
p&=&{1\over 2c}\left[{\epsoo(n^2-1)\over N}+{2\epsoo n\over N}\om{dn\over d\om}\right]{\hbar\om\over c}q\nonumber \\
&\cong&[n-1+\om{dn\over d\om}]{\hbar\om\over c}q\equiv K{\hbar\om\over c}q.
\label{for2}
\eea
This is the momentum imparted to the particle, which implies a change in {\sl field} momentum per photon equal to 
\be
{\hbar\om\over c}[1-K]\cong{\hbar\om\over c}{1\over 1+K}={\hbar\om\over n_gc}
\ee
if $|K|\ll 1$, where $n_g=(d/d\om)(n\om)$. As in the case of a two-level atom considered by Hinds and Barnett,
this corresponds to the Abraham momentum; our result simply generalizes theirs in replacing $n$ by $n_g$
in the expression for the change in photon momentum.

In the case of a polarizable particle in a host dielectric rather than in free space we obtain, from 
(\ref{for1}),
\be
p={I\over 2\epsoo c^2}[(2-{n_{bg}\over n_b})\alpha+\om\alpha'],
\label{forr2}
\ee
where the intensity $I=(1/2)c\epsoo n_b\cE^2$. If dispersion in the medium and in the polarizability of the guest particle are negligible, we can set $n_{bg}=n$ and $\alpha'=0$, and then (\ref{forr2}) reduces to 
a well known expression \cite{gordon}. However, this momentum can be large in a slow-light 
medium ($n_{bg}$ large), for example, because the gradient of the field (\ref{pulse}) responsible 
for the dipole force on the particle is large \cite{slow}; this is a consequence of the spatial compression of
a pulse in a slow-light medium. We discuss this case further in Section \ref{sec:sphere}.
But first we return to some other well known results that are relevant there.

\section{Electric Dipole Radiation Rate and Rayleigh Scattering}\label{sec:rayleigh}
A Hertz vector $\bPi(\br,\om)$ can be defined for a dielectric
medium, analogous to the case of free space \cite{bornwolf}, by writing the electric and magnetic field 
components at frequency $\om$ as
\be
\bE(\br,\om)=k_0^2[\eps_b(\om)/\epsoo]\bPi(\br,\om)+\nabla[\nabla\cdot\bPi(\br,\om)],
\label{her1}
\ee
\be
\bH(\br,\om)=-i\om\eps_b(\om)\nabla\times\bPi(\br,\om).
\label{her2}
\ee
Here $k_0=\om/c$ and we denote by $\eps_b(\om)$ the (real) permittivity of the dielectric. We will be interested here
in a dipole source inside the ``background" dielectric medium. The identifications (\ref{her1}) and (\ref{her2}) are consistent with the propagation of a wave of frequency $\om$ with the phase velocity $c/n_b(\om)$ in the 
medium [$n_b(\om)=\sqrt{\eps_b(\om)/\epsoo}$], as will be
clear in the following.

The curl of $\bE(\br,\om)$ in (\ref{her1}) is simply
\be
\nabla\times\bE(\br,\om)=k_0^2[\eps_b(\om)/\epsoo]\nabla\times\bPi(\br,\om),
\ee
since the curl of a gradient is zero. Now apply the curl operation to this equation, assuming no free currents and therefore $\nabla\times\bH(\br,\om)=-i\om\bD(\br,\om)$:
\bea
\nabla\times(\nabla\times\bE)&=&i\om\muo\nabla\times\bH=\om^2\muo\bD\nonumber \\
&=&k_0^2[\eps_b(\om)/\epsoo]\nabla\times(\nabla\times\bPi)\nonumber \\
&=&k_0^2[\eps_b(\om)/\epsoo][\nabla(\nabla\cdot\bPi)-\nabla^2\bPi],
\eea
implying
\be
\nabla^2\bPi={\epsoo\over\eps_b}{\om^2\over k_0^2}\muo\bD+\nabla(\nabla\cdot\bPi)=
-{1\over\eps_b}\bD+[\bE-{\eps_b\over\epsoo}k_0^2\bPi],
\ee
\be
\nabla^2\bPi+k^2\bPi=\bE-{1\over\eps_b}\bD, \ \ \ \ \ k^2=k_0^2\eps_b(\om)/\epsoo=n_b^2(\om)\om^2/c^2.
\ee
If $\bD(\br,\om)=\eps_b(\om)\bE(\br,\om)$, the right-hand side is zero, and all we have done is rederived
what we already know: the field propagates with phase velocity $\om/k(\om)=c/n_b(\om)$. Suppose, however,
that within the medium there is a localized source characterized by a dipole moment density $\bP_s(\br,\om)=
\bp_0(\om)\delta^3(\br)$. Then $\bD=\eps_b\bE+\bP_s$ and
\be
\nabla^2\bPi+k^2\bPi=-{1\over\eps_b}\bp_0(\om)\delta^3(\br).
\ee
The solution of this equation for $\bPi(\br,\om)$ is simply 
\be
\bPi(\br,\om)={1\over 4\pi\eps_b(\om)}\bp_0(\om){e^{ikr}\over r},
\label{her102}
\ee
and from this one obtains the electric and magnetic fields due to the source in the medium. In the far field,
assuming $\bp_0=p\hz$ and letting $\theta$ be the angle between the $z$ axis and the observation
point,
\be
E_{\theta}={k_0^2p\over 4\pi\epsoo}\sin\theta{e^{ikr}\over r},
\ee
\be
H_{\phi}={n_bk_0^2p\over 4\pi\epsoo}\sqrt{\epsoo\over\muo}\sin\theta{e^{ikr}\over r},
\ee
in spherical coordinates. The Poynting vector $\bS=\bE\times\bH$ implies the radiation rate
\be
P={n_bp^2\om^4\over 12\pi\epsoo c^3},
\label{here}
\ee
analogous to the fact that the spontaneous emission rate of an atom in a dielectric without local
field corrections is proportional to the (real) refractive index at the emission frequency.

\subsection*{Polarizability of a Dielectric Sphere}
Suppose, somewhat more generally, that the source within the medium occupies a volume $V$ and is characterized
by a permittivity $\eps_s(\om)$. Then $\bD(\br,\om)=\eps(\br,\om)\bE(\br,\om)$, where $\eps=\eps_s(\om)$ within the
volume $V$ occupied by the source and $\eps(\br,\om)=\eps_b(\om)$ outside this volume, and
\be
\nabla^2\bPi+k^2\bPi=[1-\eps(\br,\om)/\eps_b(\om)]\bE.
\ee
The solution of this equation is 
\be
\bPi(\br,\om)=-{1\over 4\pi}\left[1-{\eps_s(\om)\over\eps_b(\om)}\right]\int_V d^3r'\bE(\br',\om){e^{ik|\br-\br'|}\over|\br-\br'|}.
\label{her3}
\ee
Suppose further that the extent of the volume $V$ is sufficiently small compared to a 
wavelength that we can approximate (\ref{her3}) by
\be
\bPi(\br,\om)=-{1\over 4\pi}\left[1-{\eps_s(\om)\over\eps_b(\om)}\right]V\bE_{\rm ins}(\om){e^{ikr}\over r},
\ee
with $r$ the distance from the center of the source (at $\br=0$) to the observation point and $\bE_{\rm ins}(\om)$
the (approximately constant) electric field in the source volume $V$. This has the same form as (\ref{her102})
with $\bp_0(\om)=\eps_b(\om)[\eps_s(\om)/\eps_b(\om)-1]V\bE_{\rm ins}(\om)$. In other words, $\bPi(\br,\om)$
has the same form as the Hertz vector for an electric dipole moment
\be
\bp_0(\om)=[\eps_s(\om)-\eps_b(\om)]V\bE_{\rm ins}(\om).
\label{her4}
\ee
Consider, for example, a small dielectric sphere of radius $a$: $V=4\pi a^3/3$. The field inside such a sphere
is $\bE_{\rm ins}(\om)=[3\eps_b/(\eps_s+2\eps_b)]\bE_b(\om)$, where $\bE_b(\om)$ is the (uniform) electric field
in the medium in the absence of the source. The dipole moment (\ref{her4}) in this case is therefore related to the
external field $\bE_{\rm out}(\om)$ by $\bp_0(\om)=\alpha(\om)\bE_{\rm out}(\om)$, where the polarizability
\be
\alpha(\om)=4\pi\eps_b\left({\eps_s-\eps_b\over\eps_s+2\eps_b}\right)a^3.
\label{her5}
\ee

\subsection*{Rayleigh Attenuation Coefficient}
The cross section for Rayleigh scattering for an ideal gas of refractive index $n(\om)$ can be deduced as 
follows \cite{rayleighnote}. An electric field $\bE_0\cos\om t$ induces an electric dipole moment $\bp(t)=\alpha(\om)\bE_0\cos\om t$ in each of $N$ isotropic, polarizable particles
per unit volume, each particle having a spatial extent small compared to a wavelength. The power radiated by this 
dipole is, from Eq. (\ref{here}),
\be
{dW_{\rm rad}\over dt}=n(\om){\om^4\over 12\pi\epsoo c^3}\alpha^2(\om)\bE_0^2\equiv\sigma_{\rm R}(\om)I,
\label{rayleigh1}
\ee
where $W_{\rm rad}$ denotes energy of the radiated field, $I={1\over 2}n(\om)c\epsoo\bE_0^2$ is the intensity of the
field incident on the dipole, and 
\be
\sigma_{\rm R}(\om)={1\over 6\pi N^2}\left({\om\over c}\right)^4[n^2(\om)-1]^2
\label{rayleigh2}
\ee
is the (Rayleigh) scattering cross section. We have assumed that local field corrections are negligible and used the formula $n^2(\om)-1=N\alpha(\om)/\epsoo$ to express $\sigma_{\rm R}(\om)$ in terms of the refractive index $n(\om)$. The attenuation coefficient is then
\be
a_R=N\sigma_{\rm R}={1\over 6\pi N}\left({\om\over c}\right)^4[n^2(\om)-1]^2.
\label{ray1899}
\ee
Rosenfeld \cite{rosenfeld} obtains instead
\be
a_R=N\sigma_{\rm R}={1\over 6\pi n(\om)N}\left({\om\over c}\right)^4[n^2(\om)-1]^2,
\label{rosenfeld}
\ee
because he does not account for the factor $n(\om)$ in the dipole radiation rate (\ref{rayleigh1}). 
Rayleigh's derivation of
(\ref{ray1899}) follows essentially the one just given, but the factor $n(\om)$ appears in neither the dipole radiation
rate nor the expression for the intensity (or actually, in his derivation, the energy density) \cite{rayleigh}. In practice the difference
between (\ref{ray1899}) and (\ref{rosenfeld}) is negligible for the case assumed here of a dilute medium \cite{meas}.

\section{Force on a Dielectric Sphere}\label{sec:sphere}
The expression (\ref{forcey}) for the force on a polarizable particle in a field (\ref{pulse}) may be generalized
to allow for absorption by the particle simply by taking the polarizability $\alpha(\om)$ in (\ref{eqd}) to
be complex. Assuming again that $\cE$ is slowly varying in time compared to $\exp(-i\om t)$, and slowly varying in
space compared to $\exp(ikz)$, we obtain
\be
F={1\over 4c}[(2n_b-n_{bg})\alpha_R+n_b\om \alpha_R']{\pa\over\pa\tau}|\cE|^2
+{1\over 2}n_b{\om\over c}\alpha_I|\cE|^2,
\label{forceyy}
\ee
where $\tau=t-n_{bg}z/c$ and $\alpha_R$ and $\alpha_I$ are the real and imaginary parts, respectively, of $\alpha(\om)$. If we replace $n_{bg}$ by $n_b$ and take $\alpha_R'\cong 0$, we recover
results that may be found in many previous works when absorption is assumed to be negligible \cite{gordon}. The last term 
in (\ref{forceyy}) is the absorptive contribution to equation (7) of a paper by Chaumet and Nieto-Vesperinas 
\cite{chaumet} when the field is assumed to have the form (\ref{pulse}). 

The polarizability in the case of a dielectric sphere of radius $a$ much smaller than the wavelength of the field
is given by (\ref{her5}). Dispersion affects the force (\ref{forceyy}) both through the group index
($n_{bg}$) of the host dielectric medium and the variation of the real part of the sphere's polarizability with frequency ($\alpha_R'$). The latter
depends on both the intrinsic frequency dependence of the permittivity of the material of the sphere and the frequency dependence of the refractive index of the host medium. If these dispersive contributions to the force exceed the remaining two contributions to the force (\ref{her5}), 
\be
F\cong{1\over 4c}\left[-\alpha_Rn_{bg}+n_b\om\alpha_R'\right]{\pa\over\pa\tau}|\cE|^2.
\ee
Using (\ref{her5}) for this case, we obtain
\be
F\cong -{3\pi\epsoo a^3\over c}n_{bg}{n_s^2n_b^4\over(n_s^2+2n_b^2)^2}{\pa\over\pa\tau}|\cE|^2
\label{forceyyy}
\ee
if the dispersion of the dielectric material constituting the sphere is much smaller than that of the
host dielectric medium, i.e., if $d\eps_s/d\om\ll d\eps_b/d\om$. (Here $n_s$ is the refractive index at frequency $\om$ of the material of the sphere.) This result implies that, in the case of a slow-light host medium ($n_{bg}\gg 1$), the force on the sphere can be much larger than would be the case in a ``normally dispersive" medium, and is in the
direction opposite to that in which the field propagates.

The simple formula (\ref{forceyyy}), and similar expressions obtained in other limiting cases of (\ref{forceyy}),
obviously allow for a wide range of forces when a pulse of radiation is incident on a dielectric sphere in a host dielectric medium. Here we make only a few remarks concerning the last term in (\ref{forceyy}). Although
we have associated this contribution to the force with absorption, such a force appears even if the
sphere does not absorb any radiation of frequency $\om$. This is because there must be an imaginary
part of the polarizability simply because the sphere scatters radiation and thereby takes energy out of the
incident field. According to the optical theorem in this case of scattering by a nonabsorbing polarizable particle that
is small compared to the wavelength of the field, the imaginary part of the polarizability
is related to the complete (complex) polarizability as follows \cite{pwmetc}:
\be
\alpha_I(\om)={1\over 4\pi\epsoo}{2\om^3\over 3c^3}n_b|\alpha(\om)|^2.
\label{ot1}
\ee
Then the force proportional to $\alpha_I(\om)$ in (\ref{forceyy}) is
\be
F_{\rm scat}\equiv {1\over 2}n_b^5{\om\over c}\alpha_I|\cE|^2={8\pi\over 3}\left({\om\over c}\right)^4
{n_b^5I\over c}\left({\eps_s-\eps_b\over\eps_s+2\eps_b}\right)^2a^6,
\ee
which is just the well known ``scattering force" \cite{padgett} on a dielectric sphere in a medium with refractive index $n_b$, which may be taken to be real in the approximation in which the field is far from any absorption resonances
of the sphere.

\section{Conclusions}\label{sec:conclusions}
In this attempt to better understand the different electromagnetic momenta and the forces on electrically
polarizable particles in dispersive dielectric media, we have made several simplifications, including
the neglect of any surface effects, the treatment of the medium as a nondeformable body, and the approximation
of plane-wave fields. We have shown that conservation of momentum, even in seemingly simple examples
such as the Doppler effect, generally requires consideration not only of the
Abraham momentum and the Abraham force, but also of a contribution to the momentum of the medium due
specifically to the dispersive nature of the medium. We have generalized some well known expressions for the
forces on particles immersed in a dielectric medium to include dispersion. While we have presented arguments
in favor of the interpretation of the Abraham momentum as the momentum of the field, our simplified analyses
lead us to the conclusion that neither the Abraham nor the Minkowski expressions for momentum give the
recoil momentum of a particle in a dispersive dielectric medium. Finally we have shown that the force exerted
on a particle in a strongly dispersive medium is approximately proportional to the group index $n_{bg}$, and can therefore become very large in a slow-light medium.

\chapter{Nonstationary electromagnetics of controllably dispersive media}\label{ch:nonstationary}

\newcommand{\eq}[1]{Eq.~(\ref{#1})}


\section*{Abstract}
Recent experiments have demonstrated that it is possible to alter the dispersion of a medium without significantly altering its absorption or refractive index and that this may be done while a wave propagates through the medium.  This possibility opens up a new set of potential experiments to the field of nonstationary optics.  We consider the basic kinetics of waves propagating through a medium whose group and phase velocities are a function of position and time.  We compare the dynamics of waves propagating through two homogeneous media, one a nondispersive medium with a time dependent phase velocity, and one a dispersive medium with a time dependent group velocity and show that new dynamic effects accompany new kinetic ones.


\begin{section}{Introduction}\label{se:intro}
The terms \emph{nonstationary}, \emph{inhomogeneous}, and \emph{dispersive} describe media whose properties vary with time, position, and frequency, respectively.  In this paper, we consider transformations wrought upon waves as they propagate through nonstationary, inhomogeneous media with variable dispersion.  Our interest in this topic is motivated by experimental advances in the manipulation of dispersion through nonlinear optics.  Recent experiments have demonstrated that it is possible to achieve either strong normal dispersion (in the case ``slow-light media \cite{Hau-Harris-Dutton-Behroozi-1999,Liu-Dutton-Behroozi-Hau-2001}'') or strong anomalous dispersion (in the case of ``superluminal media \cite{Wang-Kuzmich-Dogariu-2000}'') in nearly transparent media using nonlinear optical effects.  In either case, the strength of the dispersion, as experienced by a probe beam, depends on the intensity of one or more controllable auxiliary beams.  If auxiliary intensities are changed while the probe beam is in transit, then the dispersion, as perceived by the probe beam, is time dependent.  These media, which we will refer to collectively as controllably dispersive media, thus form a newly accessible class of nonstationary media.  Although the nonstationarity of a slow light medium has been used experimentally to adiabatically transform a ``slow light'' pulse to a ``stopped'' one \cite{Liu-Dutton-Behroozi-Hau-2001,Ginsberg-Garner-Hau-2007}, an explicit connection between this application and the field of nonstationary electromagnetics has not, to our knowledge, been made.

The field of nonstationary electromagnetics, although much less developed than its inhomogeneous counterpart, is now a venerable one.  In 1958, Morgenthaler introduced propagation equations for electromagnetic waves in an isotropic nondispersive medium whose permittivity and permeability were allowed to vary with time (but not with space) \cite{Morgenthaler-1958}.  Although the propagation equations are not generally analytically soluble, he found useful solutions assuming, on the one hand, a step functional time dependence and, on the other hand, an adiabatic time dependence.  He found, among other things, that the frequency of a wave would vary with the permittivity and permeability in such a way that wavelength would be preserved.  

Over the next two decades, much of the progress of the field was in the Soviet Union, with a particular focus on the behavior of light in plasmas.  We refer the reader to Stepanov and colleagues \cite{Stepanov-1993,Kratsov-Ostrovsky-Stepanov-1974} for an introduction to this literature and highlight here two aspects of it.  First, the two simplified cases (step function and adiabatic changes) used by Morgenthaler continued to be used extensively, and were generalized from a purely nonstationary picture to one that allowed inhomogeneities to propagate at a fixed velocity according to the traveling wave law ($f=f(x-vt)$) \cite{Stepanov-1993, Stepanov-1969}.  This picture allows a single framework to unite the purely nonstationary case ($v\rightarrow \infty$) \footnote{Here we note a basic characteristic of nonstationary media: there is no maximum speed because changes in a medium at two different positions need not be causally connected to each other.}, a 1-dimensional purely inhomogeneous case ($v=0$), and the case of an inhomogeneity (for example, an ionization front) moving at any velocity in between these two extremes.  
Second, a distinction between ``kinetic'' and ``dynamic'' phenomena was found to be useful in differentiating between those results of the theory that are general to all linear wave phenomena and those that are specific to particular media \cite{Stepanov-1969, Stepanov-1993, Sorokin-Stepanov-1971}.  We employ both of these concepts in this paper.

The advent of the laser, and particularly the pulsed laser, introduced a new way to dynamically alter the characteristics of a medium.  Although they were apparently unaware of previous work in nonstationary electromagnetics, a few authors noted the possibility of using light to alter the phase velocity of propagating electromagnetic waves.  In 1977, Lampe, Ott, and Walker noted that an ionizing laser might be swept across a gas to create a superluminal ionization front that could interact with a microwave pulse in a nonstationary fashion \cite{Lampe-Ott-1977}.  In 1988, Wilks, Dawson, and Mori examined the problem of a wave propagating through a medium which is then quickly ionized by a high-intensity ultrashort pulse \cite{Wilks-Dawson-Mori-1988}.  Large frequency shifts were soon realized for waves reflecting off of a relativistically propagating ionization front \cite{Savage-Joshi-Mori-1992}.  These works founded a sub-literature surrounding nonstationary effects in plasmas \cite{Berezhiani-Mahajan-Miklaszewski-1999, Hashimony-Zigler-Papadopoulos-2001,Geltner-Avitzour-Suckewer-2002,Avitzour-Geltner-Suckewer-2005, Shvartsburg-2005,Avitzour-Shvets-2008}.  We point the reader to Shvartsburg, who reviewed some of the recent work on nonstationary effects in plasmas \cite{Shvartsburg-2005}.  We also highlight the work of Avitzour and Shvets, whose 2008 paper relies on controlled dispersion in nonstationary media.  They proposed a method for altering the spectral width of a pulse without changing its central frequency using controlled dispersion in a nonstationary magnetized plasma \cite{Avitzour-Shvets-2008}.  We will look at this method from a kinetic perspective in the next section.

In parallel with the more recent work on nonstationary plasmas, several authors have explored electromagnetic propagation through nonstationary media abstractly, in the tradition of Morgenthaler \cite{Dodonov-Klimov-Nikonov-1993, Biancalana-Amann-Uskov-Oreilly-2007,Budko-2009}.  Dodonov, Klimov, and Nikonov studied the quantization of a linear, nonstationary medium and were able to quantitatively relate temporal changes in dielectric permittivity to photon generation \cite{Dodonov-Klimov-Nikonov-1993}. Biancalana, Amann, Uskov, and Oreilly treated a 1-dimensional nondispersive nonstationary medium by introducing a transmission matrix for moving interfaces.  Using this matrix they generalized Bragg reflection to propagating interfaces and showed that temporal periodicity leads to k-vector bandgaps just as spatial periodicity leads to frequency bandgaps \cite{Biancalana-Amann-Uskov-Oreilly-2007}.  Budko has found an analogy between a nonstationary nondispersive medium and an expanding Universe \cite{Budko-2009}.

One surprising fact about recent works, both those focusing on nonstationary plasmas \cite{Lampe-Ott-1977,Wilks-Dawson-Mori-1988,Savage-Joshi-Mori-1992,Berezhiani-Mahajan-Miklaszewski-1999, Hashimony-Zigler-Papadopoulos-2001,Geltner-Avitzour-Suckewer-2002,Avitzour-Geltner-Suckewer-2005, Shvartsburg-2005,Avitzour-Shvets-2008} and those dealing with abstract, nondispersive, linear media \cite{Dodonov-Klimov-Nikonov-1993, Biancalana-Amann-Uskov-Oreilly-2007,Budko-2009}, is that they make no reference to the substantial Soviet literature from the 1960s and 1970s.  One consequence of the disjointedness of which this fact is symptomatic is that basic principles have had to be rediscovered, multiple times, in incidental and typically incomplete ways.  For example, a basic principle of wave propagation in nonstationary linear media is the importance and generality of kinetic effects. Simple spatial and temporal symmetries play an important role in determining the effects of nonstationary propagating waves.  Aspects of these symmetries have been rediscovered and used in many contexts \cite{Morgenthaler-1958, Stepanov-1969, Wilks-Dawson-Mori-1988,Biancalana-Amann-Uskov-Oreilly-2007,Avitzour-Shvets-2008,Budko-2009}.

We hope that this paper, in addition to providing a simple theoretical basis for treating a new class of nonstationary dispersive media, will demonstrate the utility of explicitly recognizing kinetic constraints and of maintaining a clear distinction between kinetic and dynamic aspects of wave behavior.  Section~\ref{Sec:symmetry} is dedicated to an exploration of the interplay between dispersion, nonstationarity, and inhomogeneity from a strictly kinetic perspective.  We introduce three basic kinetic relationships representing three basic preserved symmetries. Along with appropriate dispersion relationships, they allow for the derivation of all the subsequent kinetic results.  Although these relationships have not, to our knowledge, been explicitly presented together before, they have been implicit in many previously derived results.  We show that they lead to Snell's Law for a motionless interface and to appropriate Doppler shifts for a moving one, that they lead to the preservation of wavelength for homogeneous nonstationary media and to the preservation of frequency in inhomogeneous stationary media, how they give rise to Biancalana \emph{et al.}'s ``generalized frequency'' and give insight to Avitzour and Shvets' proposal for compressing pulse spectrum without altering carrier frequency.  We also find some effects that are described here for what may be the first time.  For example, we find that dispersion modulates the frequency response of a wave to temporal changes in the refractive index and that Doppler reflections may lead to large changes in pulse bandwidth when the moving interface approaches the group velocity.  We emphasize that although we have derived these results with controllably dispersive nonlinear optical media in mind, the results apply to any type of wave propagating through any linear medium.

Unlike kinetic effects, the dynamic aspects of wave behavior depend on the specific microscopic details of interactions between nonstationary media and fields.  In Section~\ref{Sec:comp}, we develop boundary conditions appropriate to an idealized version of the nonlinear optical media currently used to achieve exotic dispersion.  Unsurprisingly, we find that these boundary conditions are fundamentally different from those posited for nondispersive media.  We compare pulse transformations wrought by a changing phase velocity in a specific, homogeneous, nondispersive medium (originally described by Morgenthaler \cite{Morgenthaler-1958}), with those wrought by a change in group velocity in our idealized dispersive medium.  We note many interesting similarities and differences in their effects on twenty key quantities such as energy, momentum, and photon density.

\end{section}

\begin{section}{Kinetics of dispersive, inhomogeneous, nonstationary media}\label{Sec:symmetry}

\begin{figure}
	\centerline{\includegraphics[width=8.5cm]{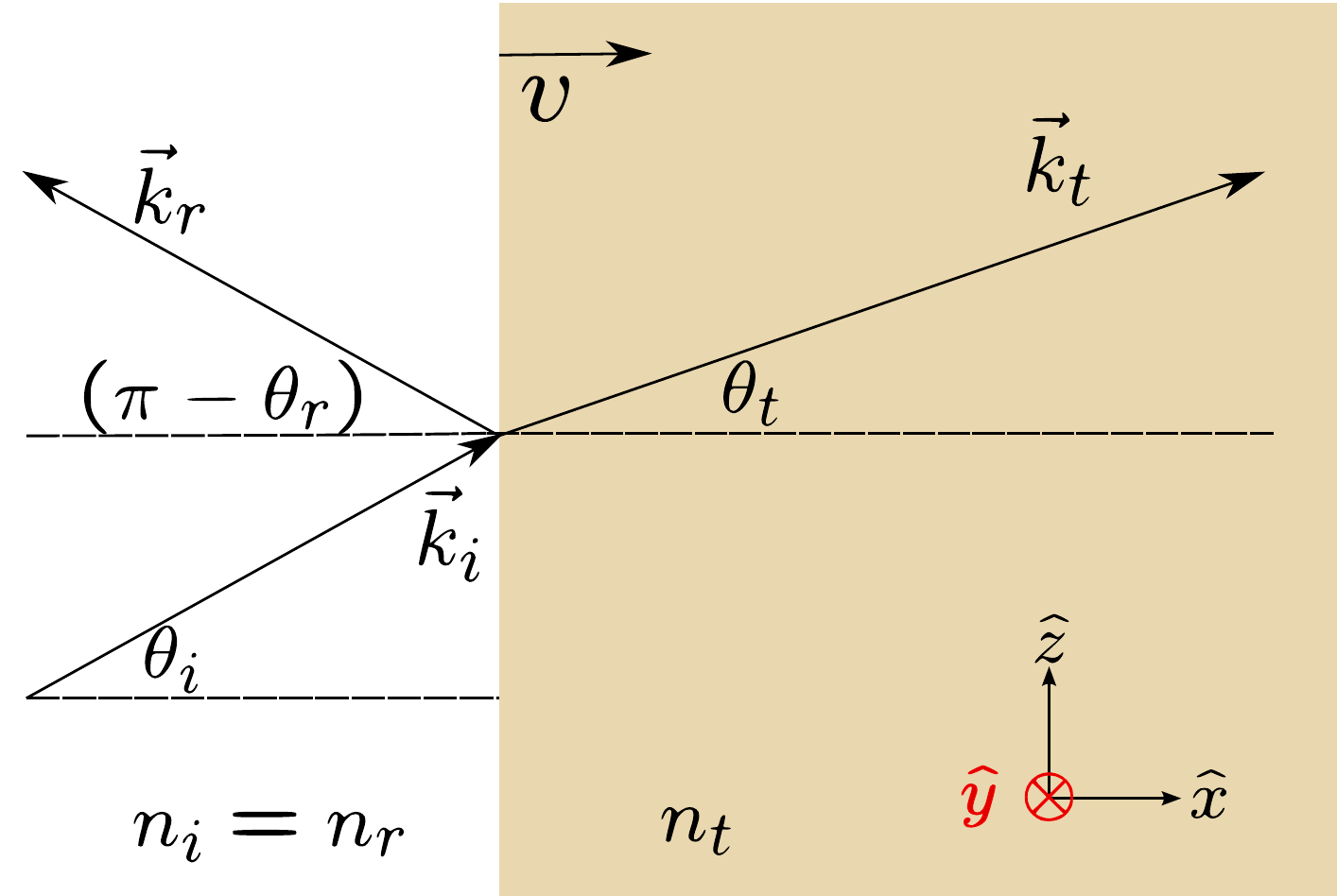}}
	\caption{\label{Fig:interface} An incident plane wave with k-vector $\vec k_i$ interacts with a planar interface.  This results in a refracted wave with k-vector $\vec k_r$ and a transmitted wave with k-vector $\vec k_t$.  For a non-dispersive and isotropic medium, the refractive index for the reflected wave ($n_r$) is the same as that for the incident wave ($n_i$).}
\end{figure}

Consider a plane wave incident on an infinite planar interface, as drawn in Fig.~\ref{Fig:interface}.  Its wave vector and frequency describe periodic translational symmetries in 3 dimensions of space and 1 of time.  A planar interface breaks symmetry in only one dimension, i.e., the one parallel to its normal.
If an interface is motionless, then its normal is purely spatial; in the geometry of Fig.~\ref{Fig:interface}, we could write $\hat n= \hat x$.  The normal of a moving interface in 3+1 dimensions is partially rotated into time; if we introduce the temporal unit vector $\hat{ct}$ and a temporal rotation angle $\phi$, then we may write $\hat n = \cos\phi \,\hat x+\sin\phi\,\hat{ct}$.  The angle $\phi$ is related to the velocity $v$ and the normalized velocity $\beta$ by
\begin{equation}\label{Eq:phi}
 \frac{v}{c}=\beta=\tan\phi.
\end{equation}
All the other translational symmetries of the incident plane wave are preserved under reflection and transmission.
Thus,
\begin{equation}\label{Eq:ky}
k_{yi}=k_{yr}=k_{yt},
\end{equation} 
\begin{equation}\label{Eq:kz}
k_{zi}=k_{zr}=k_{zt},
\end{equation} 
and
\begin{equation}\label{Eq:ktx}
\frac{\omega_i}{c}-k_{xi}\beta=\frac{\omega_r}{c}-k_{xr}\beta=\frac{\omega_t}{c}-k_{xt}\beta,
\end{equation}
where the geometry is again defined by Fig.~\ref{Fig:interface}.
These three equations, combined with appropriate dispersion relationships, provide a basis for the kinetic aspects of wave behavior.

Equations (\ref{Eq:ky}), (\ref{Eq:kz}), and (\ref{Eq:ktx}) represent quantities that are unchanged by a planar interface.  Extending this concept, we see that if these quantities are not altered by a single interface, they cannot be altered by multiple similar interfaces (``similar'' means here that they share the same velocity and the same normal).  By progressively increasing the number and variety of similar interfaces, we can similarly see that the same quantities are preserved under interaction with any pattern of medium parameters that obeys the traveling wave law, 
\begin{equation}\label{Eq:travel}
f=f(x-vt).
\end{equation}

When the planar disturbance described by the function $f$ in \eq{Eq:travel} is more complicated than a single interface, the description of the wave response in terms of a single incident, reflected, and transmitted wave becomes inadequate.  In order to express conserved quantities appropriately for this more general case, we now introduce the convention that a line appearing over a given quantity means that the quantity is invariant under planar changes in the material parameters.  Using this convention, we extend Equations (\ref{Eq:ky}), (\ref{Eq:kz}), and (\ref{Eq:ktx}) to
\begin{equation}\label{Eq:kyg}
 \overline{k_y},
\end{equation} 
\begin{equation}\label{Eq:kzg}
 \overline{k_z},
\end{equation} 
and
\begin{equation}\label{Eq:ktxg}
 \overline{\frac{\omega}{c}\cos(\phi)-k_x \sin(\phi)}.
\end{equation}
Despite their lack of an explicit equality, we refer these statements equations in acknowledgment of the fact that they are, in fact, compressed equalities.

\subsection{Snell's Law and Doppler shifts}
For a motionless interface, $\beta=0$ and \eq{Eq:ktx} reduces to the statement that frequency is preserved:
\begin{equation}\label{Eq:omega}
 \omega_i=\omega_r=\omega_t.
\end{equation}  
Combining this with \eq{Eq:kz} and the expansion $k_z=n\omega\sin\theta/c$ yields
\begin{equation}\label{Eq:Snell}
 \sin\theta_i=\sin\theta_r=\sin\theta_t,
\end{equation} 
which is Snell's Law (note that Snell's Law applies to the reflected wave as well as the transmitted wave).

If we allow the interface to move at a velocity $v$, frequency is no longer conserved, and Snell's Law will no longer hold \footnote{For a moving object in a vacuum, this is easily solved via Lorentz transformation.  For general nonstationary objects, such a transformation may be costly: isotropic media may be come anisotropic.  In addition, Lorentz transformations cannot be used to simplify the treatment of interfaces moving faster than the speed of light.}.  If both sides of the interface are dispersionless, then a simple extension of Snell's Law obtained by combining \eq{Eq:ktx} and \eq{Eq:kz} may be useful:
\begin{equation}\label{Eq:Snell2}
 \frac{n_t\sin\theta_t}{1-n_t \cos\theta_t \beta}=\frac{n_i\sin\theta_i}{1-n_i \cos\theta_i \beta}.
\end{equation} 
One interesting fact about this formula is that it shows that there are some cases where there two sets of angles and frequencies for the transmitted wave \footnote{Multiple transmitted solutions can correspond to the absence of a reflected solution.  This has been noted in the past for moving interfaces.  See \cite{Lampe-Ott-1977}}.
When dispersion is important, \eq{Eq:Snell2} loses its utility because $\theta_t$ and $n_t$ both become frequency dependent.  We may still find angle, refractive index, and frequency using both relationships
\begin{equation}
 n_t(\omega_t)\omega_t\sin\theta_t=n_i(\omega_i)\omega_i\sin\theta_i,
\end{equation} 
and
\begin{equation}
 \omega_t(1-n_t(\omega_t)\cos\theta_t\beta)=\omega_i(1-n_i(\omega_i)\cos\theta_i\beta),
\end{equation} 
combined with an explicit function for $n_t(\omega_t)$ to calculate the properties of the transmitted wave.  We will explore the effects of dispersion in a simplified case, that of the transmitted wave at normal incidence, in the next subsection.

The relationships for the reflected wave,
\begin{equation}
 n_r(\omega_t)\omega_r\sin\theta_r=n_i(\omega_i)\omega_i\sin\theta_i,
\end{equation} 
and
\begin{equation}\label{Eq:Doppler2}
 \omega_r(1-n_r(\omega_r)\cos\theta_r\beta)=\omega_i(1-n_i(\omega_i)\cos\theta_i\beta),
\end{equation}
combined with an explicitly function for $n_r(\omega_r)$ appear identical to those for the transmitted wave.  However, the behavior is different in this case because the medium that defines $n_r$ is the same as the medium that defines $n_i$ and because the sign of $\cos\theta$ is different in the reflective case.  Numerical investigation suggests that the Doppler effect grows large when $|v_g| \approx v$ for an approaching wall.  In Subsection \ref{ss:Doppler1D} we will show analytically that this is the case for a wave that is normal to the interface.  For now, we note that for normal incidence on a moving interface in a vacuum, $\cos\theta_r=-1$, $\cos\theta_i=1$, and $n_r=n_i=1$.  Then \eq{Eq:Doppler2} gives the standard reflective Doppler shift,
\begin{equation}
 \omega_r=\omega_i\frac{1-\beta}{1+\beta}.
\end{equation}  
Note that in the geometry of Fig.~\ref{Fig:interface} a positive value for $\beta$ means that the interface is receding.

\subsection{Transmission at normal incidence}
In controllably dispersive media, large changes in group velocity may be accompanied by minute changes in phase velocity.  In addition, the phase velocity changes are gradual in the sense that they occur over several carrier frequency cycles.  Under these conditions, reflection is negligible.
At normal incidence, $\cos\theta_i=\cos\theta_t=1$.  Then \eq{Eq:ktxg} simplifies to
\begin{equation}\label{Eq:gen_omega}
 \overline{\frac{\omega}{c}\cos\phi-k\sin\phi},
\end{equation}
where we have removed the subscript $x$ from $k_x$ in acknowledgment of the fact that at normal incidence $k=k_x$.
This mixture of wave vector and frequency is the conserved quantity that Biancalana \emph{et al.} found and referred to as the generalized frequency \cite{Biancalana-Amann-Uskov-Oreilly-2007}.  When $\phi=0$, the medium is stationary and frequency is preserved.  When $\phi= \pm\pi/2$, the medium is homogeneous and wave vector is preserved.  In between these extremes, a mixture between wave vector and frequency is preserved.

The preservation of the generalized frequency is accompanied by a second constraint: 
\begin{equation}\label{Eq:v_p}
 \frac{\omega}{k}=v_p,
\end{equation} 
where $v_p$ is the phase velocity.  When the phase velocity changes, the ratio $\omega/k$ must change as well.  Together, Eqs. (\ref{Eq:gen_omega}) and (\ref{Eq:v_p}) determine the frequency and wave vector responses.  We now look at the impact of dispersion upon these responses.  Consider a medium for which the spectral dependence and the time dependence of the refractive index are independent so that we may apply $n=n_o(x,t)+(\delta n/(\delta \omega))(\omega-\omega_o)$ for frequencies near $\omega_o$.
From
\begin{equation}
 \frac{d}{dn_o}\left(\frac{\omega}{c}\cos\phi-k\sin\phi\right)=0,
\end{equation} 
we obtain
\begin{equation}
 \frac{d\omega}{dn_o}=\frac{\omega \sin\phi}{\cos\phi-n_g \sin\phi},
\end{equation} 
where $n_g=n+\omega \delta n/\delta \omega$.  Since $\tan\phi=v/c$, we can also write
\begin{equation}
 \frac{d\omega}{dn_o}=\frac{\omega v/c}{1-v/v_g},
\end{equation}
which emphasizes the large frequency shifts expected when the interface velocity $v$ approaches the group velocity $v_g$.

\subsection{Preservation of a generalized bandwidth in 1+1 dimensions}\label{sse:generalized-bandwidth}
The group velocity is defined using $v_g=d\omega/dk$.  When group velocity dispersion may be neglected, either because it is small, or because we are limiting ourselves to a narrow spectrum, the approximation
\begin{equation}\label{Eq:domega-dk}
 \frac{\Delta \omega}{\Delta k} \approx v_g
\end{equation}
becomes useful. 
As formalized in Eq. ~(\ref{Eq:domega-dk}), when a narrow band pulse experiences a change in group velocity, either its spectral bandwidth $\Delta \omega$, or its spatial bandwidth $\Delta k$, or both, must change.  Extending Eq. ~(\ref{Eq:gen_omega}), we find a generalized bandwidth,
\begin{equation}\label{Eq:gen-bandwidth}
\overline{\frac{\Delta \omega}{c} \cos\phi-\Delta k \sin\phi},
\end{equation} 
that is preserved when a pulse interacts with an interface moving at the velocity defined by $\phi$ and Eq ~(\ref{Eq:phi}).  \eq{Eq:gen-bandwidth} shows that the division of change between spectral and spatial bandwidth depends on their comparative magnitudes and on the interface velocity.

The preserved generalized bandwidth provides us with a simple tool that we can use to analyze one particular aspect of the stopped light experiment of Liu, Dutton, Behroozi, and Hau \cite{Liu-Dutton-Behroozi-Hau-2001}.  In the Liu experiment, a coupling beam was used to control the group index perceived by a probe beam in real time.  However, this coupling beam had a finite speed and was copropagating with the probe beam.  Because of this, changes to the group index perceived by the probe beam were not instantaneous, but moved at nearly the speed of light.  In their report \cite{Liu-Dutton-Behroozi-Hau-2001}, and in earlier theoretical work \cite{Fleischhauer-Lukin-2000}, the speed of the coupling beam was treated as effectively infinite, because it was so much faster than the slowed advancement of the probe pulse.  Using the preserved generalized bandwidth, we now show that treating the velocity of the coupling beam as essentially infinite was justified.

When the pulse described by Liu \emph{et al.} entered the slow light medium, it crossed a spatial interface, and $\phi$ was $0$.  The preserved quantity associated with this case was $\overline{\Delta \omega/c}$, meaning that spectral bandwidth was preserved.  The spatial bandwidth ($\Delta k$) and the group index ($n_g$) were increased by 7 orders of magnitude, and Eq. ~(\ref{Eq:domega-dk}) remained satisfied.  

While the new foreshortened pulse was propagating through the slow-light medium, the coupling beam was then turned off.  The speed of the coupling beam was approximately $c$, so that $\phi$ was approximately $\pi/4$.  Thus, the preserved generalized frequency was approximately $\overline{(\Delta\omega/c-\Delta k)/\sqrt{2}}$, which appears to distribute changes evenly between spectral and spatial bandwidths.  However, the slow speed of the probe envelope was reflected by the fact that $\Delta k$ was larger than $\Delta \omega/c$ by a factor of 10$^7$.  According to the preservation of the generalized bandwidth, $\Delta \omega/c$ would have gone to zero while $\Delta k$ changed by 1 part in 10$^7$.  This is the size of the error introduced by assuming that the coupling speed was effectively infinite.

The end result of the Liu experiment was a reconstitution of the original pulse.  In the next section we use the concept of preserved generalized bandwidth to show how the pulse may be spectrally compressed or expanded using transmission in controllably dispersive media.

\subsection{Pulse compression/decompression using controlled dispersion}
\begin{figure}
  \centerline{\includegraphics[width=8.5cm]{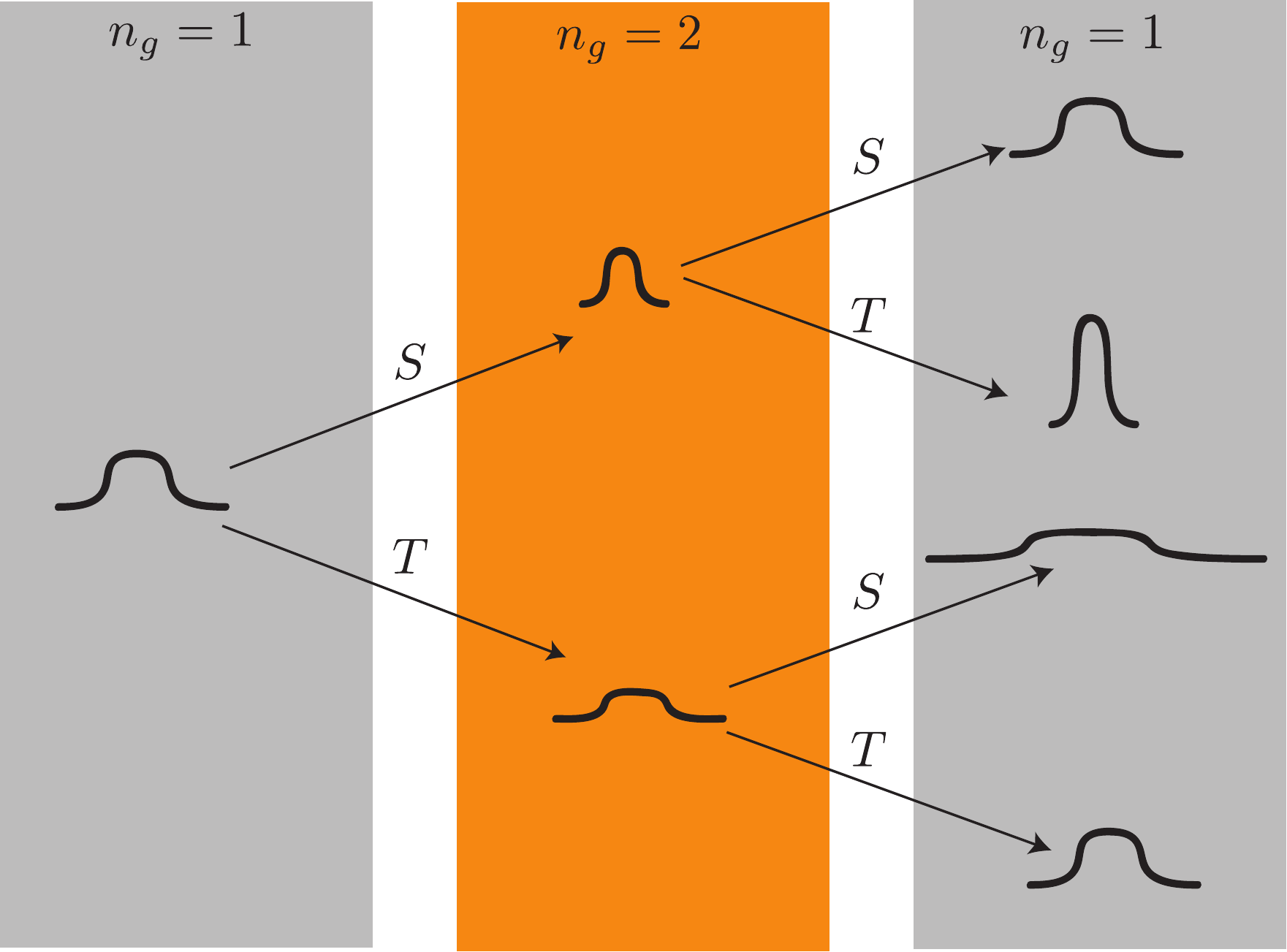}}
  \caption{\label{Fig:pulse-scaling} Field energy density verses spatial extent of a pulse.  S: when a pulse undergoes a spatial change in group index ($n_g$) it scales longitudinally but not temporally.  T: when a pulse undergoes a temporal change in ($n_g$), it changes temporally but not longitudinally.  By mixing spatial and temporal interfaces, it is possible to scale the pulse bandwidth and longitudinal extent without otherwise changing the envelope shape.}
\end{figure}

The generalized bandwidth is conserved across any number of interfaces so long as they have the same velocity.  By mixing interfaces of different velocities, we can change that bandwidth.  In Fig.~\ref{Fig:pulse-scaling}, a pulse begins in a medium with a group index $n_g$ of 1, transitions to one with $n_g=2$, and then back to $n_g=1$.  When both transitions are spatial (S corresponds to a motionless planar interface) or when both are temporal (T corresponds to time-dependent but spatially independent change to the medium--the infinite velocity case) the pulse ends up in its original form.  By mixing spatial and temporal transformations, the pulse can either be compressed or extended.  In this way the pulse bandwidth may be compressed or decompressed while the pulse envelope is scaled longitudinally but otherwise unchanged.

This mechanism for pulse compression and decompression was recently proposed and demonstrated in simulation for a plasma geometries, where the group velocity may be controlled dynamically via an applied magnetic field \cite{Avitzour-Shvets-2008}.  We note here that the Liu experiment demonstrates all the capabilities necessary to perform pulse compression or expansion using nonlinear optics \cite{Liu-Dutton-Behroozi-Hau-2001}.

\subsection{Dispersive modulation of the 1-D Doppler shift}\label{ss:Doppler1D}
In a dispersive medium, Doppler shifts are difficult to predict intuitively because of the interdependence between refractive index and frequency.  Here we find an intuitive formula for the change in frequency in the case where the group velocity differs from the phase velocity, but where the range of frequencies concerning us is sufficiently narrow that we may neglect group velocity dispersion.  Expanding \eq{Eq:ktx} for the case of reflection at normal incidence, we get 
\begin{equation}\label{Eq:1D-Dopper-general}
 \omega_i\left(1-n_i \frac{v}{c}\right)=\omega_r\left(1+n_r \frac{v}{c}\right).
\end{equation}
To derive a Doppler shift, we must introduce an explicit dispersion relationship $n=n(\omega)$.  In general, the resulting equation is analytically intractable though a numerical solution is readily found.  An interesting argument for the physical relevance of the concept of group velocity to Doppler shifts in dispersive media is that if we require that the group velocity be constant with frequency a simple analytical solution results.  Requiring that $n_g$, the group index, remain constant with frequency gives a refractive index of the form $n=n_g+b/\omega$, where $b$ can be any real constant.  Applying this dispersion relationship to Eq.~(\ref{Eq:1D-Dopper-general}) yields the exact expression,
\begin{equation}\label{Eq:Doppler-no-dispersion}
 \frac{\Delta}{\omega}=-2 n \frac{v}{c},
\end{equation}
where $\beta=v/c$, $\Delta=\omega_r-\omega_i$, $\omega_r+\omega_i=2\omega$, and $n=n(\omega)$.
Eq.~(\ref{Eq:Doppler-no-dispersion}) is attractive because of its symmetry but incomplete because it depends on the quantity $\omega$, which it does not resolve.  Rewriting Eq.~(\ref{Eq:Doppler-no-dispersion}) in terms of $\omega_i$ and $\Delta$
\begin{equation}\label{Eq:Doppler-dispersion}
 \frac{\Delta}{\omega_i}=\frac{-2 n(\omega_i) v/c}{1+v/v_g},
\end{equation} 
which clearly reveals the role of dispersion in determining the expected Doppler shift.
Eq.~(\ref{Eq:Doppler-dispersion}) has the interesting interpretation that the Doppler shift will be large when the interface is approaching (if $v_g>0$) or receding (if $v_g<0$) with a speed $|v| \approx |v_g|$.  This result is exact when the group velocity is constant with frequency.

\subsection{Dispersion and the 1-D Doppler effect on pulse bandwidth}
Intuitively, we expect that a pulse reflecting from an interface whose velocity is close to that of the pulse will be temporally stretched if the interface is receding and compressed if the interface is approaching.  We expect this temporal stretching/compressing should have a corollary decrease or increase of the spectral bandwidth, $\Delta \omega$.  We now show that this is the case.

Adapting expression~(\ref{Eq:gen-bandwidth}), we find
\begin{equation}\label{Eq:Doppler-bandwidth-1}
 \Delta \omega_i-\Delta k_i v=\Delta \omega_r+\Delta k_r v,
\end{equation} 
where $\Delta \omega_i$, $\Delta k_i$, and $\Delta \omega_r$, $\Delta k_r$ are the spectral and spatial bandwidths of the incident and reflected pulses, and $v$ is the interface velocity.  Assuming that the initial and final pulses are sufficiently narrow such that each may be associated with a specific group velocity, we may combine $c v_g=\Delta \omega/\Delta k$ with Eq.~(\ref{Eq:Doppler-bandwidth-1}) to obtain
\begin{equation}\label{Doppler-bandwidth-ratio}
\frac{\Delta \omega_r}{\Delta \omega_i}=
\frac{1-\frac{v}{v_{gi}}}{1+\frac{v}{v_{gr}}},
\end{equation} 
where $v_{gi}$ is the group velocity of the incident pulse and $v_{gr}$ is the group velocity of the receding pulse.
As always, a positive sign for $v$ means a receding interface and a positive sign for $v_g$ means a positive group velocity.  Thus, as expected, when a pulse reflects from an interface that is receding at roughly the pulse group velocity of the approaching pulse, the reflected pulse has a narrowed bandwidth.  When a pulse reflects from a boundary that approaches at roughly the group velocity of the reflected pulse, the reflected pulse is temporally compressed and has a broad bandwidth.

\subsection{How group velocity modulates the frequency response of a pulse to a time-dependent refractive index}
When interface velocity, $v$, is zero, $\phi$ is also zero and $\omega$ is conserved.  When velocity is infinite, $\phi=\pm \pi/2$.  The refractive index is a function of time only and $\vec k$ is conserved.  In this way, the interface model is easily extended to treat the case of a time-dependent (but spatially homogeneous) medium.
Since $k$ is conserved, frequency must compensate for a change in refractive index according to
\begin{equation}\label{Eq:n-omega}
\omega(t)=\frac{n_o}{n(t)}\omega_o,
\end{equation}
where we take the medium to have an initial refractive index $n_o$ and the pulse to have an initial central frequency $\omega_o$.
If $n$ were driven to $0$, $\omega$ would be up-converted by many orders of magnitude.  The changing refractive index associated with plasma generation has been proposed for use in frequency up-conversion in several works \cite{Yablonovitch-1973,Yablonovitch-1974,Wilks-Dawson-Mori-1988,Savage-Joshi-Mori-1992,Berezhiani-Mahajan-Miklaszewski-1999,Geltner-Avitzour-Suckewer-2002,Avitzour-Geltner-Suckewer-2005}.  On a more pedestrian level, frequency response to a time-dependent refractive index is also at the heart of the function of acousto-optic and electro-optic modulators.

Interestingly, the frequency response to a changing refractive index can be modulated by the group velocity of a pulse in a medium.
To see this, we begin with a dispersive medium with a nondispersive time-dependence whose refractive index takes the form $n=n_0(\omega)+t \delta n/\delta t$.  (This might be taken to model, for example, a gas that is a mixture of two atoms, one with resonance that is near $\omega$, the time-dependent frequency of the wave traveling through the medium, and another whose resonances are all spectrally distant but whose concentration changes with time.) Then, from Eq.~(\ref{Eq:n-omega}) we find 
\begin{equation}\label{Eq:ng-modded-response}
 \frac{d\omega}{dt}=\frac{\omega}{n_g}\frac{\delta n}{\delta t},
\end{equation}
where $n_g$ is the group refractive index, defined by $n_g=n+\omega \delta n/\delta\omega$.
One interpretation of Eq. ~(\ref{Eq:ng-modded-response}) is that the frequency response of a pulse to a temporal change in refractive index is proportional to the distance the pulse travels while the change occurs.
\end{section}

\begin{section}{A comparison of two nonstationary homogeneous media: kinetics and dynamics}\label{Sec:comp}

\begin{table*}[htpb]
\centerline{
\begin{tabular}{c l l c c }
Number & Quantity & Symbol & Morgenthaler & Liu\\
\hline
1&  Phase velocity      	&$v_p$ 			&$v_{pf}/v_{pi}$ 	&1 			\\
2&  Group velocity 		&$v_g$ 			&$v_{pf}/v_{pi}$ 	&$v_{gf}/v_{gi}$ 	\\
3&  Permittivity      		&$\epsilon$ 		&$\epsilon_f/\epsilon_i$&1 			\\
4&  Permeability	 	&$\mu$ 			&$\mu_f/\mu_i$ 		&1		 	\\
\hline
5&  Wave number      		&$k$ 			&1 			&1 			\\
6&  Spatial bandwidth    	&$\Delta k$ 		&1		 	&1 			\\
7&  Central frequency 		&$\omega_0$ 		&$v_{pf}/v_{pi}$ 	&1 			\\
8&  Spectral bandwidth 		&$\Delta \omega$ 	&$v_{pf}/v_{pi}$	&$v_{gf}/v_{gi}$	\\
\hline
9&   Total energy density	&$u_t$			&$v_{pf}/v_{pi}$	&1			\\
10&  Electric displacement	&$\vec D$		&1			&$\sqrt{v_{gf}/v_{gi}}$	\\
11&  Magnetic induction		&$\vec B$		&1			&$\sqrt{v_{gf}/v_{gi}}$	\\
12&  Electric field		&$\vec E$		&$\epsilon_i/\epsilon_f$&$\sqrt{v_{gf}/v_{gi}}$	\\
13&  Magnetizing field		&$\vec H$		&$\mu_i/\mu_f$		&$\sqrt{v_{gf}/v_{gi}}$	\\
\hline
14&  Photon energy		&$\hbar \omega$		&$v_{pf}/v_{pi}$	&1			\\
15&  Photon number density	&$N$			&1			&1			\\
16&  Poynting vector		&$S$			&$(v_{pf}/v_{pi})^2$	&$v_{gf}/v_{gi}$	\\
17&  Abraham momentum density	&$\vec E\times \vec H/c^2$&$(v_{pf}/v_{pi})^2$	&$v_{gf}/v_{gi}$	\\
18&  Minkowski momentum density~&$\vec D\times \vec B$	&1			&$v_{gf}/v_{gi}$	\\
19&  Canonical momentum density	&$N(n \hbar \omega/c)$ 	&1			&1			\\
20&  Angular momentum density	&$N(\hbar)$ 		&1			&1			\\
\end{tabular}}
\caption{\label{Tab:comp}A comparison of narrow-pulse transformations by two dynamic media.}
\end{table*}

In this section we seek to augment our understanding of the effects of a time-dependent group velocity in a dispersive medium by comparing them with the effects of a time-dependent \emph{phase} velocity in a \emph{non}dispersive medium.  To do so, we now define two scenarios, which we call Scenarios A and B.  In both scenarios, we imagine a spectrally narrow pulse traveling through a time-dependent medium and ask how that pulse is transformed as the medium changes with time.

In Scenario A, we take the medium to be nondispersive, spatially homogeneous, and isotropic.  We assume that changes in the medium are adiabatic, spatially homogeneous, and isotropic.  We allow $\epsilon$ and $\mu$ to vary gradually and independently over time.  We assume that loss at all times is negligible, meaning that $\epsilon$ and $\mu$ are taken to be real for the narrow spectral range of the pulse.  Because the medium is nondispersive, the group velocity is equal to the phase velocity.

In Scenario B, we take the medium to be dispersive, spatially homogeneous, and isotropic.  We assume that changes in the medium are adiabatic, spatially homogeneous, and isotropic.  We fix $\epsilon$ and $\mu$ for the central frequency, but allow their slopes to vary gradually and independently over time.  We take loss at all times to be negligible, meaning that $\epsilon$ and $\mu$ are taken to be real for the narrow spectral range of the pulse.  

\subsection{Kinetics: basic symmetries}
The spatial homogeneity of Scenarios A and B implies that
\begin{equation}
 \vec k_f=\pm \vec k_i,
\end{equation} and the adiabaticity of the two scenarios eliminates reflection, requiring that we use only the positive sign.  Spatial homogeneity also implies that
\begin{equation}
 \Delta k_f=\Delta k_i
\end{equation}
for the pulses of the two scenarios.
Since $|k|=n\omega/c$ and $\Delta k \approx n_g \Delta \omega/c$, changes in the phase and group velocity imply proportional changes in the central frequency and pulse bandwidth respectively.  Since $n_g=n+\omega\delta n/\delta \omega$, changes in $n$ imply changes in $n_g$ for a nondispersive medium and therefore changes in $n$ lead to changes in bandwidth as well.

To continue our comparison, we now specify particular boundary conditions for the two changing media.  In doing so, we lose the generality of Section~\ref{Sec:symmetry}, but gain the ability to examine the effects of changes to specific, time-dependent media on basic quantities like energy and momentum densities and reflection and refraction coefficients.

We take our two media from previous works.  The non-dispersive medium is taken from Morgenthaler \cite{Morgenthaler-1958}. The dispersive medium is an idealization of the dynamic EIT medium used in the Liu experiment \cite{Liu-Dutton-Behroozi-Hau-2001}.

\subsection{Dynamics: boundary conditions and reflection}
\subsubsection{Morgenthaler's nondispersive medium}
Morgenthaler allowed for the electric permittivity ($\epsilon$) and magnetic permeability ($\mu$) of the medium to be functions of time, but not of space or direction or frequency.  That is,
$\epsilon(\mathbf r, \mathbf k, \mathbf \omega, t)=\epsilon(t)$, and $\mu(\mathbf r, \mathbf k, \mathbf \omega, t)=\mu(t)$.  He assumed that charge and flux, and therefore the electric displacement ($\mathbf D$) and the magnetic induction ($\mathbf B$), would be conserved at an interface boundary.  Under this assumption, the reflection and transmission coefficients for the electric field are given by \cite{Morgenthaler-1958}
\begin{equation}\label{Eq:Er}
 \frac{E_r}{E_i}=\frac{1}{2}\left(\frac{\epsilon_i}{\epsilon_r}-\sqrt\frac{\mu_i\epsilon_i}{\mu_r\epsilon_r}\right),
\end{equation}
and
\begin{equation}\label{Eq:Et}
 \frac{|\mathbf E_t|}{|\mathbf E_i|}=\frac{1}{2}\left(\frac{\epsilon_i}{\epsilon_r}+\sqrt\frac{\mu_i\epsilon_i}{\mu_r\epsilon_r}\right).
\end{equation}
In any Morgenthaler medium, the magnitude of the magnetic field is given by $|\mathbf H|=|\mathbf E|/\eta$, where $\eta=\sqrt{\mu/\epsilon}$ is the impedance.  In general, there is a reflected wave.

We consider the case where changes to the medium are adiabatic and homogeneous.  In this case the reflected power vanishes.  Using the fact that what is preserved across one interface must be conserved across many parallel interfaces (see the beginning of Section~\ref{Sec:symmetry}), we find that we may take the boundary conditions across an adiabatic change to be
\begin{equation}
 \mathbf D_f=\mathbf D_i
\end{equation} 
and 
\begin{equation}
 \mathbf B_f=\mathbf B_i.
\end{equation} 
The values of $\mathbf E$ and $\mathbf H$ at any time can then be found via $\epsilon$ and $\mu$.

\subsubsection{Liu's dispersive medium}
\begin{figure}
	\centerline{\includegraphics[width=8.5cm]{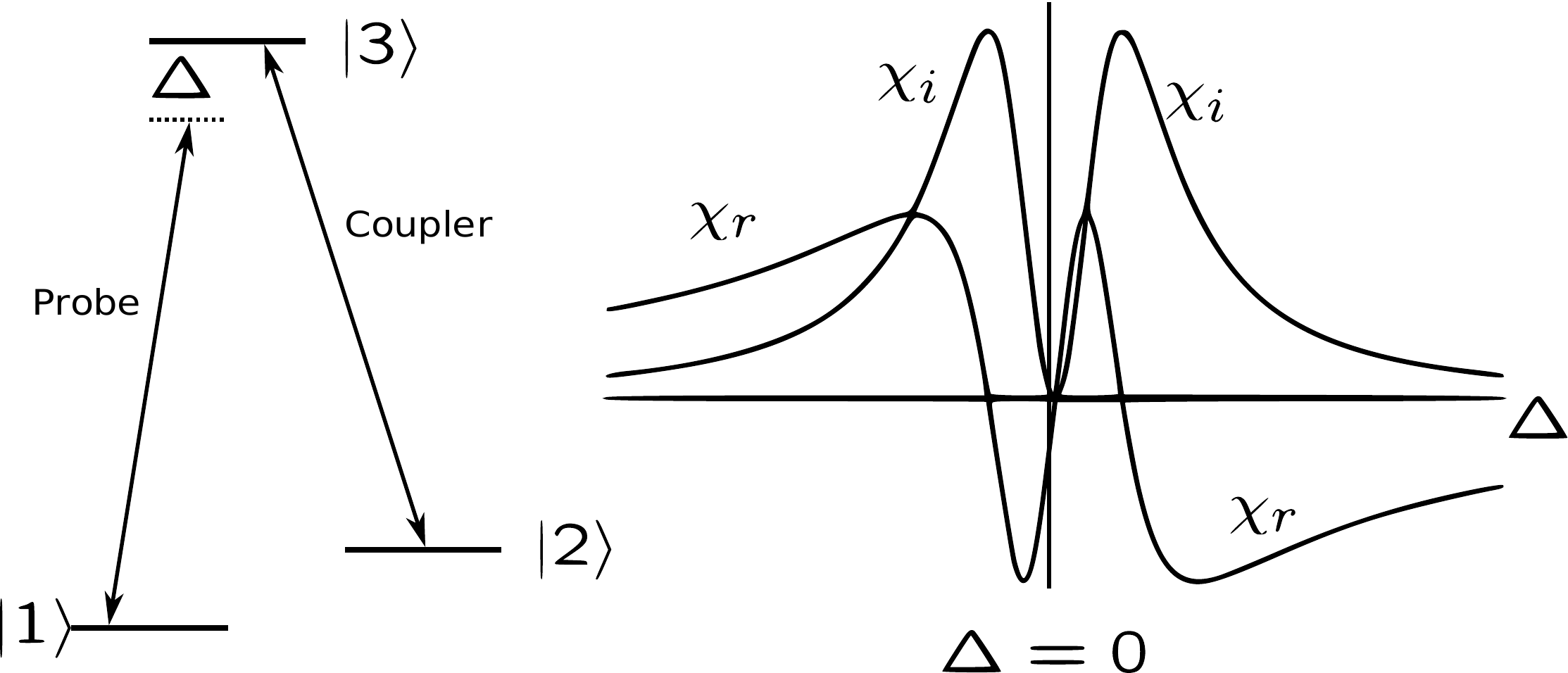}}
	\caption{\label{Fig:EIT2} EIT energy diagram.  The coupling beam and the medium may be taken together to form an effective medium as felt by the probe beam.  The properties of the effective medium depend on the state of the coupling beam.}
\end{figure}
The Liu experiment involves a pulse interacting with an EIT medium across a narrow spectral range around the EIT resonance.  The system as a whole comprises two beams of light interacting with a cloud of cold atoms.  The two beams are tuned to two connected atomic transitions which share an excited state. The stronger beam, referred to as the coupling beam in Figure~\ref{Fig:EIT2}, is considered in combination with the cloud of atoms to form the EIT medium, as seen by the weaker beam, which we call the probe beam. When EIT is established, the atoms are in a coherent dark state, so-called because it does not couple to the probe beam.  The dark state may be represented as a combination of electronic levels $|1\rangle$ and $|2\rangle$ as 
\begin{equation}
 |D\rangle=\frac{\Omega_c |1\rangle-\Omega_p|2\rangle\exp[i (\vec k_p-\vec k_c)\cdot\vec r-i(\omega_p-\omega_c)t]}{\sqrt{\Omega_c^2+\Omega_p^2}}.
\end{equation}
If the probe beam is sufficiently weak such that the atomic number density remains significantly larger than the probe photon number density, then the EIT medium will be approximately linear in probe intensity, so that the only sizable nonlinear effects of the EIT medium are the effects of the coupling beam on the probe beam.  Surprisingly, this linear regime may be maintained even as the coupling beam is completely attenuated \cite{Fleischhauer-Lukin-2000}.

For our nondispersive medium, we idealize the Liu experiment \cite{Liu-Dutton-Behroozi-Hau-2001} by neglecting decoherence and other losses and require that total pulse energy be conserved under changes to the group velocity.  Thus,
\begin{equation}
 u_{tf}=u_{ti},
\end{equation}
where $u_t$ represents the total energy density associated with the wave including not only that portion that is in field form but also that portion that is coherently stored in the medium and is a cycle-averaged quantity.  For a narrow-band pulse propagating through a lossless but dispersive medium, we may define a non-dispersive energy density ($u_n$) which is related to the dispersive energy ($u_t)$ by
\begin{equation}
 u_n=u_t\frac{n}{n_g}.
\end{equation} 
In terms of the macroscopic fields, $u_n$ is given, for the isotropic medium we are considering, by 
\begin{equation}
 u_n=\frac{1}{2}\left(\epsilon E^2+\mu H^2 \right),
\end{equation} 
where $E^2$ and $H^2$ are time averaged quantities.
(Note that $u_n$ matches the standard form for the electromagnetic energy density in a lossless, dispersionless, macroscopic medium \cite{Jackson-1998}.  When dispersion becomes negligible, $n_g\approx n$, and $u_t$ reduces to this form.)

If, as we have assumed for our comparison, changes in the group velocity occur around a constant refractive index, then we may take the quantity $u_n n_g$ to be fixed.  We will also assume that the impedance at line center is unchanged, so that the ratio between $H$ and $E$ remains fixed.  Because the refractive index at line center will also remain fixed, all macroscopic electromagnetic fields scale together.  Therefore, all terms quadratic in any pair of fields scale with the group velocity.  The direction of the fields is unchanged by the homogeneous change in the isotropic medium.  We summarize this as
\begin{equation}
 \mathbf F_f \sqrt{v_{gf}}=\mathbf F_i \sqrt{v_{gi}},
\end{equation} 
where $\mathbf F$ could be any of the four macroscopic electromagnetic fields.
\subsection{Transformations wrought under changing phase and group velocities in the `Morgenthaler' and `Liu' media}

We are now able to perform a more complete comparison between scenarios A and B.  In Table~\ref{Tab:comp}, we list 20 different quantities and show how they are scaled as the media of Scenarios A (Morgenthaler) and B (Liu) change.  Quantities 1-4 summarize the difference between the two scenarios, emphasizing that the phase velocity and group velocity change in Scenario A while only the group velocity changes in Scenario B.  The scaling of quantities 5-8 depends only on symmetry (see Section~\ref{Sec:symmetry}) and therefore are general consequences of the scaling of quantities 1-4.  The scaling of quantities 9-13 may be taken as expressions of the boundary conditions for the two media.  

The remaining quantities (14-20) can be found in simple ways from a knowledge of the scaling of the first 13 quantities.  The per photon energy (14) scales with frequency, which scales as in line 7.  That the photon number density (15) is conserved is related to our assumption of no loss.  The Poynting vector (quantity 16) scales as the square of the phase velocity for the non-dispersive case.  This quadratic dependence combines the linear dependence of the total energy density on the phase velocity with the fact that the group velocity is equal to the phase velocity for a nondispersive medium.  Because the dispersive case assumes that the phase velocity remains unchanged, the Poynting vector of the Liu medium feels only the effect of the change in group velocity.  The Minkowski momentum (18) is conserved in the nondispersive case, but is proportional to the group velocity for the nondispersive case.  This second dependence underscores the fact that the Minkowski momentum, as defined by $\mathbf D\times \mathbf B$, does not give the total momentum when dispersion is taken into account.  The canonical momentum (19), which does give the total momentum, is conserved for both media.  Because the total photon number is conserved for both media, so is the total angular momentum density associated with the pulse for both media.  We note that although the total angular momentum does not change, its distribution between field and medium does.

In concluding this section we emphasize three points.  First, to understand the effects of a particular medium on a pulse, it is necessary to specify appropriate boundary conditions that depend on the details of the way in which the changing medium interacts with the field.  Second, boundary conditions that are reasonable for a non-dispersive medium (Scenario A) may fail for a simple dispersive medium (Scenario B) for the particular reason that changes in dispersion imply a shifting of energy from field to medium.  It then becomes necessary to model different changes on a case-by-case basis.  Finally, a general comparison of the scaling of different quantities for Scenario A and Scenario B shows that the changes wrought in a controllably dispersive nonstationary medium are qualitatively different from those wrought in a nonstationary nondispersive medium.  In this sense, controllably dispersive media open up a new and potentially fruitful niche in nonstationary electromagnetics.
\end{section}
\begin{section}{Conclusion}
Controllably dispersive media are an experimental reality \cite{Wang-Kuzmich-Dogariu-2000,Liu-Dutton-Behroozi-Hau-2001, Ginsberg-Garner-Hau-2007}.  All the requirements necessary to perform interesting nonstationary experiments in controllably dispersive media have been demonstrated for one experimental system \cite{Liu-Dutton-Behroozi-Hau-2001, Ginsberg-Garner-Hau-2007}, are near at hand for another \cite{Wang-Kuzmich-Dogariu-2000}, and appear to be feasible for others \cite{Wicht-Rinkleff-Danzmann-2002,Avitzour-Shvets-2008}.  In this paper we have outlined a basic theory for the treatment of controllably dispersive media from the perspective of nonstationary electromagnetics.  

Although allowing the properties of a medium to depend not only on position but also on both time and frequency opens up a large parameter space, the behavior of waves in this space is subject to basic kinetic constraints imposed by symmetry.  We have explicitly defined the requirements of symmetry for the case of a plane wave interactive with a moving planar interface. We have justified these requirements by rederiving well-established results, such as Snell's Law and the free space Doppler shift, and used them to simply derive analytical descriptions of more esoteric phenomena, such as the effect of dispersion on the reflective Doppler shift.  Using the outlined kinetic theory, we have provided an analysis that corroborates a recent proposal \cite{Avitzour-Shvets-2008} for the compression and decompression of pulse envelopes in magnetized plasmas using strictly nonstationary effects and have shown that the proposed mechanism may be applied in any controllably dispersive medium.  We have introduced a simple expression that shows how a pulse may be significantly compressed or decompressed under Doppler reflection--even if the Doppler shifts themselves are small.  Finally, we have shown that dispersion modulates the frequency response of a wave to a temporal change in the refractive index--the response is damped for slow light media, and amplified for fast light media.

Unlike kinetic effects, which rely only on simple symmetries, dynamic effects are determined by boundary conditions which depend upon particular details of the medium at hand.  We have compared dynamic effects in two simple, homogeneous, adiabatically nonstationary media--one which is nonstationary in phase velocity and a second which is nonstationary in group velocity.  We compared the effective boundary conditions associated with these media and the changes wrought in spectrally narrow pulses propagating through them.  We have found substantial differences in both the boundary conditions and the effect of the nonstationary medium on fundamental quantities such as energy, and electromagnetic momentum.  These differences suggest that it may be useful to consider controllably dispersive media as a new and potentially fruitful category of nonstationary electromagnetic media.
\end{section}



\chapter{Dispersion and the quantum noise associated with a single cavity mode}\label{ch:cavity}

\renewcommand{\eq}[1]{Eq.~(\ref{#1})}
\renewcommand{\bE}{{\bf E}}
\renewcommand{\bD}{{\bf D}}
\renewcommand{\bH}{{\bf H}}
\renewcommand{\bB}{{\bf B}}
\renewcommand{\bS}{{\bf S}}
\renewcommand{\br}{{\bf r}}

\newcommand{\lc}{$\ell$-cavity}
\newcommand{\Lc}{$L$-cavity}
\newcommand{\llc}{$c$-cavity}

\title{Vacuum field energy and spontaneous emission in anomalously dispersive cavities}
\maketitle
\section*{Abstract}
Anomalously dispersive cavities, particularly white light cavities, have been proposed for use in LIGO-like gravity wave detectors and in ring-laser gyroscopes. In this paper we analyze the quantum noise associated with anomalously dispersive cavity modes.
The vacuum field energy associated with a particular cavity mode is proportional to the cavity-averaged group velocity of that mode.
For anomalously dispersive cavities with group index values between 1 and 0, this means that the total vacuum field energy associated with a particular cavity mode must exceed $\hbar \omega/2$.
For white light cavities in particular, the group index approaches zero and the vacuum field energy of a particular spatial mode may be significantly enhanced.
We predict enhanced spontaneous emission rates into anomalously dispersive cavity modes and broadened laser linewidths when the linewidth of intracavity emitters is broader than the cavity linewidth.

\section{Introduction}\label{se:introduction}

When the finesse of an optical cavity is improved, buildup increases but resonance bandwidth decreases.  
Resonance bandwidth may be increased without degrading buildup by decreasing cavity length.
For some applications, such as interferometry-based gravity-wave detection and Sagnac interferometry, decreasing cavity length degrades overall system performance.  
For other applications, such as fast single-photon generation \cite{Englund-Fattal-Vuckovic-2005,Moerk-2010}, decreasing cavity length may be profitable, but is not possible past a minimum length.

When actual decreases in length are either unprofitable or impossible, it is natural to consider the use of anomalous intra-cavity dispersion.  
For a given cavity geometry, buildup scales with finesse ($F$) \footnote{By buildup we refer to the resonant enhancement of intracavity intensity as normalized by the incident intensity of a mode.  For high-finesse cavities where the round trip electric field transformation is $E\rightarrow E e^{-i\phi}(1-\delta)$, the buildup is $4/\delta^2$.}, while resonance bandwidth scales inversely with quality ($Q$).  
Quality and finesse are related by $Q=F*(\nu_o L_g/c)$, where $\nu_o$ is the resonant frequency, $c$ is the speed of light, and $L_g$ is the round-trip group optical path difference.  $L_g$ itself is defined in terms of the round trip phase $\phi$ and the angular frequency $\omega$ by  $L_g \equiv c d\phi/d\omega$ \cite{Candler-1946}.
In the presence of dispersion, $L_g =\ell n_g$, where $\ell$ is the cavity round trip length and $n_g$ is an effective cavity group index.  The response of the ratio $F/Q$ to changes in $n_g$ is identical to its response to changes in $\ell$.
Unlike $\ell$, however, $n_g$ may approach zero for a particular range of frequencies \cite{Wicht-Muller-Rinkleff-Danzmann-2000,Pati-Salit-Salit-Shahriar-2007}.
Cavities where $n_g$ approaches zero at a cavity resonance have been referred to as ``white-light cavities,'' in reference to their dispersion-broadened resonances \cite{Wicht-Danzmann-Rinkleff-1997}.  

The white-light cavity literature is centered around two potential applications, gravity wave sensing \cite{Wicht-Danzmann-Rinkleff-1997, Wise-Whiting-2004,Karapetyan-2004} and rotation sensing \cite{Shahriar-Pati-Tripathy-Gopal-Messall-Salit-2007}.  LIGO-like gravity wave detectors take the form of large Michelson interferometers. 
In signal recycling, mirrors may be placed near the entrance to each Michelson arm, forming a pair of Gire-Tournois cavities.
In power recycling, a mirror is placed in advance of the central beam splitter, making the entire interferometer into an optical cavity.
In either case, the increase in signal due to enhanced buildup is gained only at the expense of bandwidth, and
white-light cavities have been considered as a way to combine large bandwidth with large buildup.

Rotation in a ring-resonator gyroscope is equivalent to cavity elongation/truncation when resonator and mode are co-/counter-rotating \cite{Shahriar-Pati-Tripathy-Gopal-Messall-Salit-2007}.  
The frequency response of a cavity mode to a change in length may be enhanced via anomalous dispersion;
as the round-trip length $\ell$ changes, the resonant frequency $\omega_o$ associated with a given mode changes according to $d\omega_o/d\ell=-n\omega_o/(\ell n_g)$, which grows large as $n_g$ approaches $0$.
Shahriar \emph{et al.} found that the increase of frequency sensitivity in a passive anomalously dispersive ring cavity would be counterbalanced by the concurrent increase in resonance bandwidth, leaving no net gain in rotation resolution
\cite{Shahriar-Pati-Tripathy-Gopal-Messall-Salit-2007}. 
However, they argued that this cancellation could be avoided by using an active interferometer in the form of a white-light ring-laser gyroscope.

The effectiveness of white-light cavities for any sensing application is partially determined by white-light cavity noise.
Some implementation-dependent sources of noise have been quantified by Wicht \emph{et al.} \cite{Wicht-Danzmann-Rinkleff-1997} for a double-lambda system and by Sun \emph{et al.} \cite{Sun-Shahriar-Zubairy-2009} for a double gain system.  
A comparison of several implementations can be found in a later work by Wicht \emph{et al.} \cite{Wicht-Rinkleff-Danzmann-2002}.
However, the white-light cavity literature does not include, to our knowledge, any work on the noise due to vacuum energy intrinsic to a white-light cavity.

In this paper, we show by simple and largely classical arguments that the vacuum field energy associated with the ground state of an anomalously dispersive cavity mode diverges as that mode approaches the ideal ($n_g\rightarrow 0, F\rightarrow \infty$) associated with a white-light cavity.  
This divergence is intrinsic to the definition of a white-light cavity and is independent of any specific implementation. It has the consequence that spontaneous emission of an excited particle into the white-light cavity mode may increase substantially.  
This, in turn, has two consequences.  
First, the width of a white-light laser line must be broadened by anomalous dispersion.  
This broadening, if applied to the configuration proposed by Shahriar \emph{et al.} would 
cancel the increased frequency sensitivity of a white-light ring-laser gyroscope.
Second, it suggests that anomalous dispersion may be useful in enhancing the quantum yield of single photon emitters where radiative decay into the mode of interest must compete with nonradiative decay or with radiation into other spatial modes.

In the following section we review the result of applying the standard classical expression for the narrow-bandwidth energy density in a lossless, dispersive medium to the quantization of a closed cavity.  We introduce a distinction between two energies, which we call the ``field energy'' and the ``total energy,'' and which are related by the ratio between cavity-averaged phase and group velocities.  The total cavity energy is quantized as a simple harmonic oscillator, but only the field energy interacts with dipoles inside the cavity.
In Section III, we explore the vacuum field noise in an anomalously dispersive cavity using an alternative coupled-cavity approach.  This approach allows us to find an expression for the vacuum fields that is independent of any specific expression for the electromagnetic energy density in a dispersive medium.  It also provides a simple view of the effect of dispersion on the power spectrum of the vacuum field energy.  We show that if we apply the assumptions of Section ~\ref{se:closed} (no loss and negligible group velocity dispersion for the spectrum of the resonance), then the results of Section ~\ref{se:closed} follow.  We then relax these assumptions using a numerical model of a physical medium and show numerically that as the round trip loss approaches zero, the quantum field noise of a white light cavity resonance diverges.  This occurs even if the direct effect of the gain of the anomalously dispersive medium is neglected.  We briefly address the relationship between this increased noise and the quantum limited laser linewidth of anomalously dispersive cavities.

\section{Spontaneous emission in a lossless, dispersive medium}\label{se:closed}
Electromagnetic field quantization in an evacuated, closed and lossless cavity proceeds as follows:  
\begin{enumerate}
\item A complete set of orthonormal electromagnetic modes is identified.  
\item The volume-integrated energy of each of these modes is represented.
\item Each mode is quantized as though it were a simple harmonic oscillator.  \footnote{Note that this is the only step that introduces a non-classical element.  It is the two classical steps that define the relationship between mode excitations and electromagnetic fields.}
\end{enumerate}

When a dispersive medium is introduced into a cavity, both steps 1 and 2 become problematic.   Step 1 becomes problematic because a dispersive medium must have loss (or gain), which couples the electromagnetic fields of a dispersive cavity to external degrees of freedom.  True electromagnetic modes can then only be found by explicitly including these external degrees of freedom. 

Step 2 becomes problematic because dispersion also complicates the concept of electromagnetic field energy density.  Whether or not energy that has been stored by a dispersive medium returns to field form depends on future interactions between medium and field \footnote{a concrete application of this concept may be found in \cite{Glasgow-Meilstrup-Peatross-Ware-2007}}.  In other words, the electromagnetic energy density of a dispersive medium is nonlocal in time.

These difficulties with steps 1 and 2 can both be avoided if, rather than seek a complete set of orthonormal electromagnetic modes we focus on a single cavity mode.  Although a causal dispersive medium must have loss (or gain), causality does not prevent this loss from being small or even vanishing for particular frequencies.  Thus, although a causal dispersive cavity must be open in general, it may be closed at particular resonances.  In addition, a perfect, closed cavity mode has a vanishing spectral width.  Although the electromagnetic energy density in a dispersive medium is, in general, an ill-defined quantity, it is well defined in the absence of loss for spectrally narrow excitations.

\subsection{Energy density of a quasimonochromatic planar wave in a lossless, dispersive medium}
The effect of dispersion on the electromagnetic energy density of a quasimonochromatic excitation in a lossless linear medium is particularly simple for planar waves.
When only radiative energy transportation is non-negligible, Poynting's theorem takes the form 
\begin{equation}
\nabla \cdot \bS=-\partial u/\partial t, 
\end{equation} 
where $\bS$ is the Poynting vector and $u$ is the electromagnetic energy density.
If absorption, scattering, and group velocity dispersion are all negligible, a direct calculation of the divergence of a spectrally narrow planar wave propagating through a dispersive medium yields
\begin{equation}\label{Eq:u-from-S}
 u(z,t)=\frac{\langle S(z,t)\rangle}{v_g},
\end{equation}
where $v_g$ is the group velocity of the planar wave, and the angle brackets denote a cycle average.

If dispersion were neglected, $\langle S\rangle/v_g$ in \eq{Eq:u-from-S} would become $\langle S\rangle /v_p$, where $v_p$ is the phase velocity.  In other words, if $u_f$ is an expression for the nondispersive energy density, then $u$ can be found from $u_f$ via
\begin{equation}\label{Eq:u-from-uf}
 u=\frac{n_g}{n} u_f,
\end{equation} 
where $n_g$, and $n$ are the refractive index and the group index.  The standard form for the cycle averaged electromagnetic energy density in a lossless, isotropic, linear, \emph{dispersionless} medium is \cite{Landau-Pitaevski-Lifshitz, Jackson-1998}
\begin{equation}\label{Eq:uf}
 u_f=\frac{\epsilon \langle E^2\rangle }{2}+\frac{\mu \langle H^2\rangle}{2}.
\end{equation}
Substituting \eq{Eq:uf} into \eq{Eq:u-from-uf} gives
\begin{equation}\label{Eq:u-planar}
 u=\frac{n_g}{n}\left(\frac{\epsilon \langle E^2\rangle }{2}+\frac{\mu \langle H^2\rangle}{2}\right).
\end{equation}

For spectrally narrow excitations in lossless media, dispersion has a simple relationship to energy storage by the dispersive medium.  If we call $u_f$ the field energy density and $u$ the total energy density then we can also define a third quantity, $u_s$ as the stored energy density, or the difference between the total energy density and the field energy density.  In normally dispersive media, $u>u_f$, and $u_s$ is positive because electromagnetic energy is dispersively stored in the medium.  In anomalously dispersive cavities, where $0\leq n_g <n$, $u<u_f$ and $u_s$ is negative.  Just as $u_s>0$ in a dispersive medium signifies temporary absorption, $u_s<0$ in an anomalously dispersive medium signifies temporary emission.  In either case, energy exchange is governed by the interaction between the instantaneous spectrum of the planar wave and the response function complex susceptibility of the dispersive medium \cite{Peatross-Ware-Glasgow-2001}.  In the limit of a white light cavity, $u_s=-u_f$ and $u=0$, meaning that all of the field energy has been donated by the medium.

The energy density given by \eq{Eq:u-from-S} has the strength that the effect of dispersion on the energy density is explicit and simple.  However, it was derived for planar waves and does not apply when there is interference between waves propagating in different directions.

\subsection{Energy of a quasimonochromatic mode in a lossless, dispersive medium}
A more general but less intuitive form for the energy density may be found by using vector identities and Maxwell's equations to make the transformation
\begin{equation}
\nabla \cdot \bS=\bE\cdot \frac{\partial\bD}{\partial t}+\bH\cdot\frac{\partial \bB}{\partial t},
\end{equation}
and then using the narrow-band nature of the excitation and linearity to obtain \cite{Landau-Pitaevski-Lifshitz, Jackson-1998}
\begin{equation}\label{Eq:u-standard}
u = Re\left[\frac{d (\omega \epsilon)}{d\omega}\right]\frac{\langle E^2\rangle}{2}+Re\left[\frac{d (\omega \mu)}{d\omega}\right]\frac{\langle H^2 \rangle}{2},
\end{equation} 
where the angle brackets denote a cycle average.

\eq{Eq:u-standard} shows that the simple relationship between dispersion and energy density given for planar waves in \eq{Eq:u-from-uf} does not generalize to non-planar waves.  For example, consider the energy density of counter-propagating waves of equal amplitude and frequency and identical polarization in a dielectric. If we assume that $\epsilon$ and $\mu$ and their first frequency derivatives are real, then we can use the identities $n=\sqrt{\epsilon\mu/(\epsilon_o\mu_o)}$ and $n_g=n+\omega dn/d\omega$ to obtain
\begin{equation}\label{Eq:ng-over-n}
 \frac{n_g}{n}=1+\frac{1}{2}\frac{d \ln \epsilon}{d\ln \omega}+\frac{1}{2}\frac{d \ln \mu}{d \ln \omega}.
\end{equation}
Since the medium is a lossless dielectric, $\omega d\epsilon/d\omega=\eta\epsilon$, and $\omega d\mu/d\omega=0$, where $\eta$ is a real number.  Applying these values to \eq{Eq:ng-over-n} gives $n_g/n=1+\eta/2$. 
In the standing wave that results from the counter-propagating waves, electric field nodes correspond with magnetic field antinodes.  At electric field nodes, $E^2=0$, and $u/u_f=1$.  At electric field anti-nodes, $H^2=0$, and $u/u_f=1+\eta$.  Thus, \eq{Eq:uf} does not generalize to non-planar waves.

However, an analogue to \eq{Eq:uf} does apply to cavity modes.  
Assuming losslessness and vanishing imaginary derivatives for $\epsilon$ and $\mu$, we rewrite
\eq{Eq:u-standard} as 
\begin{equation}\label{Eq:u-lossless}
u = \left(1+\frac{d \ln \epsilon}{d\ln \omega}\right)\frac{\epsilon \langle E^2\rangle}{2}+\left(1+\frac{d \ln \mu}{d\ln \omega}\right)\frac{\mu \langle H^2 \rangle}{2}.
\end{equation} 

We now note that the electric and magnetic contributions to the total modal energy of a nondispersive cavity mode are equal.  That is, for the field associated with a mode that satisfies closed boundary conditions and the Helmholtz equation,
\begin{equation}\label{Eq:yin=yang}
 \int \frac{\epsilon \langle E^2\rangle}{2}d^3 r=\int \frac{\mu \langle H^2\rangle}{2}d^3 r.
\end{equation}
This is equivalent to the fact that the cycle-integrated momentum and position energies of a simple harmonic oscillator must be equal.
The spatial field distribution of a mode, as governed by the Helmholtz equation and a particular set of boundary conditions, does not change when dispersion is added, so long as $\epsilon$ and $\mu$ are not changed for the modal frequency.  Thus \eq{Eq:yin=yang} remains true when dispersion is taken into account (although it no longer equates the total electric and magnetic contributions to the modal energy).

Applying \eq{Eq:yin=yang} to an integral of \eq{Eq:u-lossless} over the modal volume of a homogeneous cavity gives
\begin{equation}\label{Eq:U2}
 U=\left(1+\frac{1}{2}\frac{d \ln \epsilon}{d\ln \omega}+\frac{1}{2}\frac{d \ln \mu}{d \ln \omega}\right)
          \int u_f d^3 r,
\end{equation} 
where $U$ is the total modal energy and $u_f$ is the field energy density given by \eq{Eq:uf}.
Because we are taking $\mu$ and $\epsilon$ and their first derivatives to be real, we can substitute \eq{Eq:ng-over-n} into \eq{Eq:U2}, which gives
\begin{equation}\label{Eq:U3}
U=\frac{n_g}{n}U_f,
\end{equation} 
where the field energy $U_f$ is defined by
\begin{equation}\label{Eq:Uf}
 U_f=\int\left(\frac{\epsilon\langle E^2\rangle}{2}+\frac{\mu\langle H^2 \rangle}{2}\right)d^3 r.
\end{equation} 
\eq{Eq:U3} is a modal energy analog to \eq{Eq:u-from-uf} for planar waves.

$U_f$ is a cycle-averaged expression for the energy of a nondispersive electromagnetic mode.  \eq{Eq:U3} gives a simple description of the effect of dispersion on the total modal energy associated with a given field strength.

\subsection{The quantized electric field}
Assuming that a dispersive cavity mode may be quantized as a simple harmonic oscillator, we write its Hamiltonian $H$ in terms of an annihilation operator ($a$) and a creation operator ($a^\dagger$) as
\begin{equation}\label{Eq:Hoscillator}
 H=\hbar \omega\left(a^\dagger a+\frac{1}{2}\right).
\end{equation}
We note here that the energy associated with $H$ cannot be negative and cannot therefore be related to \eq{Eq:U3} when the ratio $n_g/n$ is negative \footnote{Interestingly, a second argument suggests the impossibility of a lossless resonance at a frequency where there is a negative group velocity.  Causality dictates that the combination of losslessness and a negative group velocity is only seen in an active medium.  If there is a frequency with a negative group velocity, then nearby are at least two frequencies that share the same wavelength as the on-resonance frequency and that are therefore also at resonance.  However, it appears that at least one of these must also be at a frequency where there is gain.  Net round trip gain at resonance in a cavity is not tenable in the steady state: the cavity must either lase or have its quality destroyed.}.
The electric field may then be written in terms of $a$ and $a^\dagger$ in the form
\begin{equation}\label{Eq:E}
 \bE(\br, t)=\mathcal{E} a {\bf f}(\br) e^{i\omega t}+\mathcal{E}^* a^\dagger {\bf f}^*(\br) e^{-i\omega t},
\end{equation} 
where ${\bf f}(\br)$ gives the spatial distribution and polarization of the electric field and $\mathcal{E}$ is a complex number whose amplitude gives a characteristic electric field strength.  
To make the effect of the mode volume ($V$) on the electric field explicit, we choose to normalize ${\bf f}$ according to the rule $\int{f^2 d^3 r}=V$. Combining
Eqs. ~(\ref{Eq:E}), ~(\ref{Eq:Hoscillator}), ~(\ref{Eq:Uf}), ~(\ref{Eq:U3}), and ~(\ref{Eq:yin=yang}) then gives
\begin{equation}\label{Eq:FS}
 |\mathcal{E}|=\sqrt{\frac{\hbar \omega n}{\epsilon n_g V}}.
\end{equation} 
If we restrict our consideration to dielectrics, then $\epsilon=n^2$, and $|\mathcal{E}|=\sqrt{\frac{\hbar \omega}{\epsilon_0 n n_g V}}$, which is the expression for the electric field strength given by Garrison and Chiao \cite{Garrison-Chiao-2008} who follow Milonni \cite{Milonni-1995}.  This expression agrees with a more general and earlier one implicit in work by Drummond \cite{Drummond-1990}.
In addition, dispersive expressions for characteristic field strengths (electrical or otherwise) may be related to expressions that neglect dispersion \cite{Glauber-Lewenstein-1990} using the rule $|\mathcal{E}_{dispersive}|/|\mathcal{E}_{nondispersive}|=\sqrt{n/n_g}$.

\eq{Eq:FS} suggests three separate ways to increase the characteristic electric field strength: the mode volume $V$ may be decreased, the impedance $n/\epsilon$ may be increased, or the group index $n_g$ may be brought close to zero.  In the ideal white light cavity, $n_g\rightarrow 0$, which implies that $|\mathcal{E}|\rightarrow \infty$.

One other interesting aspect of the ideal white light cavity is the mode energy.  As $n_g\rightarrow 0$ from the positive side, $U$ as defined in \eq{Eq:U3} also approaches zero.  This is a result of a cancellation of energy between a negative stored energy ($U_s$) and a positive field energy ($U_f$).  By analogy with the definition of $U_f$, we can define a field Hamiltonian ($H_f$) as
\begin{equation}
 H_f=\hbar \omega\frac{n}{n_g}\left(a^\dagger a+\frac{1}{2}\right),
\end{equation} 
which gives a vacuum field noise energy of
\begin{equation}
 U_v=\frac{\hbar \omega n}{2 n_g},
\end{equation} 
which becomes infinite for the closed cavity model as $n_g$ approaches zero.
In atomic vapor systems currently used to achieve anomalous dispersion, $n$, $\epsilon_r$, and $\mu_r$ are very close to $1$.  The total vacuum field noise then reduces to
\begin{equation}\label{Eq:vapor-field-energy}
 U_v=\frac{\hbar \omega}{2 n_g}.
\end{equation} 

In the next section we will present an open cavity model of a white light cavity that will reveal the effect of the group index on the cavity noise spectrum and allow us to make specific calculations based on reasonable models for a white light cavity medium.

\section{Coupled cavity approach}\label{se:open}
\begin{figure}
	\centerline{\includegraphics[width=8.5cm]{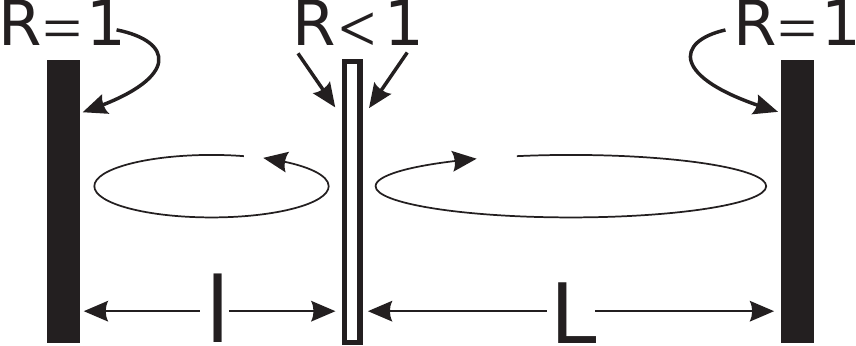}}
	\caption{\label{Fig:cavities} A small cavity of length $\ell$ is coupled to a larger cavity of length $L$ by a mirror with reflectance $R$.  The small cavity contains a lossless, dispersive medium with a group index of $n_g$.}
\end{figure}

In the previous section, we saw that the standard expression for the electromagnetic energy density in a dispersive medium implies that dispersion may scale the amplitudes of the electromagnetic fields associated with frequency eigenmodes in a closed, lossless cavity.  Because the closed cavity treatment neglects loss, the field modes in the model are delta functions in frequency and give no insight into the effect of dispersion on the power spectrum of the modes.  For the case of a white light cavity, the closed cavity treatment also suffers from the fact that the classical expression for the electromagnetic energy density of a white light mode suggests a total energy of zero.
In this section we extend and corroborate the closed cavity approach by examining the effect of dispersive on the electromagnetic fields of a lossy cavity.

Cavity loss means that the electromagnetic energy inside a cavity is coupled to some external system or systems.  
In practice, cavity modes are coupled both to external electromagnetic field modes and also to atomic systems which in turn may be coupled to other modes and systems.  Electromagnetic field modes are equivalent to simple harmonic oscillators.
In linear regimes, atomic systems may also be modeled as simple harmonic oscillators.   Thus a simple and reasonable way to model loss is to couple the system explicitly to a set of simple harmonic oscillators.  

The simplest way to do this may be to couple the lossy cavity to a larger cavity.  The lossy cavity mode then behaves as a coupled harmonic oscillator and the mode is referred to as a pseudomode.  This lossy cavity pseudomode can be probed using the modes of the overall system, which behave as uncoupled harmonic oscillators, and are often called ``true modes'' or ``universe modes \cite{Lang-Scully-Lamb-1973}.''  A nice discussion of the interrelations between pseudomodes, quasimodes, and true modes may be found in a thorough paper by Dalton and colleagues \cite{Dalton-Barnett-Garraway-2001}.  Our contribution here is to show not only that the true mode approach yields insight on the pseudomodes of a dispersive cavity, but that it does so in a way that not depend on the exact form of the electromagnetic energy density.

Figure ~\ref{Fig:cavities} depicts a simple 1 dimensional implementation of this approach involving three mirrors.  The two outer mirrors are perfect, and the combined system is closed.  
The inner mirror has a reflectance $R$, is lossless, and is frequency independent for the range of frequencies that will interest us. These three mirrors form three distinct resonant systems: a longer reservoir cavity of length $L$, a shorter cavity of length $\ell$, and a closed system comprised of the these two cavities coupled together.  From here on we will refer to these resonant systems as the \lc, the \Lc, and the \llc.  
The \Lc~is evacuated while the \lc~contains a dispersive but lossless medium with refractive index, relative permeability and relative permittivity all equal to 1 at line center.  \footnote{We choose $n=1$ not only because it is close to the refractive index value associated with many possible white light cavity implementations, but also because it allows us to make a simple distinction between total and field electromagnetic energies.}
Our object is to see how dispersion affects the pseudomodes of the \lc~by using the modes of the \llc~ as probes.

We will begin by using this model to corroborate the results of Section ~\ref{se:closed}. By neglecting absorption/gain and group velocity dispersion, we will use the open cavity model rederive \eq{Eq:vapor-field-energy} without relying on any expression for the electromagnetic energy density in a dispersive medium.  We will then extend this result by including group velocity dispersion using a specific model for an anomalously dispersive medium.

\subsection{Total field noise of an open dispersive cavity mode in the absence of loss, gain, and group velocity dispersion}
We we begin by using the model depicted in Figure ~\ref{Fig:cavities} to calculate the total field energy in a dispersive cavity pseudo-mode.  The total field energy ($U_{f\ell}$) of an \lc~pseudo-mode is a combination of contributions from many \llc~modes.  Thus
\begin{equation}
 U_{f\ell}=\ell A \sum_{\phi_\ell=-\pi}^{\phi_\ell=\pi}u_{f \ell i},
\end{equation}
where $A$ is the mode area, $u_{f\ell i}$ is the average energy density of the $i$'th \llc~mode in the \lc, and the sum is taken across one \lc~free spectral range.

For the $i$'th \llc~mode, the electric field amplitude in the \lc~ ($\bE_{\ell i}$) is related to the electric field in the \Lc~($\bE_{L i}$) by
\begin{equation}\label{Eq:field-ratios}
 \bE_{\ell i}=\bE_{L i}\frac{t}{1-\sqrt{R}e^{-i\phi_{\ell i}}},
\end{equation} 
where $t$ is the transmission coefficient of the central mirror and $\phi_{\ell i}$ is the round trip phase of the \lc~at $\omega_i$.

Using \eq{Eq:field-ratios} and noting that the ratio of \emph{field} energy densities is proportional to the squared modulus of the squared electromagnetic fields, we obtain
\begin{equation}\label{Eq:density-ratio}
 \frac{u_{f \ell i}}{u_{L i}}=\frac{1-R}{1+R-2\sqrt{R}\cos(\phi_i)},
\end{equation} 
where we use the losslessness of the mirror to get $tt*=1-R$.

\begin{equation}
 U_{f\ell}=\ell A \sum_{\phi=-\pi}^{\phi=\pi}u_{L i}\frac{1-R}{1+R-2\sqrt{R}\cos(\phi_i)}.
\end{equation} 
We know that if $R$ were $1$, the modes of the \Lc~would act as ordinary simple harmonic oscillators because we understand the quantization of electromagnetic fields in a vacuum.  The \llc~modes are perturbed by the effect of the small cavity.  However, assuming that $L$ is large and that this perturbation would therefore not change the simple harmonic oscillator character of the overall modes, we take the total vacuum energy of the overall cavity modes to be $\hbar \omega/2$.

To get an expression for $u_{L i}$, we divide by a weighted volume.  If we take the mode area in each cavity to be $A$ then the volumes of the \Lc ~and the \lc ~are $AL$ and $A\ell$.  The squared field in the \lc~ differs from the that in the~\Lc by the factor given in \eq{Eq:density-ratio}.  The energy density in the small cavity differs from that of the field energy by the factor $n_g$.  Taken together, these considerations give an \Lc~ energy density of 
\begin{equation}\label{Eq:uL}
 u_{Li}=\frac{\hbar \omega_i}{2}\frac{1}{LA+\ell^\prime_i A},
\end{equation} 
where
\begin{equation}\label{Eq:ellprime}
 \ell^\prime_i\equiv \ell n_g \frac{1-R}{1+R-2\sqrt{R}\cos(\phi_i)}.
\end{equation} 

Thus, the total field energy of the pseudo-mode may be written as
\begin{equation}
 U_{f\ell}=\ell A \sum_{\phi=-\pi}^{\phi=\pi}\frac{\hbar \omega_i}{2}\frac{1}{LA+\ell^\prime A}\frac{1-R}{1+R-2\sqrt{R}\cos(\phi_i)}.
\end{equation}

As $L$ grows large, the free spectral range of the \llc~($\Delta \omega_{L+\ell}$) falls below the minimum relevant resolution of the system and we may replace the sum over modes by a spectral integral without sacrificing accuracy.  Thus,
\begin{equation}\label{Eq:U4}
\begin{split}
 U_{f\ell}\approx \ell A \int_{\phi=-\pi}^{\phi=\pi}
                               &\frac{\hbar \omega}{2}\frac{1}{LA+\ell^\prime(\omega) A} \times \\
                  &\frac{1-R}{1+R-2\sqrt{R}\cos(\phi(\omega))}\frac{1}{\Delta\omega_{L+\ell}}d\omega.
\end{split}
\end{equation}

We can calculate $\Delta \omega_{L+\ell}$ by noting that the round-trip phase $\phi$ for any cavity must change by $2\pi$ between two adjacent resonances, giving
\begin{equation}\label{Eq:FSRexp}
 \Delta \phi=2\pi=\phi_{n+1}-\phi_{n}=\Delta \omega\frac{d\phi}{d\omega}+\frac{\Delta \omega^2}{2}\frac{d^2 \phi}{d\omega^2}+ \dots~.
\end{equation}
Since $\Delta \omega_{L+\ell}$ is small for large $L$, \eq{Eq:FSRexp} reduces to 
\begin{equation}\label{Eq:FSRapprox}
\Delta \omega_{L+\ell} \approx 2 \pi d\omega_{L+\ell}/d\phi_{L+\ell}
\end{equation} 
 when applied to the \llc.
Using the definition of the group index and the \llc~geometry, we find
\begin{equation}\label{Eq:dphi}
 \frac{d\phi_{L+\ell}}{d\omega_{L+\ell}}=\frac{2 L}{c}+\frac{2 \ell}{c} n_g\frac{1-R}{1+R-2\sqrt{R}\cos(\phi_\ell)}.
\end{equation} 
Combining Eqs. ~(\ref{Eq:FSRapprox}) and ~(\ref{Eq:dphi}) and using the definition \ref{Eq:ellprime} gives
\begin{equation}\label{Eq:FSRfinal}
 \Delta \omega_{L+\ell} \approx \frac{2 \pi c}{2(L+\ell^\prime(\omega))}.
\end{equation}

Substituting \eq{Eq:FSRfinal} into \eq{Eq:U4} gives
\begin{equation}\label{Eq:U5}
  U_{f\ell}\approx 
    \frac{\ell}{\pi c} 
    \int_{\phi=-\pi}^{\phi=\pi}
      \frac{\hbar \omega}{2}
      \frac{1-R}{1+R-2\sqrt{R}\cos(\phi(\omega))}d\omega.
\end{equation} 
We notice here that the term $L+\ell^\prime(\omega)$ resulting from the expression for $d\phi_c/d\omega_c$ cancels exactly with a similar effective length in the expression for the energy density.  This is a particular example of what may be a more general correspondence between the derivative $d\phi/d\omega$ and the electromagnetic storage capacity of lossless dispersive elements.

To simplify \eq{Eq:U5} we now make the transformation
\begin{equation}
 d\omega\rightarrow\frac{c}{2 \ell n_g} d\phi
\end{equation} 
to obtain
\begin{equation}\label{Eq:U6}
U_{f\ell}\approx 
          \frac{1}{2 \pi}
          \int_{\phi=-\pi}^{\phi=\pi}
          \frac{\hbar \omega}{2 n_g}
          \frac{1-R}{1+R-2\sqrt{R}\cos(\phi(\omega))}
          d\phi.
\end{equation} 

To further simplify \eq{Eq:U6}) we make a third approximation by pulling $\omega$ and $n_g$ out of the integral.
 Doing so gives us
\begin{equation}
U_{f\ell}\approx 
      \frac{\hbar \omega_0}{2n_g} 
          \frac{1}{2\pi}
          \int_{\phi=-\pi}^{\phi=\pi}
          \frac{1-R}{1+R-2\sqrt{R}\cos(\phi)} 
          d\phi,
\end{equation} 
and since
\begin{equation}
 \int_{\phi=-\pi}^{\phi=\pi}
          \frac{1-R}{1+R-2\sqrt{R}\cos(\phi)} 
          d\phi=2\pi,
\end{equation} 
we find that
\begin{equation}\label{Eq:noise}
U_{f\ell}\rightarrow \frac{\hbar \omega}{2 n_g},                                               
\end{equation}
which is just a restatement of \eq{Eq:vapor-field-energy}.

The preceding derivation leading to \eq{Eq:noise} may be altered as follows to avoid reliance on any particular expression for the total electromagnetic energy density in a dispersive medium.
Equation~(\ref{Eq:uL}) gives the energy density of a \llc~mode in the \Lc assuming a particular form for the energy density in the \lc.  Rewriting \eq{Eq:uL} without specifying a dispersive energy density gives
\begin{equation}\label{Eq:uL2}
 u_{Li}=\frac{\hbar \omega_i}{2}\frac{1}{LA+\ell A \eta},
\end{equation} 
where we make no assumptions on $\eta$, other than that it is finite.  As $L$ grows large, any finite value for $\eta$ eventually becomes irrelevant:
\begin{equation}\label{Eq:uLi_bigL}
 \lim_{L\to\infty}u_{Li}=\frac{\hbar \omega_i}{2}\frac{1}{LA}.
\end{equation} 
Through an identical process, the free spectral range of the \llc, given by \eq{Eq:FSRapprox}, becomes 
\begin{equation}\label{Eq:Delta_omega_bigL}
 \lim_{L\to\infty}\Delta \omega_{L+\ell} = \frac{2 \pi c}{2L}.
\end{equation} 
Limits ~(\ref{Eq:uLi_bigL}) and ~(\ref{Eq:Delta_omega_bigL}) may be used in place of Eqs. ~(\ref{Eq:uL}) and ~(\ref{Eq:FSRfinal}) with no change to the overall argument, yielding the final result given by \eq{Eq:noise} without invoking a particular form the energy density of an anomalously dispersive medium.

\eq{Eq:noise} is valid only where loss/gain and group velocity dispersion are both negligible over the cavity resonance.  For media with substantial dispersion, this limits its applicability to narrow spectral ranges.  For cavities exhibiting strong normal dispersion, this presents little difficulty because dispersion may narrow the resonance bandwidth of an already high-Q cavity.  
However, for the white light cavity in particular, $n_g\rightarrow 0$ and the vacuum field energy predicted in \eq{Eq:noise} diverges.  The open cavity model shows that this divergence comes is not due to higher buildup, but to a diverging bandwidth.  In practice, the bandwidth of a given white light resonance is limited by group velocity dispersion (GVD) and by the cavity finesse.  In the next subsection we explore the relationship between these two quantities in a specific white-light cavity realization.

\subsection{Cavity noise in a symmetric gain-doublet system}
Causality dictates that every anomalously dispersive cavity will exhibit both group velocity dispersion (GVD) and loss and/or gain.  These quantities could vanish at line center and so might be ignored if the resonance linewidth were sufficiently narrow.  As an anomalously dispersive cavity more closely approaches a white light cavity, the cavity linewidth grows.  A white light cavity would correspond to an infinitely wide linewidth.  Any physically reasonable model of a white light cavity resonance must include GVD.

A proper consideration of the total cavity noise must also account for the gain of the white light medium.  Causality dictates that a medium that is both anomalously dispersive and transparent at a frequency $\omega_o$ must exhibit gain somewhere not too distant from $\omega_o$.  This gain may be difficult to maintain in a high-finesse cavity without inducing lasing.  The difficulty may not be circumvented by making mirror reflectivity spectrally dependent without altering the anomalously dispersive nature of the cavity mode.  This gain adds noise to our system.  However, in the following calculations we will ignore its effect to show that anomalous dispersion alone is sufficient to require significant noise increases in a white light cavity.  Because we will ignore gain in the paragraphs to come, the analysis contained in them will underestimate the total noise associated with the cavity mode.

\begin{figure}
 \centerline{\includegraphics[width=9cm]{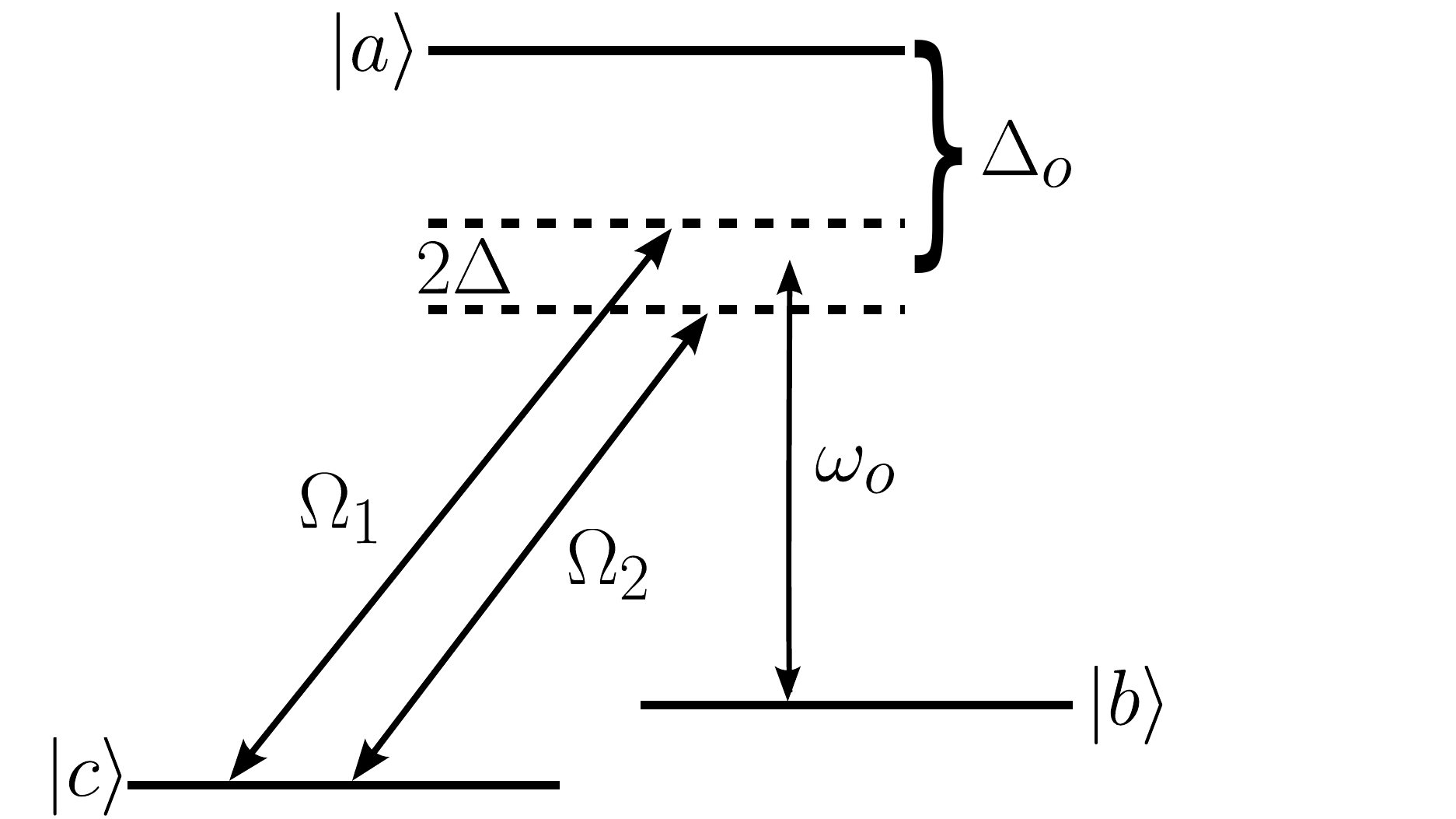}}
 \caption{\label{Fig:chi} Energy level diagram for the bichromatic Raman system.  Pump lasers represented by lines $\Omega_1$ and $\Omega_2$ provide symmetric gain lines around $\omega_o$.  Detuning from $|a\rangle$ is large ($\Delta_o >>2\Delta$).}
\end{figure}

\begin{figure}
 \centerline{\includegraphics[width=9cm]{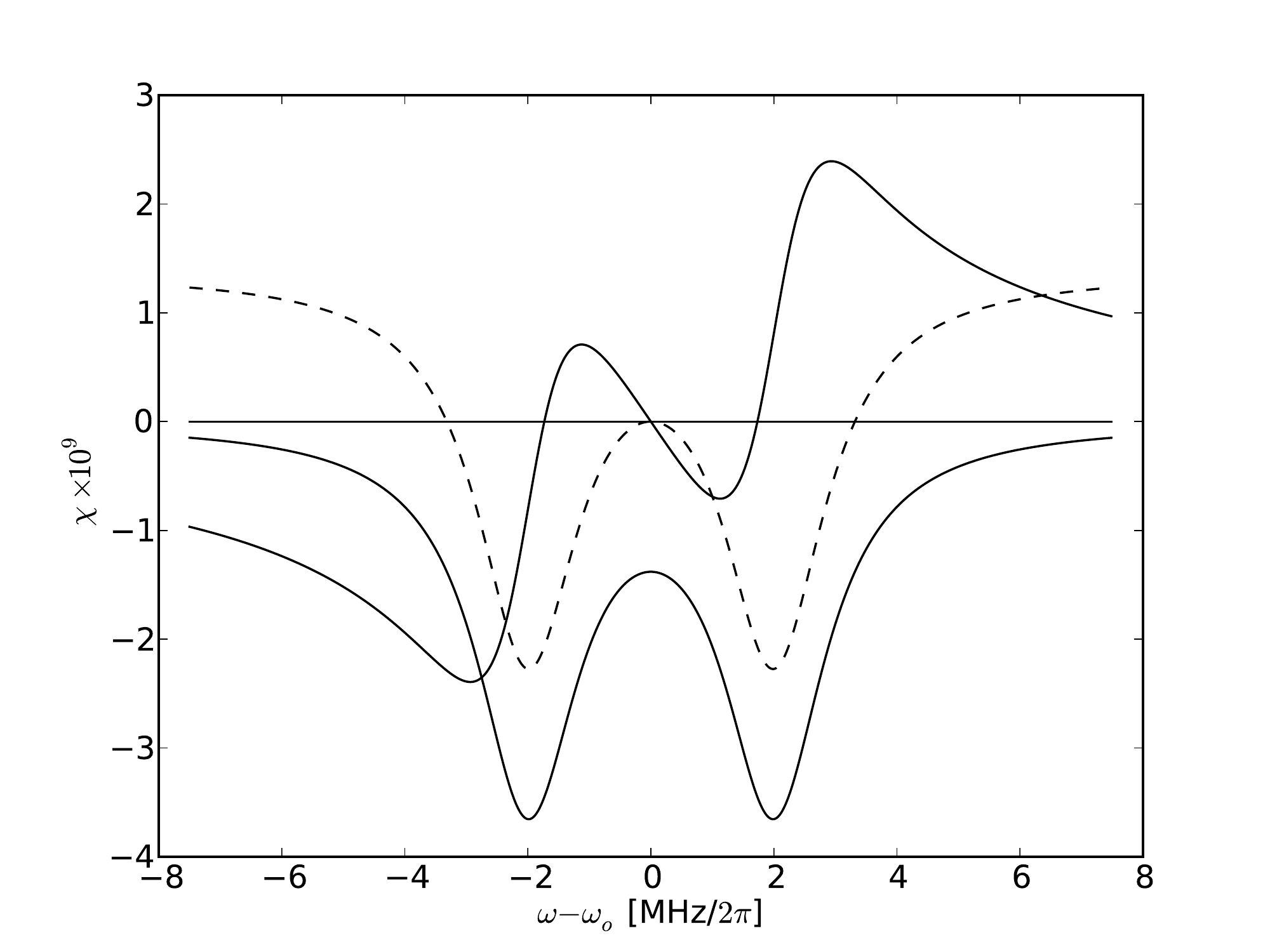}}
 \caption{\label{Fig:Raman} The susceptibility of a typical bichromatic Raman system.  The asymmetric line represents real susceptibility and the solid symmetric line represents the gain of the Raman system.  The dashed line represents the total gain in the presence of a broadband absorber whose strength is chosen to match the Raman gain at line center.}
\end{figure}

We will now solve for the pseudomode field energy given by \eq{Eq:U4} for a cavity that is homogeneously filled with a medium whose energy level diagram is depicted in Fig.~\ref{Fig:Raman}.  This medium produces a gain-doublet, as originally proposed by Steinberg and Chiao \cite{Steinberg-Chiao-1994}, accomplished through a bichromatic Raman system as experimentally realized by Wang, Kuzmich, and Dogariu \cite{Wang-Kuzmich-Dogariu-2000}.  We take the two Raman gain lines to be centered about the frequency $\omega_o$ and separated by a distance of $2\Delta$.  The value for $\Delta$ may be tuned dynamically according to experimental expedience.  The linear susceptibility ($\chi(\omega)$) may then be represented as \cite{Dogariu-Kuzmich-Wang-2001, Sun-Shahriar-Zubairy-2009} 
\begin{equation}
 \chi(\omega)=\frac{M_1}{\omega-\omega_o-\Delta+i\Gamma}+\frac{M_2}{\omega-\omega_o+\Delta+i\Gamma},
\end{equation} 
where $\Gamma$ is the Raman transition line width, and the rates $M_1$ and $M_2$ are given by
$M_j=N(|\mu_{ab}\cdot\hat e|^2/2\hbar \epsilon_o)(|\Omega_j|^2/\Delta_o^2),(j=1,2)$  \cite{Sun-Shahriar-Zubairy-2009}.
These rates $M_1$ and $M_2$ depend on $N$, the number density of participating atoms, on $\mu_{ab}\cdot \hat e$, the dipole interaction between the exciting fields and the primary Raman transition, and on the Rabi frequencies of the two exciting fields.  This last dependence means that they may be dynamically controlled by choosing the intensity of the pump lasers.  Symmetry minimizes gain and group velocity dispersion at line center when the values of $M_1$ and $M_2$ are chosen such that $M_1=M_2=M$.  For a given set of values $\omega_o$, $\Delta$, and $\Gamma$, $M$ may be chosen such that $n_g=0$ at $\omega_o$.

Figure~\ref{Fig:chi} plots the susceptibility as a function of frequency for a typical bichromatic Raman system where $M_1=M_2=M$ is chosen to minimize $n_g$ at line center. 
The dashed line indicates that this system may be combined with a broad-band absorber to roughly eliminate the net gain at line center.  
This effect was approximated in the original experiment by Wang, Kuzmich, and Dogariu \cite{Wang-Kuzmich-Dogariu-2000}.  
Loss similarly played an important role in the more recent experiments reported by Pati \emph{et al.} \cite{Pati-Salit-Salit-Shahriar-2007}.

\begin{figure}
 \centerline{\includegraphics[width=9cm]{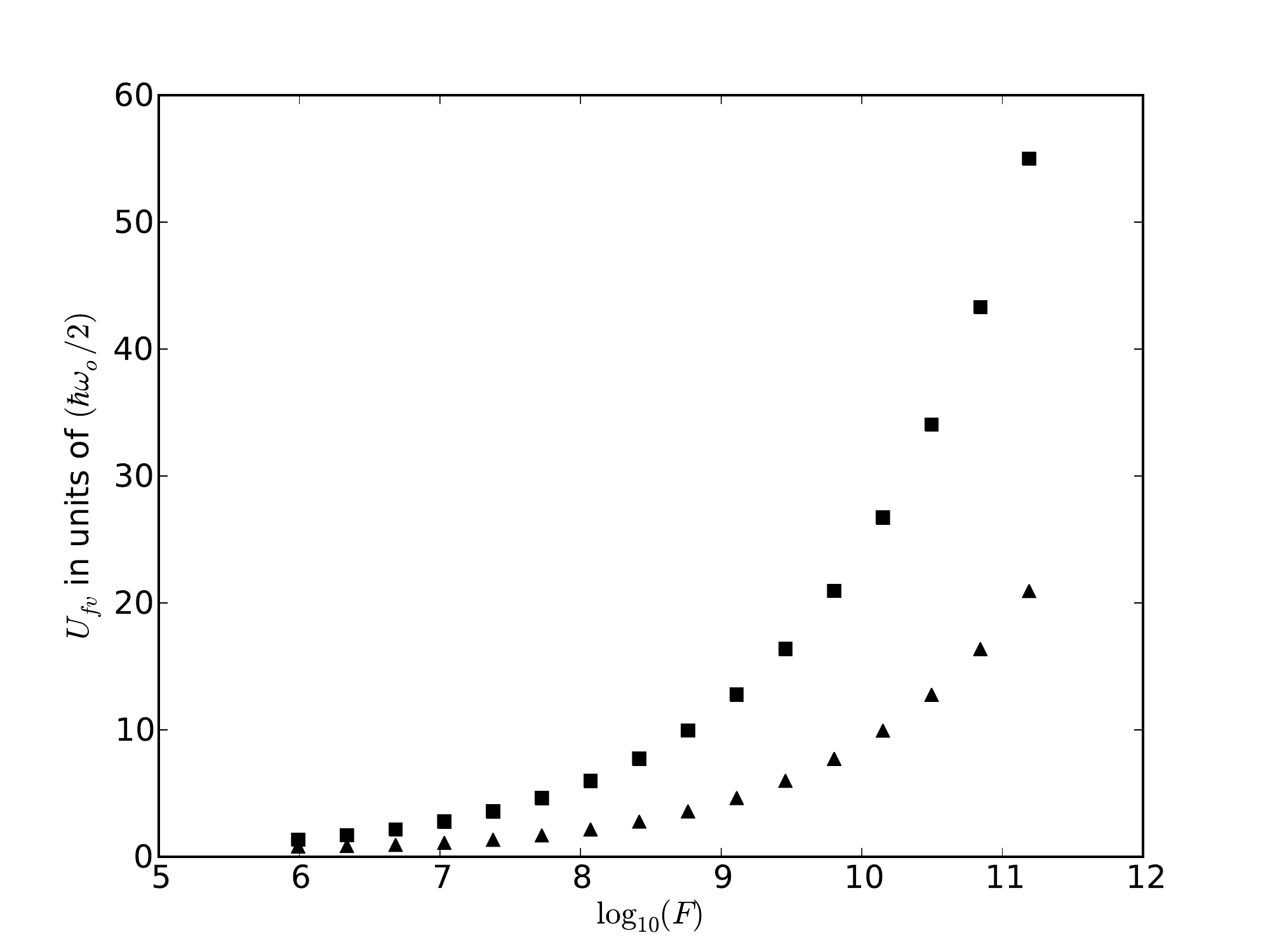}}
 \caption{\label{Fig:finesse} $U_{fv}$ is calculated for cavity lengths of 1m (squares) and 0.25m (triangles) for varying values of finesse using \eq{Eq:U5}}
\end{figure}

Figure~\ref{Fig:finesse} shows total cavity vacuum field noise for increasing finesse values as calculated from \eq{Eq:U5}.  
For both series of points, we used the values: $\Gamma=1\times 10^6$Hz/$2\pi$, $\Delta=2\times 10^6$Hz/$2\pi$, $M\approx 3.45\times 10^{-3}$Hz/$2\pi$, $\omega_o\approx 2.42\times 10^15$Hz/$2\pi$, which were taken from the literature \cite{Pati-Salit-Salit-Shahriar-2007,Sun-Shahriar-Zubairy-2009} and then adapted to avoid negative group velocities near the gain lines (Negative group velocities near the gain lines lead to extra resonances in the wings.  Causality dictates that these resonances exist at frequencies where the gain is higher or the loss lower than at line center and leads to unnecessarily large noise terms.  In many cases, it is also expected to lead to lasing.  This effect rules out the steady state stability of a large class of anomalously dispersive cavities.) We see that for both cavities the noise increases as higher finesse confines resonance to frequencies where group velocity dispersion has less effect.  As finesse becomes infinite, so does $U_{fv}$ despite the fact that \eq{Eq:U5} neglects any contribution of the gain of the active medium to the survival factor.

\subsection{Anomalous dispersion, spontaneous emission, and laser linewidth}
For a dipole emitter coupled to a large number of electromagnetic modes, the rate of spontaneous emission into one particular mode, for example the cavity mode, is proportional to the square of the electric field associated with that cavity mode.  Then \eq{Eq:E} suggests that the spontaneous emission rate into a particular cavity mode will scale with the squared electric field, which scales inversely with the group index.  When a narrow-band approximation applies, this is the case.  Thus, \eq{Eq:noise} provides a simple explanation for line narrowing Agarwal predicted in the case of lasing without inversion \cite{Agarwal-1991}; the line width is already sufficiently narrow compared to the linewidth of the emitting particle such that further narrowing provides a simple effect on the spontaneous emission rate and the laser linewidth will scale with the group velocity.  This result, although not explicitly applied in Agarwal's paper, can explain his results. However, the result is not new. For example, the role of the group velocity in determining laser linewidth was understood by Henry \cite{Henry-1982}, who framed the laser linewidth in terms of a ratio between the spontaneous emission rate and the number of photons in a mode and got an expression for the laser linewidth which is implicitly proportional to the group velocity.  

This same result applies to anomalously dispersive resonances, provided that the cavity linewidth is still narrow compared to the line shape of the gain medium.  If this condition is not met, then the broadening due to anomalous dispersion is limited by the line shape of the gain medium.  In any case, anomalous dispersion leads to an increase in laser linewidth just as it does in the linewidth of an empty cavity.

The difference between our prediction for the effect of anomalous dispersion on laser linewidth and that of Shahriar \emph{et al.} can be remedied if we redo their calculation while taking into account the effect of anomalous dispersion on the cavity photon lifetime.  Following previous authors\cite{Dorschner-Haus-Statz-1980}, they used a formula for the quantum limited linewidth that is equivalent to
\begin{equation}
 \Delta \omega_{laser}=\frac{1}{\tau_c}\sqrt{n},
\end{equation}
where $\tau_c$ was a photon lifetime and $n$.  They took $\tau_c$ to be the photon lifetime of the evacuated cavity.  However, the photon lifetime for the anomalously dispersive cavity differs from that of the evacuated cavity in their case by the factor $1/n_g$, where $n_g$ is a cavity averaged group index.  If this extra factor is taken into account, then their prediction agrees with ours.

\section{Conclusion}
We have shown that the benefits achieved through intracavity anomalous dispersion, i.e. high bandwidth combined with high buildup, come only at the cost of increased electromagnetic field noise.  An infinite finesse combined with a perfect white light mode would lead to an infinite amount of field energy in the ground state of the cavity mode.  Relaxing the infinite finesse condition to allow for cavity loss broadens the mode and relaxes the divergence, but still suggests a substantial increase in quantum field noise.
\chapter{Summary and Outlook}
\begin{quote}
 Give me a place to stand on, and I will move the Earth.
\end{quote}
--Archimedes

According to Pappus \cite{Pappus-340}, Archimedes used the phrase in the above quote to express his enthusiasm for the principle of leverage.  This dissertation is premised on the idea that the availability of controllable dispersion combined with only moderate absorption or gain provides us with a new physical lever.  The range of physically viable group index values in relatively lossless media is now so extreme that it is hard to express this range succinctly.  The group index can be made many orders of magnitude larger or smaller than its vacuum value.  Its sign may be changed and many more orders of magnitude are accessible on the negative side.  The question that we have posed is whether this lever can have interesting theoretical and experimental consequences.  Chapters ~\ref{ch:momentum}, ~\ref{ch:nonstationary}, and ~\ref{ch:cavity} can be seen as explorations of this question.  To answer these questions well, we had to come to a basic understanding of the meaning of dispersion.  Hopefully, such an understanding was partially developed through Chapters ~\ref{ch:group-velocity} and ~\ref{ch:energy-density}.

Dispersion is, at its heart, a geometrical concept.  The definition of the group velocity, $v_g=d\omega/dk$, is completely geometrical, and has geometrical consequences.  In Chapter ~\ref{ch:group-velocity}, we found that the spectrally averaged group index may be used to find the velocity of the planes of constant phase difference associated with copropagating plane waves.  Since the position of a pulse is the consequence of interference between plane waves, the group index drives the velocity of the pulse.  We joined with Rayleigh, Craven, and Candler of old in noting the important role that dispersion plays in spatial interferometers, and noted that dispersion alters the map between changes in frequency and changes in wavelength.  We briefly overviewed some of the most basic techniques by which dispersion may be controlled and combined with limited loss and gain and vanishing group velocity dispersion for a narrow spectrum.

Although it is possible to combine strong dispersion with minimal loss or gain, this is only possible to do for a narrow band of frequencies.  For such a narrow band of frequencies and in the absence of permanent loss or gain, we found ourselves able to make simple statements about the relationships between dispersion and the distribution of energy between field and medium.  In Chapter ~\ref{ch:energy-density}, we found that when strong dispersion is combined with negligible loss, the form of the electromagnetic energy density must be altered to account for dispersive energy exchange between the field and the medium.  In the case of anomalous dispersion, the total energy is negative.  We noted that this negative energy density is just a representation of energy that has left the medium but that will later return to it.  Following Peatross, Ware, and Glascow \cite{Peatross-Ware-Glasgow-2001}, we invoked the instantaneous spectrum as an explanation for this exchange.  

Energy exchange must be accompanied by momentum exchange.
In Chapter~\ref{ch:momentum} we saw that because of the effect of dispersion on the division of wave energy between field and medium, the nondispersive forms of the momenta associated with light propagating through a medium need to be adapted if dispersion is taken into account.  We saw how the non-scattering force of an electromagnetic wave in a dispersive medium on an embedded particle may be partitioned into two components, one that is proportional to a temporal gradient and another that is proportional to a spatial gradient.  By modulating the amount of dispersion in a system, we may modulate the ratio between temporal and spatial gradients, and so modulate the force and displacement of an embedded particle.

Not only is it possible to set the spectral derivative of the refractive index to a particular value, but it is possible to alter this value in real time.  
In Chapter~\ref{ch:nonstationary} we explored the transformations wrought upon a pulse by a nonstationary medium that is also dispersive.  We introduced a set of rules based on a simple symmetry that can be seen as a generalization of Snell's Law and of the reflecting Doppler effect.  Using this set of rules, we showed that dispersion may be used to modulate the frequency response of a pulse to a time dependent refractive index and the Doppler shift of a pulse reflecting off of a moving interface.  We also showed how time-dependent dispersion may be coupled with spatially dependent dispersion to scale a given pulse longitudinally and temporally (and so spectrally) over several orders of magnitude without otherwise altering the shape of the pulse envelope.  We then introduced specific boundary conditions that might be associated with slow changes to a medium and compared, for those boundary conditions, the effects of altering phase velocity in a nondispersive medium with the effects of altering group velocity in a dispersive medium on basic quantities like energy density, photon number, Canonical momentum density and Abraham momentum density and listed the comparison in Table~\ref{Tab:comp}.

In Chapter~\ref{ch:cavity}, we explored the effect of dispersion on the total vacuum field energy associated with a single cavity mode.  We argued first on the basis of the dispersive energy density that was discussed in Chapter~\ref{ch:energy-density}, and then based only on the geometric properties of a dispersive medium that the vacuum field energy is modulated by dispersion using a modes-of-the-Universe approach.  When the group index is increased by a factor of $2$ and other optical quantities are left unchanged, the spectral width of the resonance decreases by a factor of $2$ while the peak of the resonance is unchanged.  The result is a 2-fold decrease in the total quantum field noise associated with that mode.  In Appendix~\ref{appen:Lg}, we extend this idea to other features of a cavity resonance.  We see that in many respects the effect of scaling the group index associated with a particular cavity mode is exactly the effect that we would expect if we could scale the cavity length while preserving modal resonance at a particular frequency.

As was mentioned in Chapter \ref{ch:introduction-and-outline}, Professor Robert Boyd's group at the University of Rochester is currently planning experiments to test some aspects of the theory developed in Chapter~\ref{ch:momentum}.  The consideration of dispersion adds clarity to the ongoing discussion of optical momentum in a medium and I believe that there is room for additional contributions relating to the interplay between dispersion and momentum in a medium.  When dispersion is considered, two experimentally relevant formulations for the momentum of a single photon appear, one that depends on the group index and the other that depends on the phase index.  These two formulations for the momentum are related to two separable concepts, those of impulse transfer, and energy transport.

It may be interesting to use dispersion as a lever to increase or dampen the frequency response of a pulse to the temporal changes in the refractive index of a medium as was proposed in Chapter~\ref{ch:nonstationary}.  Because of the bandwidth limitations associated with non-lossy media exhibiting controlled dispersion, it is unlikely that temporal changes in the group velocity of a medium will lead to the generation of either extremely long or extremely short pulses.  However, it should be feasible to change the bandwidth and longitudinal extent of a pulse over one or two orders of magnitude while leaving the pulse envelope otherwise unchanged.  This would be interesting to pursue experimentally.  In addition, many of the results in Chapter~\ref{ch:nonstationary} are predicated on plane waves interacting with a surface that is sufficiently large and flat such that Snell's Law would hold if the surface were stationary.  For surfaces approaching the wavelength of light, it will be necessary to consider diffractive effects.  Similar ``temporally diffractive'' effects are to be expected for interactions between surfaces and pulses that are limited in temporal extent.  The spectral broadening associated with such effects may depend on the relative velocities of a given surface and wave in interesting ways.

Finally, the effect of the dispersion on the quantum field noise associated with a single cavity mode needs to be checked experimentally, particularly for cavities where the average group index approaches zero.  For this case, we predict enhanced spontaneous emission into the cavity mode so long as the empty cavity resonance is narrower than the free space line shape of the emitting particle.  This could have practical application where there is a need for fast emission into a particular spatial mode and merits exploration.
\chapter*{Appendices}
\appendix
\chapter{Forms of the group index}\label{appen:ng}
Throughout this dissertation, many formulations of the $n_g$ are used.  This appendix is dedicated to justifying and collecting these formulations.  

Although group velocity is not well defined over a finite spectrum in dispersive media, it has a sensible definition over an infinitesimal one,
\begin{equation}\label{vg}
v_g=\frac{d\omega}{dk}.
\end{equation}
Thus, the lack of a definition over a finite spectrum is just a manifestation of the fact that $v_g$ is a function of frequency.  The group index is defined, in analogy with the phase index, as \cite{Craven-1945}
\begin{equation}
n_g=\frac{c}{v_g}.
\end{equation}
Using Eq. \ref{vg}, we can write
\begin{equation}
n_g=c\frac{dk}{d\omega}.
\end{equation}
Remembering that $k=n \omega/c$, we can write
\begin{equation}
n_g=\frac{d}{d\omega}(n\omega).
\end{equation}
Expanding, we get
\begin{equation}
 n_g=n+\omega \frac{dn}{d\omega}.
\end{equation} 
We can rewrite the second term as a partial logarithmic derivative, giving
\begin{equation}
 n_g=n+\frac{dn}{d\ln(\omega)}.
\end{equation} 
One interesting thing about the logarithmic derivative is that constant multipliers are irrelevant.  Thus, $d\ln(\omega)=d\ln(\nu)=d\ln(E)=d\ln(\sigma_0)$, where $E$ here refers to the energy per photon, $\hbar \omega$.  Another interesting thing is that an inversion corresponds to a sign change.  Thus, $d\ln(\omega)=-d\ln(\lambda_0)$.

This explains a few other ways of writing $n_g$:
\begin{eqnarray}
n_g=n+\nu \frac{dn}{d\nu},\\
n_g=n+E \frac{dn}{dE},\\
n_g=n+\sigma \frac{dn}{d\sigma_0},\\
n_g=n-\lambda_0\frac{dn}{d\lambda_0},
\end{eqnarray}
Each of these relates simply to a derivative expression:
\begin{eqnarray}
n_g=\frac{d}{d\nu}(n\nu)\label{ngnu}\\
n_g=\frac{d}{dE}(n E),\\
n_g=\frac{d}{d\sigma_0}(n\sigma_0),\\
n_g=-\lambda_0^2\frac{d}{d\lambda_0}(n/\lambda_0).\\
\end{eqnarray}

All of these expressions are essentially in terms of the refractive index versus the photon (or polariton) energy.  It is also illuminating to understand $n_g$ in terms of what it does to the interaction between the photon energy and the wavelength.  We begin by expanding $d\lambda/d\nu$.
\[
\frac{d\lambda}{d\nu}=\frac{d}{d\nu}\left(\frac{c}{n\nu}\right)
\]
Recalling Eq. \ref{ngnu}, we find
\[
\frac{d\lambda}{d\nu}=-\frac{c n_g}{(n\nu)^2}.
\]
We rewrite this as
\begin{equation}\label{lnln}
n_g=-n\frac{d\ln\lambda}{d\ln\nu}.
\end{equation}
This highlights the way in which dispersion alters the mapping between changes in spectral and spatial frequencies.

The total phase associated with the propagation of a wave with spatial frequency $k$ through a length $\ell$ of a homogeneous medium is given by $\phi=k \ell$.  The total phase difference accrued by two neighboring frequencies gives another way to view the group index.
\[
\frac{\phi}{d\omega}=\ell \frac{d k}{d \omega}=\frac{l n_g}{c}.
\]
Thus,
\begin{equation}
 n_g=\frac{c}{l}\frac{d \phi}{d \omega}.
\end{equation} 
A higher group index means that phase differences between frequency components increase more quickly with space.

The quantity $n_g/n$ is in some sense a better measure of dispersion than $n_g$ itself.  Since $n_g=n+\omega dn/d\omega$, we may write
\begin{equation}
 \frac{n_g}{n}=1+\frac{d\ln n }{d\ln \omega}.
\end{equation}  

In an absorptionless, isotropic medium, $n=\sqrt{\epsilon_r \mu_r}$.  Because a logarithmic derivative does not change with its arguements are scaled by constants, we may then write
\[
\frac{n_g}{n}=1+\frac{d\ln (\epsilon \mu)}{d \ln \omega}.
\]
This may be expanded as
\begin{equation}\label{Eq:ng-over-n}
 \frac{n_g}{n}=1+\frac{1}{2} \frac{d\ln\epsilon}{d\ln\omega}+\frac{1}{2} \frac{d\ln \mu}{d\ln\mu}.
\end{equation}

\chapter{Simulating a long cavity with a short one}\label{appen:Lg}

The optical characteristics of a cavity can be modified over a narrow frequency range upon the introduction of a medium that exhibits electromagnetically induced transparency (EIT). Cavity characteristics of free spectral range, resonance bandwidth, quality, photon lifetime, and mode volume all change in a consistent fashion. In each case, the property is changed as though the cavity were increased in length by a factor of the group index $n_g$. With group indices of $\sim 10^{7}$ possible by EIT media, a 100~$\mu$m cavity so filled would exhibit the optical properties of a 1 km cavity.

\section{Introduction}
Many important parameters of an optical cavity, including free spectral range, resonance bandwidth, quality, photon lifetime, and mode volume \cite{Gerry-Knight-2005,Scully-Zubairy-1997}, depend on the cavity length. We point out that when a cavity is filled with a medium that exhibits electromagnetically induced transparency (EIT)~\cite{Fleischhauer-Imamoglu-Marangos-2005}, the vacuum versions of the expressions for these properties for frequencies near EIT resonance must be altered in a way that amounts to introducing a new effective length of the cavity. Consistently, the new length becomes the evacuated length multiplied by the group index of the medium. The cavity properties change accordingly. Using EIT, researchers have generated group indices greater than $10^7$ \cite{Hau-Harris-Dutton-Behroozi-1999}. A 100\,$\mu$m cavity filled with such a medium would then assume the characteristics of a 1\,km cavity. In this Appendix, we introduce the relevant features of EIT and then discuss each of the mentioned parameters in turn, showing how the effective length in each case is the evacuated length scaled by the group index.

\section{EIT and the group index}

Optically, an ideal EIT\footnote{EIT is a phenomenon where a medium ordinarily opaque at a particular frequency can be made transparent on application of a laser at a second frequency. Typically, the effect is created by connecting three discrete levels through two one-photon transitions, each stimulated by a laser. A ``probe'' excites the normally opaque transition, and a ``drive'' completes an overall two-photon process to the third level.  We consider the drive-beam frequency to be fixed and resonant with the transition it drives.  Further, we take the medium as manipulated by the drive to be the EIT medium. Our concern is then with the properties of the EIT medium at the probe frequency. \emph{Resonance} refers to the case where the probe is tuned to its corresponding one-photon resonance and the medium is transparent. \emph{Detuning} will mean detuning of the probe beam only, and imperfect transparency.} medium \cite{Fleischhauer-Imamoglu-Marangos-2005}
is like a dispersive vacuum. In a narrow frequency band about the EIT resonance, the absorption coefficient ($\alpha$) and the refractive index ($n$) are near their vacuum values of 0 and 1, respectively, while the group index ($n_g$) can be arbitrarily large \cite{Harris-Field-Kasapi-1992,Hau-Harris-Dutton-Behroozi-1999, Liu-Dutton-Behroozi-Hau-2001}.  The phase velocity approaches the vacuum speed of light ($c$) while the group velocity ($v_g$) can be much slower than $c$, according to $v_g=c/n_g$.  

A pulse flowing from a vacuum into an EIT medium will be spatially compressed along the axis of propagation by a factor $n_g$. The envelope of the pulse will change abruptly in slope at the interface but will be otherwise continuous because of the identical refractive indices on either side of the boundary. No energy is reflected at the interface, for example; all is transmitted. That the pulse can be foreshortened and unchanged in field intensity is owed to the fact that an EIT medium stores and exchanges energy with the propagating electromagnetic field. The total energy density, $\mu$, in an EIT medium is apportioned in general by \cite{Harris-Hau-1999,Peatross-Ware-Glasgow-2001}   
\begin{equation}\label{energy_density}
  \mu = \mu_m + \mu_f = n_g\mu_f,
\end{equation}
where $\mu_f$ is the energy density of the electromagnetic field, and $\mu_m$ is the density of energy stored by the medium.  Thus, an EIT-filled cavity may hold much more energy than is evident in the internal intensity alone.  

The foreshortening of the pulse highlights another effect of $n_g$ that is relevant to cavities.  Pulse propagation can be seen as the propagation of a set of phase relationships between its frequency constituents. On entering a high $n_g$ medium, the pulse compresses because these phase relationships change $n_g$ times more quickly for each increment of displacement.  Thus an EIT-filled cavity not only holds $n_g$ times more energy for a given $\mu_f$ but also brings about changes in phase relationships between different frequencies as though the cavity were $n_g$ times as long.  These two related effects are at the heart of the ability of an EIT cavity of length $L$ to mimic an evacuated cavity of length $n_g L$.

\section{Free spectral range}

The free spectral range, $\Delta \nu_{\textrm{FSR}}$, of a cavity is the frequency difference between successive longitudinal cavity resonances and is usually given by \cite{Verdeyen-1994,Guenther-1990}
\begin{equation}\label{eqn-fsr-nodisp}
	\Delta \nu_{\textrm{FSR}}=\frac{c}{nL},
\end{equation}
where $L$ is the round trip length of the cavity and $n$ is presumed constant. To take the EIT medium into account, we constitute $\Delta \nu_{\textrm{FSR}}$ afresh.

A cavity resonance occurs when the total phase $\phi$ accrued by light propagating over $L$ is an integer multiple of $2\pi$, or when $kL=q 2\pi$, where $k = 2 \pi/\lambda_m$ is the angular wavenumber, $\lambda_m$ is the wavelength of the light in the medium, and $q$ is an integer. The difference in $k$ between adjacent resonances at $q+1$ and $q$ is $2\pi/L$, which in terms of frequency $\nu = k c/(2 \pi n)$ becomes
\begin{equation}\label{res-diff}
	\nu_{q+1}n(\nu_{q+1})-\nu_q n(\nu_q)=c/L.
\end{equation}
A Taylor expansion of Eq.~\ref{res-diff} with
\begin{eqnarray*}
    \nu_{q+1} &=& \nu_q+\Delta\nu_{\textrm{FSR}}\\
    n(\nu_{q+1}) &=& n(\nu_q)+ \Delta\nu_{\textrm{FSR}}\frac{\delta n}{\delta \nu}+ \dots
\end{eqnarray*}
leads to the first-order approximation
\begin{equation}\label{eqn-fsr-eit}
	\Delta\nu_{\textrm{FSR}}=\frac{c}{n_g L}
\end{equation}
since $n_g(\nu_q) = \nu_q n^\prime\left(\nu_q\right) + n(\nu_q)$. When $n_g \rightarrow n$, as is ordinarily the case in transparent media, the more generally applicable Eq.~\ref{eqn-fsr-eit} reduces to Eq.~\ref{eqn-fsr-nodisp}. 
For an empty cavity $n = 1$ and $\nu_{q+1}$ exceeds $\nu_q$ by $\Delta\nu_{\textrm{FSR}} = c/L$ as seen from Eqs. \ref{eqn-fsr-nodisp} and \ref{res-diff}. Use of the resonance condition and a vacuum dispersion relation of $d\phi/d\nu = 2\pi(L/c)$ yields the same result. 
When filled with an EIT medium, $\phi$ instead changes by $d\phi/d\nu = 2\pi n_g (L/c)$. The difference $\nu_{q+1} - \nu_{q}$ becomes $c/(n_g L)$, and Eq. \ref{eqn-fsr-eit} is recovered, which shows $\Delta\nu_{\textrm{FSR}}$ to be that of an empty cavity of $L$ stretched by $n_g$, a factor equal to the increased dispersion of the EIT medium.         

\section{Resonance bandwidth}
The resonance bandwidth, $\Delta \nu_{r}$, of a cavity is given by 
\begin{equation}\label{eq-resbandwidth}
	\Delta \nu_{r} = \frac{\Delta \nu_{\textrm{FSR}}}{F}, 
\end{equation}
where $F$ is the finesse of the cavity \cite{Verdeyen-1994,Guenther-1990}. The finesse depends only on mirror reflectivities and loss within the medium and is unaffected by changes in $L$ to the extent that losses remain constant. Assuming a fixed finesse, we see from Eqs.~\ref{eqn-fsr-eit} and ~\ref{eq-resbandwidth} that insertion of an EIT medium leads to
\begin{equation}\label{eq-resbandwidth_EIT}
	\Delta \nu_{r} = \frac{c/F}{n_g L}, 
\end{equation}
that is, the equivalently sharp ${\Delta \nu_{r}}$ of an evacuated cavity of length $n_g L$.

The reduction of $\Delta \nu_{r}$, and consequent improvement in spectral discrimination, of a cavity through EIT was predicted by Lukin \textit{et al.} \cite{Lukin-Fleischhauer-Velichansky-1998}. Wang \textit{et al.} \cite{Wang-Goorskey-Xiao-2000} later demonstrated a narrowing of $\Delta \nu_{r}$ and suggested potential uses of intracavity EIT in areas of high-resolution spectroscopy, laser-frequency stabilization, and the generation of non-classical light.

\section{Quality}
The cavity quality $Q$ for a resonant mode can be expressed by 
\[\label{eq_first_Q}
	Q = \nu/\Delta\nu_{r},
\]
which through use of Eq.~\ref{eq-resbandwidth_EIT} leads to  
\begin{equation}\label{eq_Q_EIT}
	Q=F \frac{n_g L}{\lambda}
\end{equation}
in the case of a cavity filled with an EIT medium. 

The quality may also be cast as the ratio between the steady total cavity energy $U$ and the cavity energy lost per optical cycle according to
\[\label{Q-energy}
	Q=2 \pi \nu\frac{U}{-dU/dt}.
\]      
From Eq.~\ref{energy_density}, $U \sim n_g\mu_f$ for a cavity of fixed volume. The loss rate of energy, however, scales with $\mu_f$ in the medium and at the mirrors, giving $dU/dt \sim \mu_f$. The quality then increases by $n_g$ because the energy stored by the EIT medium cannot be dissipated or transmitted by mechanisms of optical loss, which determine $dU/dt$.
\section{Photon lifetime}
The lifetime $\tau$ of a photon\footnote{For a highly dispersive medium like EIT, where the energy is no longer only in an electromagnetic form, we might replace the term \emph{photon} with \emph{quantum of energy}.} in a cavity is proportional to $Q$, which by Eq.~\ref{eq_Q_EIT} leads to $\tau \sim n_g L$. A separate consideration of $\tau$ better clarifies its association with the equivalent length $n_g L$.

It suffices to define $\tau$ by the round-trip time for a photon multiplied by the expected number of circulations. For an evacuated cavity we have $\tau= (L/c)(1-s)^{-1}$, where $s$ is the probability the photon will survive one round trip, and $(1-s)^{-1}$ is the number of round trips associated with a $1/e$ photon lifetime in a high-finesse cavity \cite{Verdeyen-1994}. Insertion of an ideal EIT medium will not change $s$, but will increase the round-trip time to give $\tau= (n_g L/c)(1-s)^{-1}$.  This time is equivalent to the round-trip time of an evacuated cavity with a physical length of $n_g L$.

\section{Mode volume}

In an evacuated cavity, the mode volume $V$ is the volume available for a photon mode and is proportional to the cavity length. We may figure the mode volume from $U$, the total energy within the mode, and $\mu$, the energy density at a particular point, from $V=U/\mu$ \cite{Gerry-Knight-2005,Scully-Zubairy-1997}. The mode volume finds practical use in determining the average field amplitude $A\propto\sqrt{U/V}$ of a mode given the energy $U$ it contains. When the cavity is filled with an EIT medium, however, we must bear in mind from Eq.~\ref{energy_density} that $\mu$ is divided between $\mu_m$ and $\mu_f$, leading us to change $V$ and $A$ from their empty-cavity expressions to $V=n_gU/\mu$ and $A\propto\sqrt{U/(n_gV)}$. It is natural to factor the mode volume into components of area and length. In doing so, we are led to define a new effective mode volume $V_{eff} = n_gV$ whose area is unchanged but whose length is scaled by the group index.

We may view the expanded $V$ and reduced $A$ of intracavity EIT in terms of where the photon, or equivalent quantum of energy, resides. When that energy takes the form of an electromagnetic field, we consider the field's spatial extent to define the mode volume. However, for $n_g$-times longer than present as a photon, the energy is stored within the medium. The optical-field amplitude available to affect other atoms, for example to bring about the decay of excited atoms that do not participate in EIT, must be based on $\sqrt{U/n_g}$, or equivalently a mode volume extended by $n_g$.

\section{Caveats}
While the effective length $n_gL$ controls the cavity characteristics discussed, there are other considered uses and properties of a cavity for which the length remains $L$. We draw attention to two of these. First, the effective length does not offer the extrinsic benefits of physical size. A long cavity has an enhanced sensitivity, through detectable changes to its FSR or finesse, to the refractive index or absorption of a trace gas that fills its volume. This advantage is not conferred on a short cavity through EIT. Second, the number of nodes in the cavity is unchanged by the EIT medium. Although EIT increases $n_g$ it does not alter $n$ enough to disturb the node spacing of $\lambda_m/2$ significantly.

EIT introduces a severe bandwidth constraint for high $n_g$ operation, with $n_g \gg 1$ and constant only over a narrow range of frequency $\delta\nu$ about the EIT resonance. Also, $n_g$ and $\delta\nu$ are in opposition; high $n_g$ associates with narrow $\delta\nu$. Some additional frequency range can be gained by detuning the EIT drive laser and accordingly detuning the probe to recover the overall two-photon resonance, a process that essentially shifts the center frequency of $\delta\nu$. Large tuning ranges, however, would likely be obtained only by using different energy-level configurations in the same or another medium.

Finally, in an EIT-enhanced cavity, a stable effective length will require both mechanical stability and a steady drive-laser intensity. A positive aspect of the latter dependence, however, is that one can scan the effective length of an EIT cavity over many orders of magnitude without using any moving parts.\footnote{The drive-laser intensity and $n_g$ are inversely related, with $n_g>1$ for EIT.}

\section{Conclusion}
With respect to many optical parameters, an EIT cavity of length $L$ is identical to a vacuum cavity of length $n_g L$. Because $n_g$ can be made extremely large we can simulate the properties of very large cavities with minute ones. We can narrow the resonance bandwidth well below that set by the empty-cavity finesse, for example. As $n_g$ is not difficult to control dynamically, it should be possible to sweep the effective length of a cavity quickly and easily over a large dynamic range. It is also worth considering media beyond those that exhibit EIT, where $n_g$ can approach zero and even assume negative values. In these cases, artificial cavities are created with vanishing to negative effective lengths, something impossible to do with a vacuum cavity by any manipulation. On a practical level, the relation $\Delta \nu_{\textrm{FSR}} \sim n^{-1}_{g}$ points out a simple cavity-based approach to accurate measurements of $n_g$.


\begin{thebibliography}{10}
\bibitem{rayleighnote} A proper derivation must, of course, include the particle density fluctuations,
but this leads to the same results and conclusions in the case of an ideal gas.
\end{thebibliography}

\end{document}